%% file: main.tex
\documentclass[acmtog, authorversion, nonacm]{acmartarxiv}

\acmSubmissionID{xxx}

\copyrightyear{2026}
\acmYear{2026}
\setcopyright{rightsretained}

\title{Generalized non-exponential Gaussian splatting}
\author{Sébastien Speierer}
\email{speierers@meta.com}
\orcid{}
\affiliation{%
    \institution{Meta}
    \country{Switzerland}
}

\author{Adrian Jarabo}
\email{ajarabo@meta.com}
\orcid{}
\affiliation{%
    \institution{Meta}
    \country{Spain}
}

\usepackage[capitalize,nameinlink]{cleveref}
\usepackage{graphicx,tikz,pgfplots,array}
\usepackage{svg}

\pgfplotsset{compat=1.18}

\input{src00_macros}

\begin{document}


\begin{teaserfigure}
    \centering
    \includegraphics[width=\linewidth]{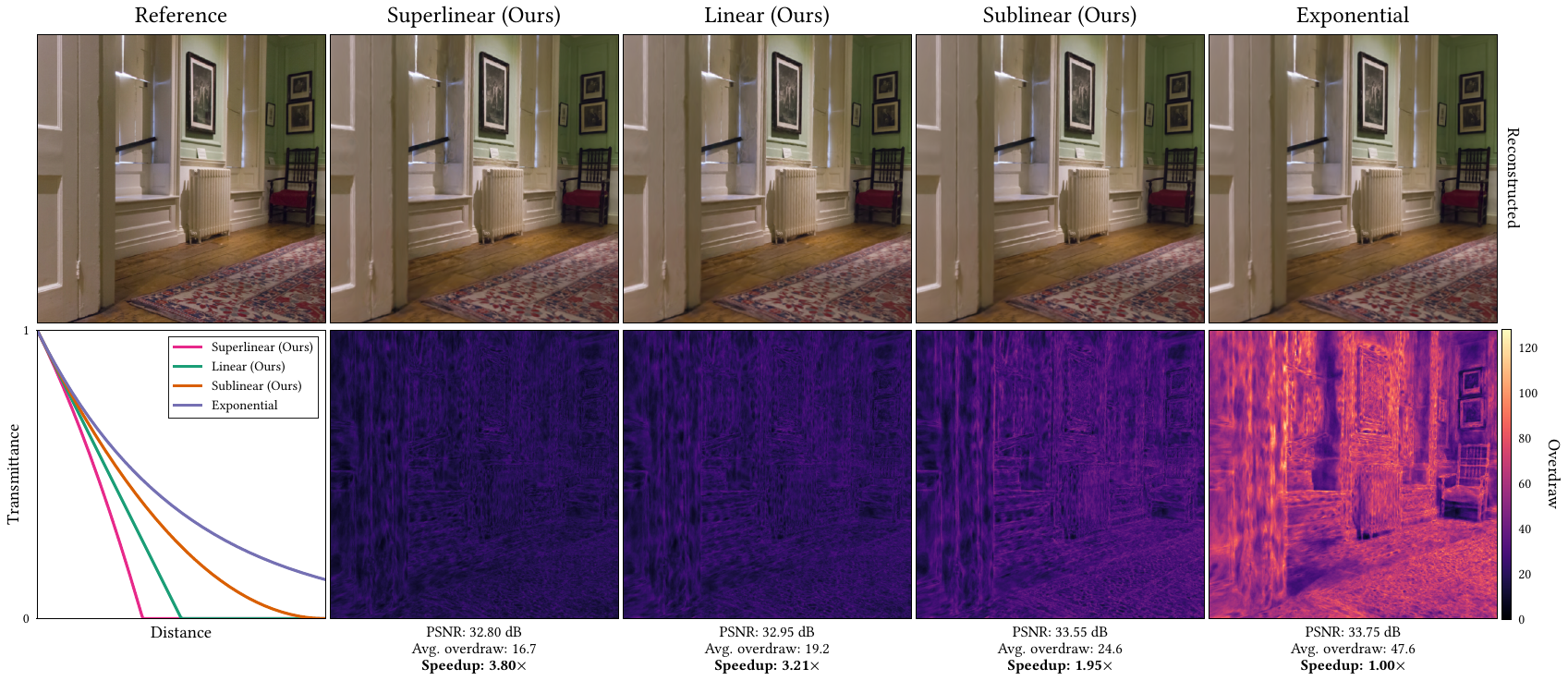}%
    \caption{%
        \textbf{Non-exponential transmittance for 3D Gaussian Splatting (3DGS).} We compare the baseline exponential transmittance model implicit in 3DGS image formation model, against our new non-exponential transmittance models on the \emph{Dr\,Johnson} scene, refined from a pretrained 3DGS asset. 
        \emph{Top row:} Ground-truth reference (left) and reconstructed images for each transmittance model. 
        \emph{Bottom row:} A plot of the transmittance decay functions (left) and per-pixel overdraw count visualizations.
        Our non-exponential models achieve comparable image quality to the baseline model, while reducing overdraws by $3$--$4\times$, directly translating to faster rendering.
      }
    \label{fig:teaser}%
\end{teaserfigure}


\begin{abstract}
In this work we generalize 3D Gaussian splatting (3DGS) to a wider family of physically-based alpha-blending operators. 3DGS has become the standard de-facto for radiance field rendering and reconstruction, given its flexibility and efficiency. At its core, it is based on alpha-blending sorted semitransparent primitives, which in the limit converges to the classic radiative transfer function with exponential transmittance. Inspired by recent research on non-exponential radiative transfer, we generalize the image formation model of 3DGS to non-exponential regimes. Based on this generalization, we use a quadratic transmittance to define sub-linear, linear, and super-linear versions of 3DGS, which exhibit faster-than-exponential decay. We demonstrate that these new non-exponential variants achieve similar quality than the original 3DGS but significantly reduce the number of overdraws, which result on speed-ups of up to $4\times$ in complex real-world captures, on a ray-tracing-based renderer. 
\end{abstract}



\maketitle

\input{src1_intro}

\input{src2_rw}
\input{src3_3dgs}

\input{src4_gen3dgs}

\input{src7_results}
\input{src6_discussion}








\bibliographystyle{ACM-Reference-Format}
\bibliography{sample}

\appendix

\renewcommand{\theequation}{A.\arabic{equation}}
\setcounter{equation}{0}

\input{srcA_3dgs2rte}

\input{srcB_adjoints}

\end{document}

%% file: src00_macros.tex
\usepackage{xcolor}

\newcommand{\Eq}[1]{Equation~\eqref{#1}}

\newcommand{\px}{\mathbf{x}}
\newcommand{\bomega}{\boldsymbol{\omega}}
\newcommand{\clampzero}[1]{\left[#1\right]_+}

%% file: src1_intro.tex
\section{Introduction}
Recent methods on radiance field rendering and reconstruction \cite{mildenhall2020nerf} have allowed photorealistic rendering of real-world scenes and humans. In particular, \emph{3D Gaussian splatting} (3DGS) \citep{kerbl20233d} has emerged as the most versatile representation, and its flexibility and efficiency have made it particularly attractive for applications including photorealistic avatars modeling~\cite{qian2024gaussianavatars,saito2024relightable}, photorealistic rendering in virtual reality~\cite{jiang2024vr-gs}, scene understanding \citep{zhou2024hugs,jang2025identity}, or generative three-dimensional graphics \citep{tang2023dreamgaussian}.

3DGS uses trainable point-based volumetric semitransparent emissive primitives, whose contribution is accumulated in each pixel using multiplicative alpha-blending. While several alternatives over the building blocks of the original technique have been proposed, including alternative emissive functions \citep{diolatzis2024n,liu2025deformable} or kernel primitives \citep{condor2025don,liu2025deformable,hamdi2024ges,huang2025deformable}, the image formation model essentially stays the same. 

In particular, the multiplicative-alpha-blending-based image formation model is a discrete form of, and in the limit converges to, the classic radiative transfer equation (RTE) \citep{lommel1889photometrie,chandrasekhar2013radiative}, which predicts an exponential transmittance under the assumption of volumes with uncorrelated particles. This contrast with nature and real-life matter, where correlation has been observed in most appearances, including clouds \citep{davis2004photon}, vegetation \citep{knyazikhin1998influence}, porous \citep{bellet2009rdfi} and granular \citep{meng2015multi} materials, or periodic crystal-like structures \citep{caglioti2003distribution}. 
This have fostered the development of new radiative transfer frameworks \citep{bitterli2018radiative,jarabo2018radiative}, that generalized the classic RTE to non-exponential regimes. 

In this work we generalize 3DGS to non-exponential transmittance regimes, by first relating it with the classic RTE, and then deriving its generalized non-exponential counterpart. 
Armed with this physically-based framework, we then derive a set of new 3DGS image formation models for the cases of linear- and quadratic-like transmittance functions, including its forward and differential modes using path-replay backpropagation~\citep{vicini2021path}. We show that these faster-than-exponential transmittance modes allow to reconstruct photorealistic scenes with similar quality than the exponential model, with the additional benefit of significantly less overdraws, which translates into speed-ups of up-to around $4\times$ in real-world scenes in our ray-tracing-based renderer compared to the baseline exponential model, as exampled in \cref{fig:teaser}. 


%% file: src2_rw.tex
\section{Related Work}

\paragraph{3D Gaussian splatting}
A vast literature exists using and extending 3D Gaussian splatting:
From the original proposal of \citet{kerbl20233d}, several works have been published improving on certain aspects of the method, including new kernels beyond Gaussians \citep{condor2025don, hamdi2024ges,huang2025deformable, liu2025deformable}, new directional emission functions replacing spherical harmonics \citep{diolatzis2024n,liu2025deformable}, spatially-varying textured emission \citep{xu2024_texturegs}, or alternative primitives including linear primitives \citep{vonlutzow2025linprim}, 3D convexes \citep{held2025convex}, triangles \citep{held2025triangle} and meshes \citep{held2025meshsplatting}. 
Other works have focused on increasing the problem dimensionality to the temporal domain \citep{wu2024_4dgs,yang2023_4dgs}, improving the robustness of the optimization process \citep{sabour2025_spotlesssplats,kulhanek2024_wildgaussians,yin2025_trackersplat,rota2024revising}, or providing scalable solutions for large scenes \citep{ren2025octreegs,feng2025flashgs}. 
All these works are based on the same exponential image formation model; our work is orthogonal to these efforts, proposing a generalized image formation model based on novel alpha-blending compositing techniques of the splats.

\paragraph{Non-exponential light transport}
Non-exponential or generalized transport have been studied in fields including radiative transfer, optics and neutron transport (see \citep{deon2022hitchhiker} for an in-depth discussion). In graphics, the first work exploring non-exponential regimes was the work by \citet{moon2007rendering} in the context of granular media, from a phenomenological approach. A more formal definition of the problem was introduced by \citet{deon2014heritage}, though practical rendering of non-exponential media was not possible until the works of \citet{jarabo2018radiative} and \citet{bitterli2018radiative}, which explored the problem from two alternative perspectives: Jarabo et al. analyzed the problem from a purely radiative perspective, while Bitterli et al. developed a path-integral-based framework. Both works proposed different non-exponential transmittance functions based on the statistical properties of the media, similar to the one used by \citet{wrenninge2017path}; this model was later extended by \citet{guo2019fractional} introducing a more general transmittance model based on fractional Gaussian fields. 
\citet{vicini2021non} exploited the additional representation power of non-exponential media to improve reconstruction of volumetric assets, which was also used by \citet{condor2022learned} to accelerate rendering of complex luminaries. Closer to us, \citet{zhou2024unified} used a linear transmittance model for closed-form integration of volumetric 3D Gaussians for physics-based rendering. 
In contrast, our work leverages non-exponential light transport for defining new alpha-blending operators in the context of infinitesimal splats.

\paragraph{Alpha blending and transparency}
From the initial work of Smith and Catmull \citep{smith1995history} on alpha compositing, alpha blending has been a fundamental tool in graphics. Porter and Duff´s \shortcite{porter1984compositing} seminal work established the mathematical foundation for compositing digital images. \citet{blinn2002theory,blinn2002practice} provided new derivations of the \texttt{over} operator and implementation considerations for various pixel formats. \citet{glassner2015interpreting} revisited the interpretation of alpha, providing a consistent and physically meaningful model of pixel structure, which we use as foundation for our model. 
Further works have generalized and formalized alpha compositing: \citet{willis2006projective} proposed a projective alpha color model, which unifies various compositing approaches using an algebraic perspective, while \citet{duff2017deep} analyzed deep compositing using Lie algebras, providing a rigorous mathematical framework for advanced compositing operations. 
Our work can be seen as a generalization of the alpha blending operator defined by an arbitrary asymptotic transmittance of semitransparent layers. 

%% file: src3_3dgs.tex
\section{Gaussian splatting}
\label{sec:3dgs}
3D Gaussian splatting~\cite{kerbl20233d} uses a set of ellipsoidal volumetric semi-transparent primitives (\emph{Gaussians})\footnote{For consistency with the related literature, during the text we will use the term \emph{Gaussian} to name the primitives, though the kernel used might be different.} to model the radiance emitted in a scene. Each primitive is represented by its directional emission $E_i(\bomega)$, center $\mathbf{c}_i$, covariance $\Sigma_i$, and opacity $\aleph_i$. The contribution of each primitive to a pixel is defined by a kernel $K(\mathbf{x} |  \{\mathbf{x}_i, \Sigma_i\})$ (usually a Gaussian, but others exist), which we use to compute the pixel opacity $\alpha_i(\mathbf{y}) = \aleph_i\cdot K(\mathbf{y} | \{\mathbf{x}_i, \Sigma_i\})$. 

Then, for a point $\px$ in space (e.g., a sensor´s pixel), the incident radiance $L(\px,\bomega)$ from direction $\bomega$ is computed as the sum of emitted radiance by a set of $N$ volumetric primitives projected to $\px$ from direction $\bomega$ as,
\begin{align}
    L_0(\px,\bomega) & = \sum_{i=1}^N E_i(\bomega) \cdot \alpha_i(\px_i) \cdot \prod_{j<i} \left( 1 - \alpha_j(\px_j) \right) \nonumber\\
    &+ L_b(\px_b,\bomega) \cdot\prod_{i=1}^N \left( 1 - \alpha_i(\px_i) \right),
    \label{eq:gaussian_splatting}
\end{align}
with $\px_i = \px-\bomega\cdot t_i$, $t_i$ the distance of the primitive $i$ to $\px$, and $L_b$ the contribution from the background. The first summand is the contribution of the Gaussians, and the second is the contribution from the background due to transparency; this second term is generally omitted, but we include it here for completeness. In the following, we remove it for simplicity. 
This formulation assumes that Gaussians are projected as billboards oriented towards the camera (\emph{splats}), which is the common image formation model in Gaussian splatting. Note that while other works have proposed working with truly-volumetric primitives~\cite{condor2025don,kheradmand2025stochasticsplats,celarek2025does}, in this work we focus on the splat-based image formation model. The splats are sorted front-to-bottom, so that $\forall i<j$ then $t_i < t_j$. In the following, we remove the spatial and directional dependency for clarity. 

\paragraph{Physical interpretation}
The image formation model in \Eq{eq:gaussian_splatting} is a multiplicative alpha-blending. In this context, we can interpret $\alpha_i$ as the area coverage of occluding matter normalized by the unit volume \citep{porter1984compositing}. This has the physical unit of m$^{-1}$, and describes the probability of being occluded in the differential thickness of the splat, analogous to what in linear transport is the \emph{differential probability of extinction} \citep{deon2022hitchhiker}. 

It is trivial to see that \Eq{eq:gaussian_splatting} is the discrete counterpart of the radiative transfer equation (RTE) \cite{chandrasekhar2013radiative} for emissive volumes, so that is equivalent to write (see proof in \cref{sec:3dgs2rte})
\begin{equation}
    L_0 = \int_0^\infty Q(\px_t) \cdot \sigma(\px_t) \cdot \overbrace{e^{-\tau_t}}^{T(\tau_t)} \, dt,
    \label{eq:rte}
\end{equation}
with $\px_t = \px - \bomega\cdot t$ with $t$ the length along the ray, $\tau_t=\int_0^t \sigma(\px_s) ds$ the optical depth, $T(\tau_t)$ the exponential transmittance predicted by the Beer-Boulder-Lambert law, and $\sigma(\px_s)$ and $Q(\px_s)$ the density and emission fields, respectively derived from the splats along the ray as 
\begin{equation}
\sigma(\px_t) = \sum_{i=1}^{N} \delta(t-t_i)\,\alpha_i \quad \text{and} \quad Q(\px_t) = \sum_{i=1}^{N} \delta(t-t_i)\,E_i,
\label{eq:sigma_and_q}
\end{equation}
with $\delta$ the Dirac delta. 

\begin{figure}[h]
  \centering
  \includegraphics[width=1\columnwidth]{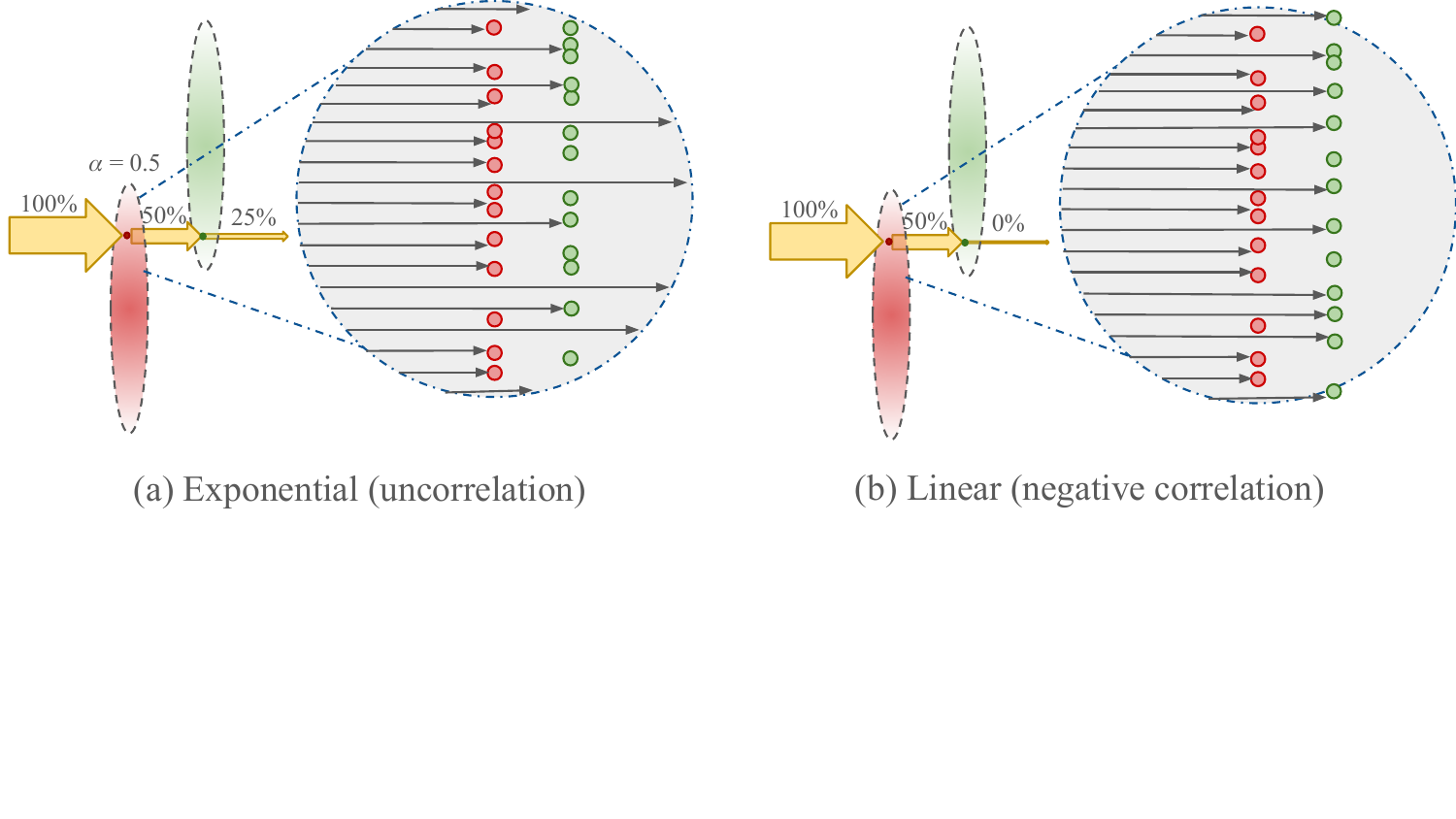} 
  \vspace{-2.25cm}
  \caption{\textbf{Physical interpretation of splats.} Splats can be thought as disk consisting of small random occluding particles. When light passes through an slab, some light is lost when hitting these particles. For uncorrelated splats (a), light occlusion in each splat is an independent stochastic process. For negatively-correlated splats (b), on the other hand, all light that passes through the first slab is occluded by the second one. }
  \label{fig:correlation}
\end{figure}

Going deeper on the physical foundations of this model, a key assumption leading us to an exponential-like behavior in \cref{eq:rte} is that the probability of extinction at each splat is an independent statistical process (see \citep{kostinski2001extinction,kostinski2002extinction}). For this to happen, the underlying coverage at each differential point in a splat needs to be fully uncorrelated from all the other splats contributing to a pixel. 
Intuitively, we can think of a splat as a disk formed by infinitesimally small random particles uniformly distributed in 2D, occluding light with probability equal to its density $\alpha_i$. Assuming two uncorrelated splats occluding $50\%$ of the light each, occlusion at each splat will be an statistically-independent process, so the total unoccluded light after passing through the two splats will be $(1-0.5)\times(1-0.5)=0.25$ (\cref{fig:correlation}, a). However, if the particles in the second splat are fully negatively correlated to the particles from the first one, all light that passes through the first one will hit particles in the second one, and the total light passing through the two splats will be $1-0.5-0.5 = 0$, forming a waterproof surface occluding all light (\cref{fig:correlation}, b). 

%% file: src4_gen3dgs.tex
\section{Non-exponential splatting}
In this section, we generalize the image formation model of 3DGS to arbitrary transmittance regimes. For that, we exploit the relationship between continuous and discrete media, and leverage the generalization of the RTE to non-exponential regimes to derive our new non-exponential splatting method. Then, we propose and evaluate a new set of non-exponential image formation models. 
\input{transmittances}

\subsection{Background: The generalized Boltzmann equation}
To account for regimes where light transport does not exhibit exponential transmittance, \citet{Larsen2007generalized} proposed the generalized Boltzmann equation (GBE), which in its integral form for emissive media reads
\begin{equation}
    L_0 = \int_0^\infty Q(\px_t) \cdot \sigma(\px_t,\tau_t) \cdot T(\tau_t) \, dt.
    \label{eq:gbe}
\end{equation}
For emissive media, the GBE is almost identical to the classic RTE with a key difference: The differential extinction probability $\sigma(\px_t,\tau_t)$ is not constant, but depends on the distance traveled by photons before being extincted. This has a major implication in the rest of the terms governing the light transport, since $\sigma(\px_t,\tau_t)$, the path-length distribution\footnote{The path-length distribution, also known as probability of extinction or chord-length distribution, is the probability of being extincted after traveling certain optical depth. Do not confuse it with the \emph{differential} extinction probability $\sigma(\px_t,t)$, which is the probability of extinction when traveling a differential step.} $p(\tau_t)$, and the transmittance $T(\tau_t)$ are directly related by 
\begin{equation}
    \sigma(\px_t,\tau_t) = \frac{p(\tau_t)}{T(\tau_t)}, \quad \text{with \quad} p(\tau_t)=-\frac{d}{dt} T(\tau_t)=-\sigma(\px_t)\cdot \frac{d}{d\tau} T(\tau_t).
    \label{eq:gbe_sigma}
\end{equation}
Note that since we are modeling a non-scattering medium Larsen´s GBE holds in our case; if we would be also considering scattering we would need to account for a more complex form to maintain reciprocity~\cite{jarabo2018radiative,bitterli2018radiative,dEon2018reciprocal}. 

\subsection{Our method}
Equipped with \cref{eq:gbe,eq:gbe_sigma} we can just simplify the GBE as the product of the path-length distribution and emission, following:
\begin{equation}
    L_0 = \int_0^\infty Q(\px_t) \cdot p(\tau_t) \, dt,
    \label{eq:gbe_p}
\end{equation}
which by using the relationship between continuous and discrete media described in \cref{eq:sigma_and_q}, we can just write the contribution of each Gaussian as a function of their emission and their probability of extinction. Including back the background contribution, the generalized Gaussian splatting image formation model becomes
\begin{equation}
    L_0 = \overbrace{\sum_{i=1}^N E_i \cdot \widebar{p}_i}^{\text{Splats}} + \overbrace{L_b \cdot \widebar{T}_{N+1}}^{\text{Background}} \,,
    \label{eq:generalized_gaussian}
\end{equation}
%
%
where $\widebar{p}_i$ is the discrete probability of extinction for Gaussian $i$ which depends on the \emph{mother} transmittance function $T(\tau_t)$ chosen for defining the model, and with $\widebar{T}_i = 1 - \sum_{j<i} \widebar{p}_j$ the discrete transmittance operator. 

\newcommand{\cellsize}{1.55cm}  
\newcommand{\datapath}{figures/transmittance_experiments/}  
\input{figures/transmittance_experiments/transmittance_experiments}

This equation is general, and works with any arbitrary base transmittance function as soon it is physically correct. We detail later how to choose such base transmittance functions, but a key aspect that needs to be considered is that it always needs to be in greater or equal to zero. With that constraint, we can just define the discrete probability of extinction of each Gaussian as
\begin{equation}
    \widebar p_i =
    \begin{cases}
      -\frac{d}{dt}T(\widebar\tau_i) & i < k, \\
      1 - \sum_{j<k}\widebar p_j & i = k, \\
      0  & \text{otherwise, } \\
    \end{cases}
    \label{eq:nexp_p}
\end{equation}
with $\widebar\tau_i=\sum_{j<i}\alpha_j$ the discrete optical depth, and $k$ the primitive that saturates transmittance so that $\sum_{i=0}^k p_i \geq 1$ and $\sum_{i=0}^{k-1} p_i < 1$. Note that allowing for a primitive to saturate transmittance is a key benefit from the generalized image formation model: Given the multiplicative nature of the original Gaussian splatting, transmittance never saturates to zero, and thus it is impossible to create fully-opaque appearances except for splats with $\alpha_i=1$. In contrast, in our model this fully-opaque behavior can be easily achieved by choosing the appropriate base transmittance function.

\subsection{Choosing a transmittance function}
Our image formation model allows to use any arbitrary base transmittance function $T(\tau_t)$ to define the blending of the different splats. However, for the transmittance to be physically meaningful there are two conditions that must be taken into account: Given the physical definition of transmittance $T(\tau_t) = 1 - \int_0^t p(\tau_s) ds$, and since the path-length distribution $p(\tau_t)$ is a probability distribution, then a) transmittance is always in the interval $T(\tau_t) \in [0,1]$ with $T(0)=1$, and b) it is always non-increasing. 
In addition to these two physical constraints, c) we impose that the slope of the transmittance function at $\tau_t=0$ is $p(0) = \frac{d}{dt}T(0) = \frac{d}{dt}\tau_t=\sigma(0)$, which effectively makes the contribution of the first splat being always its opacity $p_1 = \alpha_1$. Thus, the transmittance function only plays a role when blending several splats.

With these three constraints, we propose six transmittance functions, listed in \cref{tab:transmittances}, for which we analyze their performance in toy-examples. Later in the radiance field experiments (\cref{sec:results}) we focus on the \emph{linear} and \emph{quadratic} mother transmittance functions. 

\renewcommand{\datapath}{figures/combined_experiments/}  
\input{figures/combined_experiments/combined_experiments}

\subsection{Evaluation}
\paragraph{Transmittance} We first evaluate the emerging transmittance when blending a large set of low-opacity Gaussians. This allows us to validate that the discrete transmittance $\bar T_i=1-\sum_{j<i}p_j$ converges to the mother continuous transmittance function $T(\tau_t)$. \Cref{fig:transmittance} shows the results for transmittance functions with increasing effective decay rate. As the effective decay rate increases, the appearance gets more opaque, with transmittance quickly saturating to zero. This results in significantly less overdraws for dense areas.

\paragraph{Blending} \Cref{fig:blending} shows the results of blending three colored Gaussians using different mother transmittance functions $T(\tau_t)$, in two scenarios. In both cases, the results are similar, showing sharper blending of the back Gaussians when using faster-than-exponential transmittance functions, while at the same time obtaining more opaque appearances, as shown in the inverse transmittance images.

%% file: transmittances.tex
\begin{table*}[t]
\caption{Summary of the mother transmittance functions $T(\tau_t)$ and the corresponding discrete probabilities $\bar{p}_{i<k}$ for $i<k$ used in our experiments. The \emph{linear} transmittance was proposed by \citet{jarabo2018radiative} to stochastically model negatively-correlated opaque surfaces. The \emph{quadratic} is an empirical transmittance, that generalizes the linear transmittance providing sublinear and superlinear decays. The \emph{blended} transmittance reformulates the blended transmittance proposed by \citet{vicini2021non}, which is a linear combination of linear and exponential transmittance. Finally, the \emph{power-law} model was proposed by \citet{davis2004photon} to model fractal fluctuations in positively-correlated media. The operator $\clampzero{x}=\max(x,0)$ is the clamping operator. }
\label{tab:transmittances}
\centering
\begin{tabular}{|l|l|c|c|}
\hline
\textbf{Name} & Parameters & \( T(\tau_t) \) & \( \widebar{p}_{i<k} \) \\
\hline
Exponential & N/A &\( \exp(-\tau_t) \) & \( \alpha_i \cdot \prod_{j<i}(1-\alpha_j) \) \\
Linear & N/A & \( \clampzero{1-\tau_t} \) & \( \alpha_i \) \\
Quadratic & $c\in[-0.5, \infty)$ & \( \clampzero{1 - \tau_t - \frac{c}{2} \cdot \tau_t^2} \) & \( \alpha_i \cdot (1 + c \cdot \widebar{\tau}_i) \) \\
Blended & $\gamma\in[0,1]$ & $\clampzero{\text{lerp}(1-\tau_t, \exp(-\tau_t),\gamma)}  $ & $\alpha_i \cdot (1-\gamma +\gamma \cdot \prod_{j<i}(1-\alpha_j))$ \\
\citep{vicini2021non} & $\gamma\in[0,1]$ & $\text{lerp}(\clampzero{1-\tau_t}, \exp(-\tau_t),\gamma)  $ & $\text{lerp}( \widebar{p}_{i<k}^{\text{lin}}, \widebar{p}_{i<k}^{\text{exp}}, \gamma)$ \\
Power-Law & $v\in(-1,\infty]$ & $\clampzero{(1+\tau_t\cdot v)^{-1/v}}$ & $\alpha_i \cdot (1 + \tau_i \cdot v)^{-\frac{1+v}{v}}$\\
\hline
\end{tabular}
\end{table*}

%% file: figures/transmittance_experiments/transmittance_experiments.tex
%
%

\begin{figure*}[htbp]
\centering
\setlength{\tabcolsep}{0pt}
\begin{tabular}{@{}cc@{}c@{}c@{}c@{}c@{}c@{}c@{}c@{}c@{}}
 & \shortstack{\textbf{\tiny Power-Law}\\\textbf{\tiny ($v=2$)}} & \textbf{\tiny Exponential} & \shortstack{\textbf{\tiny Quadratic}\\\textbf{\tiny ($c=-0.5$)}} & \shortstack{\textbf{\tiny \citet{vicini2021non}}\\\textbf{\tiny ($\gamma=0.5$)}} & \shortstack{\textbf{\tiny Blend}\\\textbf{\tiny ($\gamma=0.5$)}} & \textbf{\tiny Linear} & \shortstack{\textbf{\tiny Quadratic}\\\textbf{\tiny ($c=0.5$)}} & \shortstack{\textbf{\tiny Quadratic}\\\textbf{\tiny ($c=1$)}} & \shortstack{\textbf{\tiny Power-Law}\\\textbf{\tiny ($v=-1$)}} \\[1ex]
\raisebox{\dimexpr\cellsize/2-\height/2\relax}{\rotatebox{90}{\scriptsize\textbf{$1 - \bar T_{N+1}$}}} & \includegraphics[width=\cellsize,height=\cellsize]{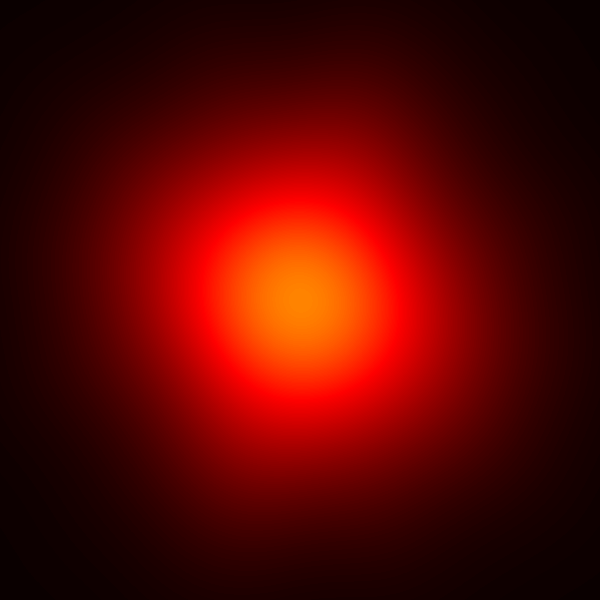} & \includegraphics[width=\cellsize,height=\cellsize]{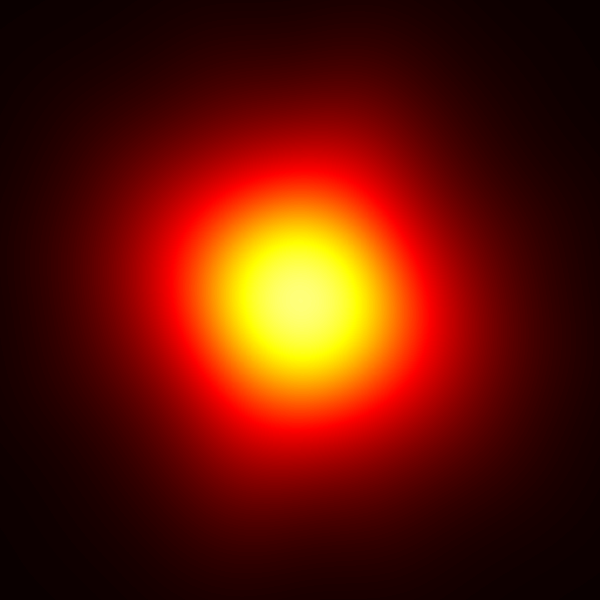} & \includegraphics[width=\cellsize,height=\cellsize]{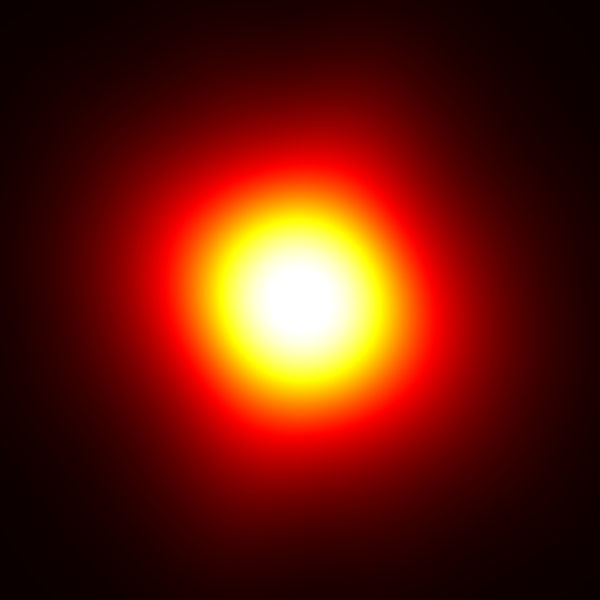} & \includegraphics[width=\cellsize,height=\cellsize]{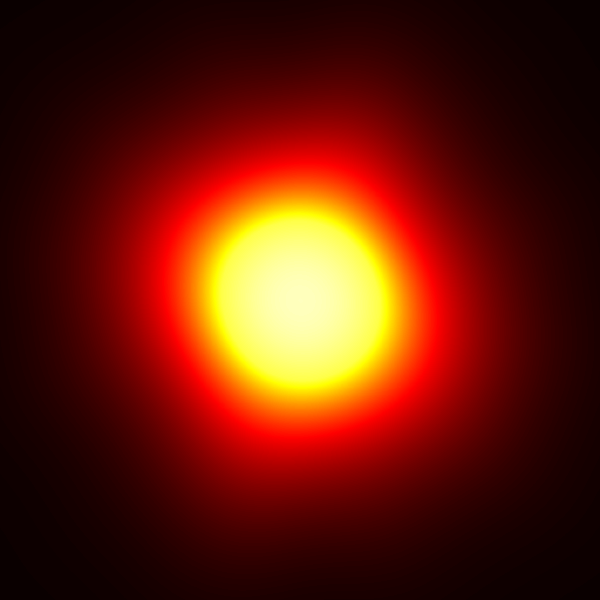} & \includegraphics[width=\cellsize,height=\cellsize]{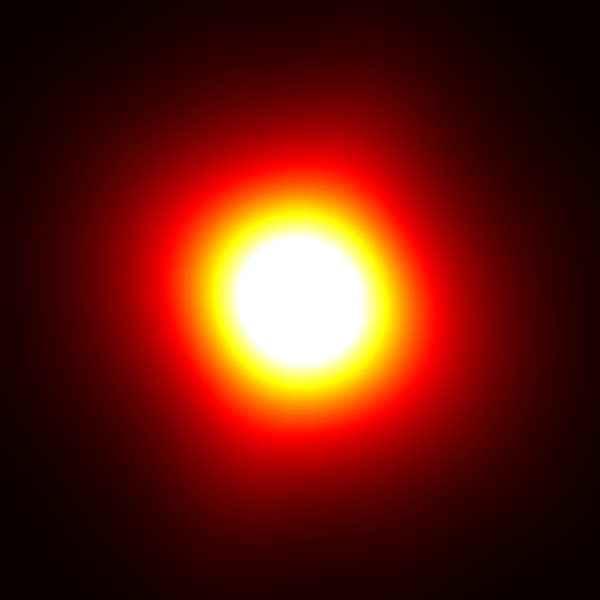} & \includegraphics[width=\cellsize,height=\cellsize]{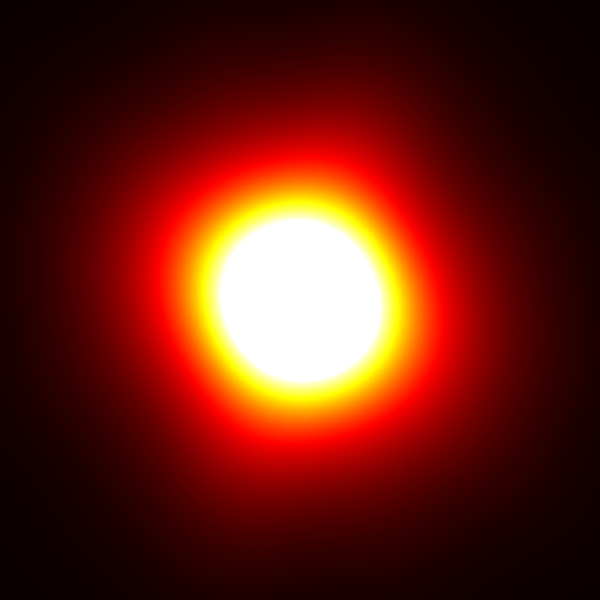} & \includegraphics[width=\cellsize,height=\cellsize]{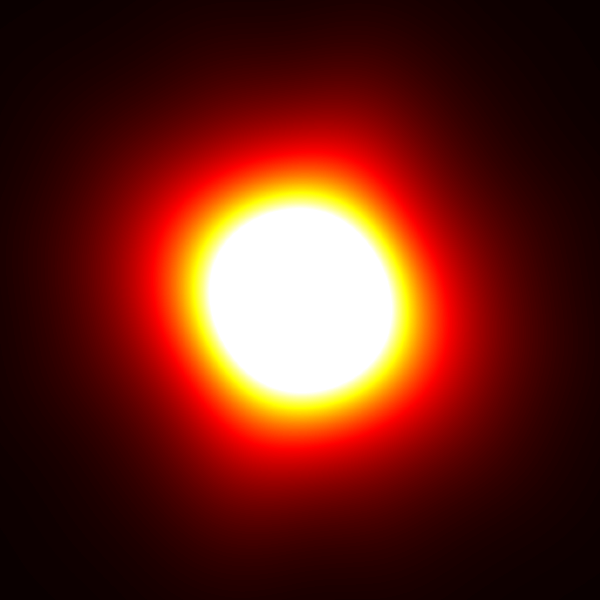} & \includegraphics[width=\cellsize,height=\cellsize]{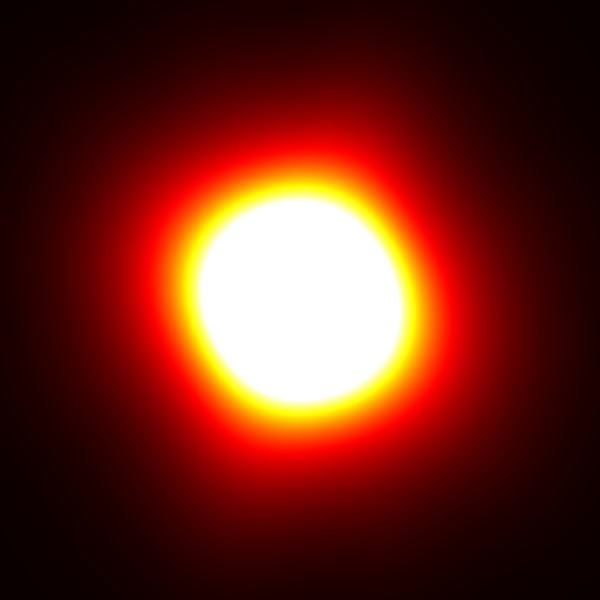} & \includegraphics[width=\cellsize,height=\cellsize]{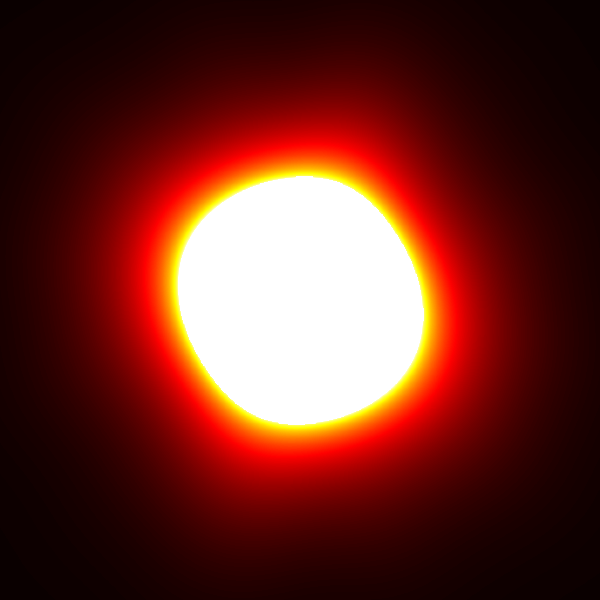} \\[0.5ex]
\raisebox{\dimexpr\cellsize/2-\height/2\relax}{\rotatebox{90}{\scriptsize\textbf{Overdraws}}} & \includegraphics[width=\cellsize,height=\cellsize]{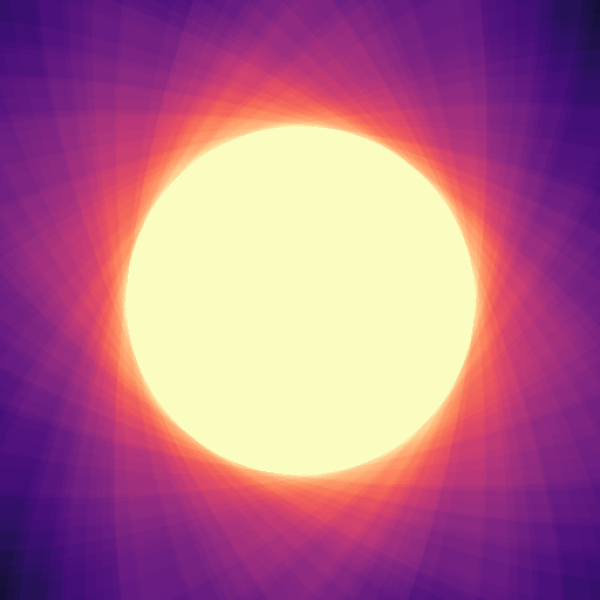} & \includegraphics[width=\cellsize,height=\cellsize]{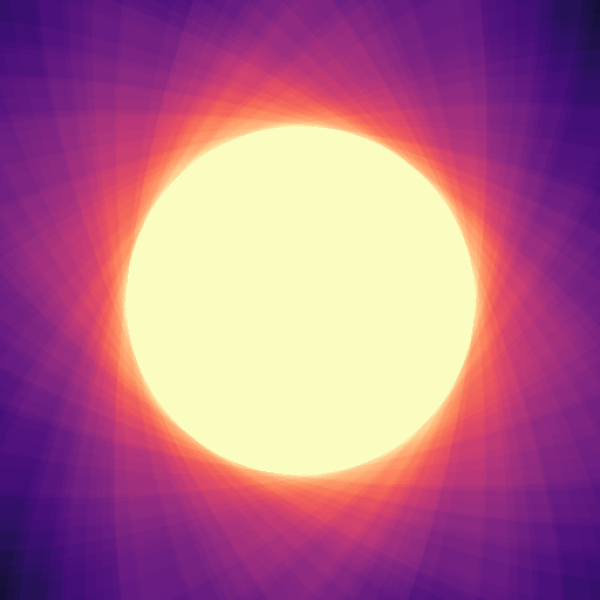} & \includegraphics[width=\cellsize,height=\cellsize]{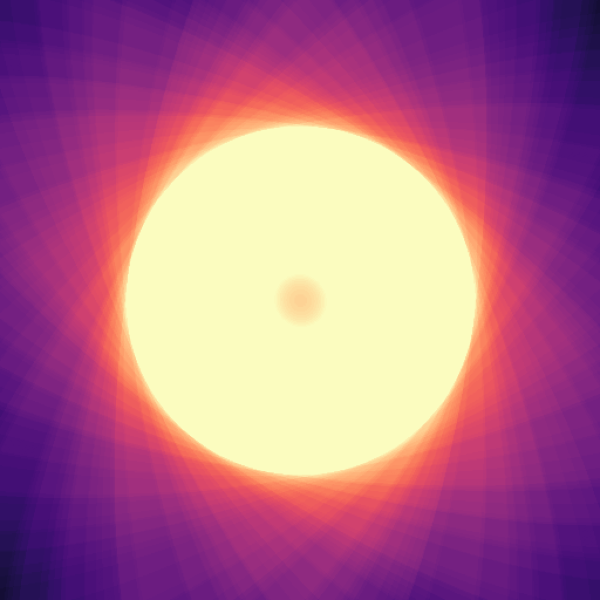} & \includegraphics[width=\cellsize,height=\cellsize]{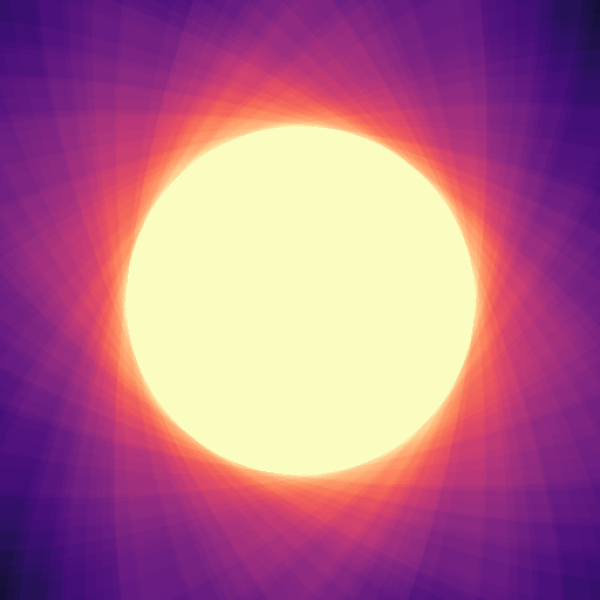} & \includegraphics[width=\cellsize,height=\cellsize]{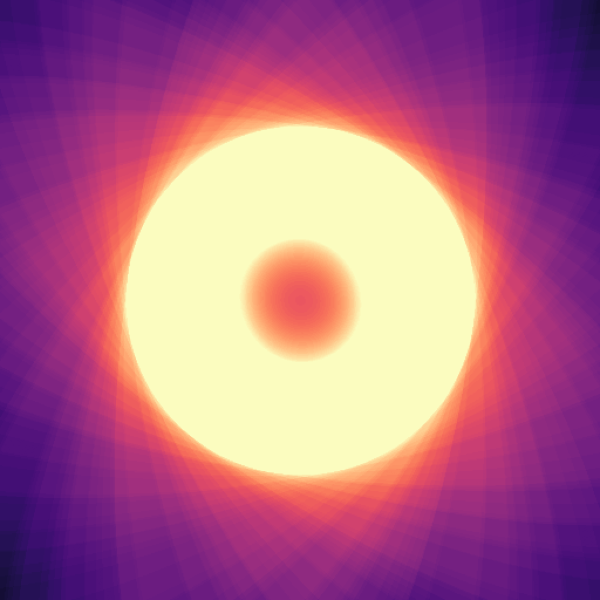} & \includegraphics[width=\cellsize,height=\cellsize]{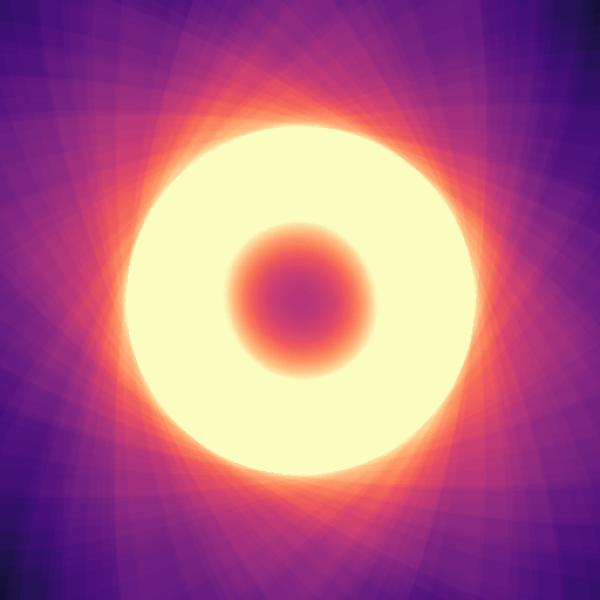} & \includegraphics[width=\cellsize,height=\cellsize]{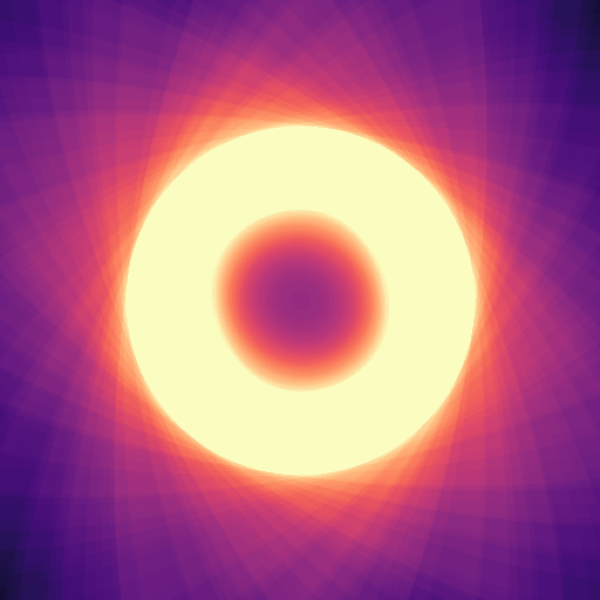} & \includegraphics[width=\cellsize,height=\cellsize]{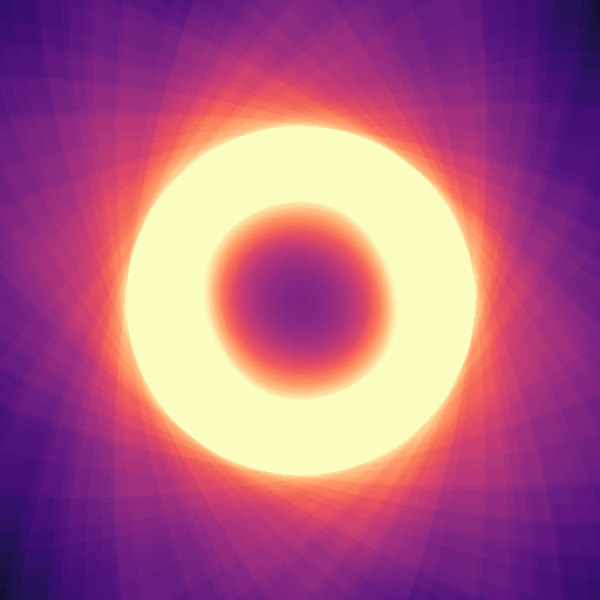} & \includegraphics[width=\cellsize,height=\cellsize]{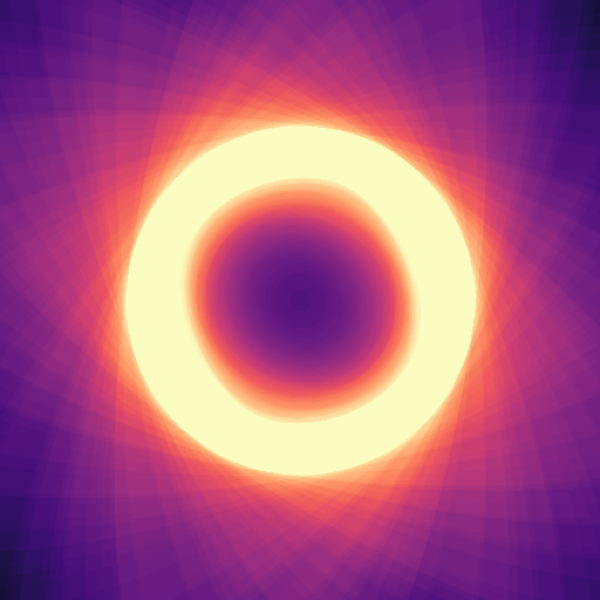} \\[0.5ex]
\raisebox{\dimexpr\cellsize/2-\height/2+0.7cm\relax}{\rotatebox{90}{\scriptsize\textbf{Transmittance}}} & \input{\datapath n100_random_center_prob_PowermLaw_v2.tex} & \input{\datapath n100_random_center_prob_Exponential.tex} & \input{\datapath n100_random_center_prob_Quadratic_cm0p5.tex} & \input{\datapath n100_random_center_prob_citetvicini2021non_gamma0p5.tex} & \input{\datapath n100_random_center_prob_Blend_gamma0p5.tex} & \input{\datapath n100_random_center_prob_Linear.tex} & \input{\datapath n100_random_center_prob_Quadratic_c0p5.tex} & \input{\datapath n100_random_center_prob_Quadratic_c1.tex} & \input{\datapath n100_random_center_prob_PowermLaw_vm1.tex}
\end{tabular}
\vspace{-0.4cm}
\caption{\textbf{Transmittance comparison.} We render 100 elliptical Gaussians with random rotation with opacity $\aleph = 0.02$, distributed uniformly in depth, for nine different transmittance functions, and compute the inverse transmittance $1 - \bar T_{N+1}$ per pixel (top) and the number of overdraws until transmittance saturates (middle, colormap goes from black to yellow). 
The bottom plots show the theoretical mother transmittance $T(\tau_t)$ of each model (green) compared with the baseline exponential (dashed red), and the discrete transmittance $\bar{T_i}$ at the rendered splats (blue dots). Our predicted discrete transmittance matches the theoretical one, which validates that our generalized model for splatting converges to the continuous generalized RTE. 
The faster is the decay, the less overdraws are required to saturate, and the more opaque is the reconstructed appearance. }
\label{fig:transmittance}
\end{figure*}

%% file: figures/transmittance_experiments/n100_random_center_prob_PowermLaw_v2.tex
\begin{tikzpicture}
\begin{axis}[width=\cellsize,height=\cellsize,scale only axis,xlabel={$\tau$},xlabel near ticks,xticklabels={},yticklabels={},xmin=0,xmax=2.02,ymin=0,ymax=1.05,grid=major,ytick={0,0.25,0.5,0.75,1}]
\addplot[red, thick, dashed, opacity=0.7] coordinates { (0.0000,1.00000000) (0.0412,0.95961368) (0.0824,0.92085842) (0.1237,0.88366834) (0.1649,0.84798023) (0.2061,0.81373343) (0.2473,0.78086973) (0.2886,0.74933328) (0.3298,0.71907047) (0.3710,0.69002986) (0.4122,0.66216209) (0.4535,0.63541981) (0.4947,0.60975754) (0.5359,0.58513168) (0.5771,0.56150036) (0.6184,0.53882343) (0.6596,0.51706234) (0.7008,0.49618009) (0.7420,0.47614121) (0.7833,0.45691162) (0.8245,0.43845864) (0.8657,0.42075091) (0.9069,0.40375833) (0.9482,0.38745202) (0.9894,0.37180426) (1.0306,0.35678845) (1.0718,0.34237908) (1.1131,0.32855165) (1.1543,0.31528266) (1.1955,0.30254955) (1.2367,0.29033069) (1.2780,0.27860530) (1.3192,0.26735346) (1.3604,0.25655604) (1.4016,0.24619469) (1.4429,0.23625179) (1.4841,0.22671045) (1.5253,0.21755445) (1.5665,0.20876823) (1.6078,0.20033685) (1.6490,0.19224598) (1.6902,0.18448187) (1.7314,0.17703133) (1.7727,0.16988168) (1.8139,0.16302079) (1.8551,0.15643698) (1.8963,0.15011907) (1.9376,0.14405631) (1.9788,0.13823841) (2.0200,0.13265547) };
\addplot[blue, only marks, mark=*, mark size=0.5pt] coordinates { (0,1.00000000) (0.1,0.91044159) (0.2,0.84113421) (0.3,0.78544138) (0.4,0.73941989) (0.5,0.70056002) (0.6,0.66717836) (0.7000000000000001,0.63809803) (0.8,0.61246815) (0.9,0.58965629) (1.0,0.56918121) (1.1,0.55066940) (1.2,0.53382594) (1.3,0.51841438) (1.4000000000000001,0.50424272) (1.5,0.49115315) (1.6,0.47901461) (1.7,0.46771726) (1.8,0.45716822) (1.9000000000000001,0.44728838) (2.0,0.43800983) };
\addplot[green!60!yellow, thick, opacity=0.9] coordinates { (0.0000,1.00000000) (0.0412,0.96116131) (0.0824,0.92652269) (0.1237,0.89537789) (0.1649,0.86717650) (0.2061,0.84148231) (0.2473,0.81794457) (0.2886,0.79627757) (0.3298,0.77624596) (0.3710,0.75765385) (0.4122,0.74033670) (0.4535,0.72415515) (0.4947,0.70899028) (0.5359,0.69473989) (0.5771,0.68131563) (0.6184,0.66864063) (0.6596,0.65664770) (0.7008,0.64527778) (0.7420,0.63447873) (0.7833,0.62420433) (0.8245,0.61441342) (0.8657,0.60506924) (0.9069,0.59613882) (0.9482,0.58759249) (0.9894,0.57940350) (1.0306,0.57154761) (1.0718,0.56400283) (1.1131,0.55674917) (1.1543,0.54976836) (1.1955,0.54304372) (1.2367,0.53655996) (1.2780,0.53030303) (1.3192,0.52426000) (1.3604,0.51841897) (1.4016,0.51276891) (1.4429,0.50729966) (1.4841,0.50200175) (1.5253,0.49686644) (1.5665,0.49188557) (1.6078,0.48705155) (1.6490,0.48235730) (1.6902,0.47779622) (1.7314,0.47336213) (1.7727,0.46904923) (1.8139,0.46485212) (1.8551,0.46076570) (1.8963,0.45678518) (1.9376,0.45290607) (1.9788,0.44912415) (2.0200,0.44543540) };
\end{axis}
\end{tikzpicture}

%% file: figures/transmittance_experiments/n100_random_center_prob_Exponential.tex
\begin{tikzpicture}
\begin{axis}[width=\cellsize,height=\cellsize,scale only axis,xlabel={$\tau$},xlabel near ticks,xticklabels={},yticklabels={},xmin=0,xmax=2.02,ymin=0,ymax=1.05,grid=major,ytick={0,0.25,0.5,0.75,1}]
\addplot[red, thick, dashed, opacity=0.7] coordinates { (0.0000,1.00000000) (0.0412,0.95961368) (0.0824,0.92085842) (0.1237,0.88366834) (0.1649,0.84798023) (0.2061,0.81373343) (0.2473,0.78086973) (0.2886,0.74933328) (0.3298,0.71907047) (0.3710,0.69002986) (0.4122,0.66216209) (0.4535,0.63541981) (0.4947,0.60975754) (0.5359,0.58513168) (0.5771,0.56150036) (0.6184,0.53882343) (0.6596,0.51706234) (0.7008,0.49618009) (0.7420,0.47614121) (0.7833,0.45691162) (0.8245,0.43845864) (0.8657,0.42075091) (0.9069,0.40375833) (0.9482,0.38745202) (0.9894,0.37180426) (1.0306,0.35678845) (1.0718,0.34237908) (1.1131,0.32855165) (1.1543,0.31528266) (1.1955,0.30254955) (1.2367,0.29033069) (1.2780,0.27860530) (1.3192,0.26735346) (1.3604,0.25655604) (1.4016,0.24619469) (1.4429,0.23625179) (1.4841,0.22671045) (1.5253,0.21755445) (1.5665,0.20876823) (1.6078,0.20033685) (1.6490,0.19224598) (1.6902,0.18448187) (1.7314,0.17703133) (1.7727,0.16988168) (1.8139,0.16302079) (1.8551,0.15643698) (1.8963,0.15011907) (1.9376,0.14405631) (1.9788,0.13823841) (2.0200,0.13265547) };
\addplot[blue, only marks, mark=*, mark size=0.5pt] coordinates { (0,1.00000000) (0.1,0.90392080) (0.2,0.81707281) (0.3,0.73856910) (0.4,0.66760797) (0.5,0.60346473) (0.6,0.54548432) (0.7000000000000001,0.49307462) (0.8,0.44570040) (0.9,0.40287786) (1.0,0.36416968) (1.1,0.32918055) (1.2,0.29755314) (1.3,0.26896447) (1.4000000000000001,0.24312258) (1.5,0.21976356) (1.6,0.19864885) (1.7,0.17956283) (1.8,0.16231057) (1.9000000000000001,0.14671590) (2.0,0.13261956) };
\addplot[green!60!yellow, thick, opacity=0.9] coordinates { (0.0000,1.00000000) (0.0412,0.95961368) (0.0824,0.92085842) (0.1237,0.88366834) (0.1649,0.84798023) (0.2061,0.81373343) (0.2473,0.78086973) (0.2886,0.74933328) (0.3298,0.71907047) (0.3710,0.69002986) (0.4122,0.66216209) (0.4535,0.63541981) (0.4947,0.60975754) (0.5359,0.58513168) (0.5771,0.56150036) (0.6184,0.53882343) (0.6596,0.51706234) (0.7008,0.49618009) (0.7420,0.47614121) (0.7833,0.45691162) (0.8245,0.43845864) (0.8657,0.42075091) (0.9069,0.40375833) (0.9482,0.38745202) (0.9894,0.37180426) (1.0306,0.35678845) (1.0718,0.34237908) (1.1131,0.32855165) (1.1543,0.31528266) (1.1955,0.30254955) (1.2367,0.29033069) (1.2780,0.27860530) (1.3192,0.26735346) (1.3604,0.25655604) (1.4016,0.24619469) (1.4429,0.23625179) (1.4841,0.22671045) (1.5253,0.21755445) (1.5665,0.20876823) (1.6078,0.20033685) (1.6490,0.19224598) (1.6902,0.18448187) (1.7314,0.17703133) (1.7727,0.16988168) (1.8139,0.16302079) (1.8551,0.15643698) (1.8963,0.15011907) (1.9376,0.14405631) (1.9788,0.13823841) (2.0200,0.13265547) };
\end{axis}
\end{tikzpicture}

%% file: figures/transmittance_experiments/n100_random_center_prob_Quadratic_cm0p5.tex
\begin{tikzpicture}
\begin{axis}[width=\cellsize,height=\cellsize,scale only axis,xlabel={$\tau$},xlabel near ticks,xticklabels={},yticklabels={},xmin=0,xmax=2.02,ymin=0,ymax=1.05,grid=major,ytick={0,0.25,0.5,0.75,1}]
\addplot[red, thick, dashed, opacity=0.7] coordinates { (0.0000,1.00000000) (0.0412,0.95961368) (0.0824,0.92085842) (0.1237,0.88366834) (0.1649,0.84798023) (0.2061,0.81373343) (0.2473,0.78086973) (0.2886,0.74933328) (0.3298,0.71907047) (0.3710,0.69002986) (0.4122,0.66216209) (0.4535,0.63541981) (0.4947,0.60975754) (0.5359,0.58513168) (0.5771,0.56150036) (0.6184,0.53882343) (0.6596,0.51706234) (0.7008,0.49618009) (0.7420,0.47614121) (0.7833,0.45691162) (0.8245,0.43845864) (0.8657,0.42075091) (0.9069,0.40375833) (0.9482,0.38745202) (0.9894,0.37180426) (1.0306,0.35678845) (1.0718,0.34237908) (1.1131,0.32855165) (1.1543,0.31528266) (1.1955,0.30254955) (1.2367,0.29033069) (1.2780,0.27860530) (1.3192,0.26735346) (1.3604,0.25655604) (1.4016,0.24619469) (1.4429,0.23625179) (1.4841,0.22671045) (1.5253,0.21755445) (1.5665,0.20876823) (1.6078,0.20033685) (1.6490,0.19224598) (1.6902,0.18448187) (1.7314,0.17703133) (1.7727,0.16988168) (1.8139,0.16302079) (1.8551,0.15643698) (1.8963,0.15011907) (1.9376,0.14405631) (1.9788,0.13823841) (2.0200,0.13265547) };
\addplot[blue, only marks, mark=*, mark size=0.5pt] coordinates { (0,1.00000000) (0.1,0.90200000) (0.2,0.80900000) (0.3,0.72100000) (0.4,0.63800000) (0.5,0.56000000) (0.6,0.48700000) (0.7000000000000001,0.41900000) (0.8,0.35600000) (0.9,0.29800000) (1.0,0.24500000) (1.1,0.19700000) (1.2,0.15400000) (1.3,0.11600000) (1.4000000000000001,0.08300000) (1.5,0.05500000) (1.6,0.03200000) (1.7,0.01400000) (1.8,0.00100000) (1.9000000000000001,0.00000000) (2.0,0.00000000) };
\addplot[green!60!yellow, thick, opacity=0.9] coordinates { (0.0000,1.00000000) (0.0412,0.95920037) (0.0824,0.91925048) (0.1237,0.88015031) (0.1649,0.84189988) (0.2061,0.80449917) (0.2473,0.76794819) (0.2886,0.73224694) (0.3298,0.69739542) (0.3710,0.66339363) (0.4122,0.63024157) (0.4535,0.59793923) (0.4947,0.56648663) (0.5359,0.53588376) (0.5771,0.50613061) (0.6184,0.47722720) (0.6596,0.44917351) (0.7008,0.42196955) (0.7420,0.39561533) (0.7833,0.37011083) (0.8245,0.34545606) (0.8657,0.32165102) (0.9069,0.29869571) (0.9482,0.27659013) (0.9894,0.25533428) (1.0306,0.23492815) (1.0718,0.21537176) (1.1131,0.19666510) (1.1543,0.17880816) (1.1955,0.16180096) (1.2367,0.14564348) (1.2780,0.13033574) (1.3192,0.11587772) (1.3604,0.10226943) (1.4016,0.08951087) (1.4429,0.07760204) (1.4841,0.06654294) (1.5253,0.05633357) (1.5665,0.04697393) (1.6078,0.03846401) (1.6490,0.03080383) (1.6902,0.02399338) (1.7314,0.01803265) (1.7727,0.01292166) (1.8139,0.00866039) (1.8551,0.00524885) (1.8963,0.00268705) (1.9376,0.00097497) (1.9788,0.00011262) (2.0200,0.00010000) };
\end{axis}
\end{tikzpicture}

%% file: figures/transmittance_experiments/n100_random_center_prob_citetvicini2021non_gamma0p5.tex
\begin{tikzpicture}
\begin{axis}[width=\cellsize,height=\cellsize,scale only axis,xlabel={$\tau$},xlabel near ticks,xticklabels={},yticklabels={},xmin=0,xmax=2.02,ymin=0,ymax=1.05,grid=major,ytick={0,0.25,0.5,0.75,1}]
\addplot[red, thick, dashed, opacity=0.7] coordinates { (0.0000,1.00000000) (0.0412,0.95961368) (0.0824,0.92085842) (0.1237,0.88366834) (0.1649,0.84798023) (0.2061,0.81373343) (0.2473,0.78086973) (0.2886,0.74933328) (0.3298,0.71907047) (0.3710,0.69002986) (0.4122,0.66216209) (0.4535,0.63541981) (0.4947,0.60975754) (0.5359,0.58513168) (0.5771,0.56150036) (0.6184,0.53882343) (0.6596,0.51706234) (0.7008,0.49618009) (0.7420,0.47614121) (0.7833,0.45691162) (0.8245,0.43845864) (0.8657,0.42075091) (0.9069,0.40375833) (0.9482,0.38745202) (0.9894,0.37180426) (1.0306,0.35678845) (1.0718,0.34237908) (1.1131,0.32855165) (1.1543,0.31528266) (1.1955,0.30254955) (1.2367,0.29033069) (1.2780,0.27860530) (1.3192,0.26735346) (1.3604,0.25655604) (1.4016,0.24619469) (1.4429,0.23625179) (1.4841,0.22671045) (1.5253,0.21755445) (1.5665,0.20876823) (1.6078,0.20033685) (1.6490,0.19224598) (1.6902,0.18448187) (1.7314,0.17703133) (1.7727,0.16988168) (1.8139,0.16302079) (1.8551,0.15643698) (1.8963,0.15011907) (1.9376,0.14405631) (1.9788,0.13823841) (2.0200,0.13265547) };
\addplot[blue, only marks, mark=*, mark size=0.5pt] coordinates { (0,1.00000000) (0.1,0.90196040) (0.2,0.80853640) (0.3,0.71928455) (0.4,0.63380399) (0.5,0.55173236) (0.6,0.47274216) (0.7000000000000001,0.39653731) (0.8,0.32285020) (0.9,0.25143893) (1.0,0.18208484) (1.1,0.16459027) (1.2,0.14877657) (1.3,0.13448224) (1.4000000000000001,0.12156129) (1.5,0.10988178) (1.6,0.09932443) (1.7,0.08978141) (1.8,0.08115529) (1.9000000000000001,0.07335795) (2.0,0.06630978) };
\addplot[green!60!yellow, thick, opacity=0.9] coordinates { (0.0000,1.00000000) (0.0412,0.95919460) (0.0824,0.91920472) (0.1237,0.87999743) (0.1649,0.84154113) (0.2061,0.80380549) (0.2473,0.76676140) (0.2886,0.73038093) (0.3298,0.69463727) (0.3710,0.65950473) (0.4122,0.62495860) (0.4535,0.59097521) (0.4947,0.55753183) (0.5359,0.52460666) (0.5771,0.49217875) (0.6184,0.46022804) (0.6596,0.42873525) (0.7008,0.39768188) (0.7420,0.36705020) (0.7833,0.33682316) (0.8245,0.30698442) (0.8657,0.27751831) (0.9069,0.24840978) (0.9482,0.21964438) (0.9894,0.19120825) (1.0306,0.17839423) (1.0718,0.17118954) (1.1131,0.16427582) (1.1543,0.15764133) (1.1955,0.15127478) (1.2367,0.14516535) (1.2780,0.13930265) (1.3192,0.13367673) (1.3604,0.12827802) (1.4016,0.12309734) (1.4429,0.11812589) (1.4841,0.11335522) (1.5253,0.10877722) (1.5665,0.10438411) (1.6078,0.10016842) (1.6490,0.09612299) (1.6902,0.09224094) (1.7314,0.08851566) (1.7727,0.08494084) (1.8139,0.08151039) (1.8551,0.07821849) (1.8963,0.07505953) (1.9376,0.07202815) (1.9788,0.06911920) (2.0200,0.06632773) };
\end{axis}
\end{tikzpicture}

%% file: figures/transmittance_experiments/n100_random_center_prob_Blend_gamma0p5.tex
\begin{tikzpicture}
\begin{axis}[width=\cellsize,height=\cellsize,scale only axis,xlabel={$\tau$},xlabel near ticks,xticklabels={},yticklabels={},xmin=0,xmax=2.02,ymin=0,ymax=1.05,grid=major,ytick={0,0.25,0.5,0.75,1}]
\addplot[red, thick, dashed, opacity=0.7] coordinates { (0.0000,1.00000000) (0.0412,0.95961368) (0.0824,0.92085842) (0.1237,0.88366834) (0.1649,0.84798023) (0.2061,0.81373343) (0.2473,0.78086973) (0.2886,0.74933328) (0.3298,0.71907047) (0.3710,0.69002986) (0.4122,0.66216209) (0.4535,0.63541981) (0.4947,0.60975754) (0.5359,0.58513168) (0.5771,0.56150036) (0.6184,0.53882343) (0.6596,0.51706234) (0.7008,0.49618009) (0.7420,0.47614121) (0.7833,0.45691162) (0.8245,0.43845864) (0.8657,0.42075091) (0.9069,0.40375833) (0.9482,0.38745202) (0.9894,0.37180426) (1.0306,0.35678845) (1.0718,0.34237908) (1.1131,0.32855165) (1.1543,0.31528266) (1.1955,0.30254955) (1.2367,0.29033069) (1.2780,0.27860530) (1.3192,0.26735346) (1.3604,0.25655604) (1.4016,0.24619469) (1.4429,0.23625179) (1.4841,0.22671045) (1.5253,0.21755445) (1.5665,0.20876823) (1.6078,0.20033685) (1.6490,0.19224598) (1.6902,0.18448187) (1.7314,0.17703133) (1.7727,0.16988168) (1.8139,0.16302079) (1.8551,0.15643698) (1.8963,0.15011907) (1.9376,0.14405631) (1.9788,0.13823841) (2.0200,0.13265547) };
\addplot[blue, only marks, mark=*, mark size=0.5pt] coordinates { (0,1.00000000) (0.1,0.90196040) (0.2,0.80853640) (0.3,0.71928455) (0.4,0.63380399) (0.5,0.55173236) (0.6,0.47274216) (0.7000000000000001,0.39653731) (0.8,0.32285020) (0.9,0.25143893) (1.0,0.18208484) (1.1,0.11459027) (1.2,0.04877657) (1.3,0.00000000) (1.4000000000000001,0.00000000) (1.5,0.00000000) (1.6,0.00000000) (1.7,0.00000000) (1.8,0.00000000) (1.9000000000000001,0.00000000) (2.0,0.00000000) };
\addplot[green!60!yellow, thick, opacity=0.9] coordinates { (0.0000,1.00000000) (0.0412,0.95919460) (0.0824,0.91920472) (0.1237,0.87999743) (0.1649,0.84154113) (0.2061,0.80380549) (0.2473,0.76676140) (0.2886,0.73038093) (0.3298,0.69463727) (0.3710,0.65950473) (0.4122,0.62495860) (0.4535,0.59097521) (0.4947,0.55753183) (0.5359,0.52460666) (0.5771,0.49217875) (0.6184,0.46022804) (0.6596,0.42873525) (0.7008,0.39768188) (0.7420,0.36705020) (0.7833,0.33682316) (0.8245,0.30698442) (0.8657,0.27751831) (0.9069,0.24840978) (0.9482,0.21964438) (0.9894,0.19120825) (1.0306,0.16308810) (1.0718,0.13527117) (1.1131,0.10774521) (1.1543,0.08049847) (1.1955,0.05351967) (1.2367,0.02679800) (1.2780,0.00032306) (1.3192,0.00000000) (1.3604,0.00000000) (1.4016,0.00000000) (1.4429,0.00000000) (1.4841,0.00000000) (1.5253,0.00000000) (1.5665,0.00000000) (1.6078,0.00000000) (1.6490,0.00000000) (1.6902,0.00000000) (1.7314,0.00000000) (1.7727,0.00000000) (1.8139,0.00000000) (1.8551,0.00000000) (1.8963,0.00000000) (1.9376,0.00000000) (1.9788,0.00000000) (2.0200,0.00000000) };
\end{axis}
\end{tikzpicture}

%% file: figures/transmittance_experiments/n100_random_center_prob_Linear.tex
\begin{tikzpicture}
\begin{axis}[width=\cellsize,height=\cellsize,scale only axis,xlabel={$\tau$},xlabel near ticks,xticklabels={},yticklabels={},xmin=0,xmax=2.02,ymin=0,ymax=1.05,grid=major,ytick={0,0.25,0.5,0.75,1}]
\addplot[red, thick, dashed, opacity=0.7] coordinates { (0.0000,1.00000000) (0.0412,0.95961368) (0.0824,0.92085842) (0.1237,0.88366834) (0.1649,0.84798023) (0.2061,0.81373343) (0.2473,0.78086973) (0.2886,0.74933328) (0.3298,0.71907047) (0.3710,0.69002986) (0.4122,0.66216209) (0.4535,0.63541981) (0.4947,0.60975754) (0.5359,0.58513168) (0.5771,0.56150036) (0.6184,0.53882343) (0.6596,0.51706234) (0.7008,0.49618009) (0.7420,0.47614121) (0.7833,0.45691162) (0.8245,0.43845864) (0.8657,0.42075091) (0.9069,0.40375833) (0.9482,0.38745202) (0.9894,0.37180426) (1.0306,0.35678845) (1.0718,0.34237908) (1.1131,0.32855165) (1.1543,0.31528266) (1.1955,0.30254955) (1.2367,0.29033069) (1.2780,0.27860530) (1.3192,0.26735346) (1.3604,0.25655604) (1.4016,0.24619469) (1.4429,0.23625179) (1.4841,0.22671045) (1.5253,0.21755445) (1.5665,0.20876823) (1.6078,0.20033685) (1.6490,0.19224598) (1.6902,0.18448187) (1.7314,0.17703133) (1.7727,0.16988168) (1.8139,0.16302079) (1.8551,0.15643698) (1.8963,0.15011907) (1.9376,0.14405631) (1.9788,0.13823841) (2.0200,0.13265547) };
\addplot[blue, only marks, mark=*, mark size=0.5pt] coordinates { (0,1.00000000) (0.1,0.90000000) (0.2,0.80000000) (0.3,0.70000000) (0.4,0.60000000) (0.5,0.50000000) (0.6,0.40000000) (0.7000000000000001,0.30000000) (0.8,0.20000000) (0.9,0.10000000) (1.0,0.00000000) (1.1,0.00000000) (1.2,0.00000000) (1.3,0.00000000) (1.4000000000000001,0.00000000) (1.5,0.00000000) (1.6,0.00000000) (1.7,0.00000000) (1.8,0.00000000) (1.9000000000000001,0.00000000) (2.0,0.00000000) };
\addplot[green!60!yellow, thick, opacity=0.9] coordinates { (0.0000,1.00000000) (0.0412,0.95877551) (0.0824,0.91755102) (0.1237,0.87632653) (0.1649,0.83510204) (0.2061,0.79387755) (0.2473,0.75265306) (0.2886,0.71142857) (0.3298,0.67020408) (0.3710,0.62897959) (0.4122,0.58775510) (0.4535,0.54653061) (0.4947,0.50530612) (0.5359,0.46408163) (0.5771,0.42285714) (0.6184,0.38163265) (0.6596,0.34040816) (0.7008,0.29918367) (0.7420,0.25795918) (0.7833,0.21673469) (0.8245,0.17551020) (0.8657,0.13428571) (0.9069,0.09306122) (0.9482,0.05183673) (0.9894,0.01061224) (1.0306,0.00000000) (1.0718,0.00000000) (1.1131,0.00000000) (1.1543,0.00000000) (1.1955,0.00000000) (1.2367,0.00000000) (1.2780,0.00000000) (1.3192,0.00000000) (1.3604,0.00000000) (1.4016,0.00000000) (1.4429,0.00000000) (1.4841,0.00000000) (1.5253,0.00000000) (1.5665,0.00000000) (1.6078,0.00000000) (1.6490,0.00000000) (1.6902,0.00000000) (1.7314,0.00000000) (1.7727,0.00000000) (1.8139,0.00000000) (1.8551,0.00000000) (1.8963,0.00000000) (1.9376,0.00000000) (1.9788,0.00000000) (2.0200,0.00000000) };
\end{axis}
\end{tikzpicture}

%% file: figures/transmittance_experiments/n100_random_center_prob_Quadratic_c0p5.tex
\begin{tikzpicture}
\begin{axis}[width=\cellsize,height=\cellsize,scale only axis,xlabel={$\tau$},xlabel near ticks,xticklabels={},yticklabels={},xmin=0,xmax=2.02,ymin=0,ymax=1.05,grid=major,ytick={0,0.25,0.5,0.75,1}]
\addplot[red, thick, dashed, opacity=0.7] coordinates { (0.0000,1.00000000) (0.0412,0.95961368) (0.0824,0.92085842) (0.1237,0.88366834) (0.1649,0.84798023) (0.2061,0.81373343) (0.2473,0.78086973) (0.2886,0.74933328) (0.3298,0.71907047) (0.3710,0.69002986) (0.4122,0.66216209) (0.4535,0.63541981) (0.4947,0.60975754) (0.5359,0.58513168) (0.5771,0.56150036) (0.6184,0.53882343) (0.6596,0.51706234) (0.7008,0.49618009) (0.7420,0.47614121) (0.7833,0.45691162) (0.8245,0.43845864) (0.8657,0.42075091) (0.9069,0.40375833) (0.9482,0.38745202) (0.9894,0.37180426) (1.0306,0.35678845) (1.0718,0.34237908) (1.1131,0.32855165) (1.1543,0.31528266) (1.1955,0.30254955) (1.2367,0.29033069) (1.2780,0.27860530) (1.3192,0.26735346) (1.3604,0.25655604) (1.4016,0.24619469) (1.4429,0.23625179) (1.4841,0.22671045) (1.5253,0.21755445) (1.5665,0.20876823) (1.6078,0.20033685) (1.6490,0.19224598) (1.6902,0.18448187) (1.7314,0.17703133) (1.7727,0.16988168) (1.8139,0.16302079) (1.8551,0.15643698) (1.8963,0.15011907) (1.9376,0.14405631) (1.9788,0.13823841) (2.0200,0.13265547) };
\addplot[blue, only marks, mark=*, mark size=0.5pt] coordinates { (0,1.00000000) (0.1,0.89800000) (0.2,0.79100000) (0.3,0.67900000) (0.4,0.56200000) (0.5,0.44000000) (0.6,0.31300000) (0.7000000000000001,0.18100000) (0.8,0.04400000) (0.9,0.00000000) (1.0,0.00000000) (1.1,0.00000000) (1.2,0.00000000) (1.3,0.00000000) (1.4000000000000001,0.00000000) (1.5,0.00000000) (1.6,0.00000000) (1.7,0.00000000) (1.8,0.00000000) (1.9000000000000001,0.00000000) (2.0,0.00000000) };
\addplot[green!60!yellow, thick, opacity=0.9] coordinates { (0.0000,1.00000000) (0.0412,0.95835065) (0.0824,0.91585156) (0.1237,0.87250275) (0.1649,0.82830421) (0.2061,0.78325594) (0.2473,0.73735793) (0.2886,0.69061020) (0.3298,0.64301274) (0.3710,0.59456556) (0.4122,0.54526864) (0.4535,0.49512199) (0.4947,0.44412561) (0.5359,0.39227951) (0.5771,0.33958367) (0.6184,0.28603811) (0.6596,0.23164282) (0.7008,0.17639779) (0.7420,0.12030304) (0.7833,0.06335856) (0.8245,0.00556435) (0.8657,0.00000000) (0.9069,0.00000000) (0.9482,0.00000000) (0.9894,0.00000000) (1.0306,0.00000000) (1.0718,0.00000000) (1.1131,0.00000000) (1.1543,0.00000000) (1.1955,0.00000000) (1.2367,0.00000000) (1.2780,0.00000000) (1.3192,0.00000000) (1.3604,0.00000000) (1.4016,0.00000000) (1.4429,0.00000000) (1.4841,0.00000000) (1.5253,0.00000000) (1.5665,0.00000000) (1.6078,0.00000000) (1.6490,0.00000000) (1.6902,0.00000000) (1.7314,0.00000000) (1.7727,0.00000000) (1.8139,0.00000000) (1.8551,0.00000000) (1.8963,0.00000000) (1.9376,0.00000000) (1.9788,0.00000000) (2.0200,0.00000000) };
\end{axis}
\end{tikzpicture}

%% file: figures/transmittance_experiments/n100_random_center_prob_Quadratic_c1.tex
\begin{tikzpicture}
\begin{axis}[width=\cellsize,height=\cellsize,scale only axis,xlabel={$\tau$},xlabel near ticks,xticklabels={},yticklabels={},xmin=0,xmax=2.02,ymin=0,ymax=1.05,grid=major,ytick={0,0.25,0.5,0.75,1}]
\addplot[red, thick, dashed, opacity=0.7] coordinates { (0.0000,1.00000000) (0.0412,0.95961368) (0.0824,0.92085842) (0.1237,0.88366834) (0.1649,0.84798023) (0.2061,0.81373343) (0.2473,0.78086973) (0.2886,0.74933328) (0.3298,0.71907047) (0.3710,0.69002986) (0.4122,0.66216209) (0.4535,0.63541981) (0.4947,0.60975754) (0.5359,0.58513168) (0.5771,0.56150036) (0.6184,0.53882343) (0.6596,0.51706234) (0.7008,0.49618009) (0.7420,0.47614121) (0.7833,0.45691162) (0.8245,0.43845864) (0.8657,0.42075091) (0.9069,0.40375833) (0.9482,0.38745202) (0.9894,0.37180426) (1.0306,0.35678845) (1.0718,0.34237908) (1.1131,0.32855165) (1.1543,0.31528266) (1.1955,0.30254955) (1.2367,0.29033069) (1.2780,0.27860530) (1.3192,0.26735346) (1.3604,0.25655604) (1.4016,0.24619469) (1.4429,0.23625179) (1.4841,0.22671045) (1.5253,0.21755445) (1.5665,0.20876823) (1.6078,0.20033685) (1.6490,0.19224598) (1.6902,0.18448187) (1.7314,0.17703133) (1.7727,0.16988168) (1.8139,0.16302079) (1.8551,0.15643698) (1.8963,0.15011907) (1.9376,0.14405631) (1.9788,0.13823841) (2.0200,0.13265547) };
\addplot[blue, only marks, mark=*, mark size=0.5pt] coordinates { (0,1.00000000) (0.1,0.89600000) (0.2,0.78200000) (0.3,0.65800000) (0.4,0.52400000) (0.5,0.38000000) (0.6,0.22600000) (0.7000000000000001,0.06200000) (0.8,0.00000000) (0.9,0.00000000) (1.0,0.00000000) (1.1,0.00000000) (1.2,0.00000000) (1.3,0.00000000) (1.4000000000000001,0.00000000) (1.5,0.00000000) (1.6,0.00000000) (1.7,0.00000000) (1.8,0.00000000) (1.9000000000000001,0.00000000) (2.0,0.00000000) };
\addplot[green!60!yellow, thick, opacity=0.9] coordinates { (0.0000,1.00000000) (0.0412,0.95792578) (0.0824,0.91415210) (0.1237,0.86867897) (0.1649,0.82150637) (0.2061,0.77263432) (0.2473,0.72206281) (0.2886,0.66979184) (0.3298,0.61582141) (0.3710,0.56015152) (0.4122,0.50278217) (0.4535,0.44371337) (0.4947,0.38294511) (0.5359,0.32047738) (0.5771,0.25631020) (0.6184,0.19044357) (0.6596,0.12287747) (0.7008,0.05361191) (0.7420,0.00000000) (0.7833,0.00000000) (0.8245,0.00000000) (0.8657,0.00000000) (0.9069,0.00000000) (0.9482,0.00000000) (0.9894,0.00000000) (1.0306,0.00000000) (1.0718,0.00000000) (1.1131,0.00000000) (1.1543,0.00000000) (1.1955,0.00000000) (1.2367,0.00000000) (1.2780,0.00000000) (1.3192,0.00000000) (1.3604,0.00000000) (1.4016,0.00000000) (1.4429,0.00000000) (1.4841,0.00000000) (1.5253,0.00000000) (1.5665,0.00000000) (1.6078,0.00000000) (1.6490,0.00000000) (1.6902,0.00000000) (1.7314,0.00000000) (1.7727,0.00000000) (1.8139,0.00000000) (1.8551,0.00000000) (1.8963,0.00000000) (1.9376,0.00000000) (1.9788,0.00000000) (2.0200,0.00000000) };
\end{axis}
\end{tikzpicture}

%% file: figures/transmittance_experiments/n100_random_center_prob_PowermLaw_vm1.tex
\begin{tikzpicture}
\begin{axis}[width=\cellsize,height=\cellsize,scale only axis,xlabel={$\tau$},xlabel near ticks,xticklabels={},yticklabels={},xmin=0,xmax=2.02,ymin=0,ymax=1.05,grid=major,ytick={0,0.25,0.5,0.75,1}]
\addplot[red, thick, dashed, opacity=0.7] coordinates { (0.0000,1.00000000) (0.0412,0.95961368) (0.0824,0.92085842) (0.1237,0.88366834) (0.1649,0.84798023) (0.2061,0.81373343) (0.2473,0.78086973) (0.2886,0.74933328) (0.3298,0.71907047) (0.3710,0.69002986) (0.4122,0.66216209) (0.4535,0.63541981) (0.4947,0.60975754) (0.5359,0.58513168) (0.5771,0.56150036) (0.6184,0.53882343) (0.6596,0.51706234) (0.7008,0.49618009) (0.7420,0.47614121) (0.7833,0.45691162) (0.8245,0.43845864) (0.8657,0.42075091) (0.9069,0.40375833) (0.9482,0.38745202) (0.9894,0.37180426) (1.0306,0.35678845) (1.0718,0.34237908) (1.1131,0.32855165) (1.1543,0.31528266) (1.1955,0.30254955) (1.2367,0.29033069) (1.2780,0.27860530) (1.3192,0.26735346) (1.3604,0.25655604) (1.4016,0.24619469) (1.4429,0.23625179) (1.4841,0.22671045) (1.5253,0.21755445) (1.5665,0.20876823) (1.6078,0.20033685) (1.6490,0.19224598) (1.6902,0.18448187) (1.7314,0.17703133) (1.7727,0.16988168) (1.8139,0.16302079) (1.8551,0.15643698) (1.8963,0.15011907) (1.9376,0.14405631) (1.9788,0.13823841) (2.0200,0.13265547) };
\addplot[blue, only marks, mark=*, mark size=0.5pt] coordinates { (0,1.00000000) (0.1,0.89559428) (0.2,0.77746824) (0.3,0.63816857) (0.4,0.45923585) (0.5,0.13606878) (0.6,0.00000000) (0.7000000000000001,0.00000000) (0.8,0.00000000) (0.9,0.00000000) (1.0,0.00000000) (1.1,0.00000000) (1.2,0.00000000) (1.3,0.00000000) (1.4000000000000001,0.00000000) (1.5,0.00000000) (1.6,0.00000000) (1.7,0.00000000) (1.8,0.00000000) (1.9000000000000001,0.00000000) (2.0,0.00000000) };
\addplot[green!60!yellow, thick, opacity=0.9] coordinates { (0.0000,1.00000000) (0.0412,0.95788884) (0.0824,0.91383918) (0.1237,0.86755580) (0.1649,0.81865993) (0.2061,0.76665188) (0.2473,0.71084887) (0.2886,0.65027467) (0.3298,0.58344508) (0.3710,0.50789682) (0.4122,0.41893938) (0.4535,0.30505938) (0.4947,0.10301575) (0.5359,0.00000000) (0.5771,0.00000000) (0.6184,0.00000000) (0.6596,0.00000000) (0.7008,0.00000000) (0.7420,0.00000000) (0.7833,0.00000000) (0.8245,0.00000000) (0.8657,0.00000000) (0.9069,0.00000000) (0.9482,0.00000000) (0.9894,0.00000000) (1.0306,0.00000000) (1.0718,0.00000000) (1.1131,0.00000000) (1.1543,0.00000000) (1.1955,0.00000000) (1.2367,0.00000000) (1.2780,0.00000000) (1.3192,0.00000000) (1.3604,0.00000000) (1.4016,0.00000000) (1.4429,0.00000000) (1.4841,0.00000000) (1.5253,0.00000000) (1.5665,0.00000000) (1.6078,0.00000000) (1.6490,0.00000000) (1.6902,0.00000000) (1.7314,0.00000000) (1.7727,0.00000000) (1.8139,0.00000000) (1.8551,0.00000000) (1.8963,0.00000000) (1.9376,0.00000000) (1.9788,0.00000000) (2.0200,0.00000000) };
\end{axis}
\end{tikzpicture}

%% file: figures/combined_experiments/combined_experiments.tex
%
%

\begin{figure*}[htbp]
\centering
\setlength{\tabcolsep}{0pt}
\begin{tabular}{@{}cc@{}c@{}c@{}c@{}c@{}c@{}c@{}c@{}c@{}}
 & \shortstack{\textbf{\tiny Power-Law}\\\textbf{\tiny ($v=2$)}} & \textbf{\tiny Exponential} & \shortstack{\textbf{\tiny Quadratic}\\\textbf{\tiny ($c=-0.5$)}} & \shortstack{\textbf{\tiny \citet{vicini2021non}}\\\textbf{\tiny ($\gamma=0.5$)}} & \shortstack{\textbf{\tiny Blend}\\\textbf{\tiny ($\gamma=0.5$)}} & \textbf{\tiny Linear} & \shortstack{\textbf{\tiny Quadratic}\\\textbf{\tiny ($c=0.5$)}} & \shortstack{\textbf{\tiny Quadratic}\\\textbf{\tiny ($c=1$)}} & \shortstack{\textbf{\tiny Power-Law}\\\textbf{\tiny ($v=-1$)}} \\[1ex]
\raisebox{\dimexpr\cellsize/2-\height/2\relax}{\rotatebox{90}{\scriptsize\textbf{RGB}}} & \includegraphics[width=\cellsize,height=\cellsize]{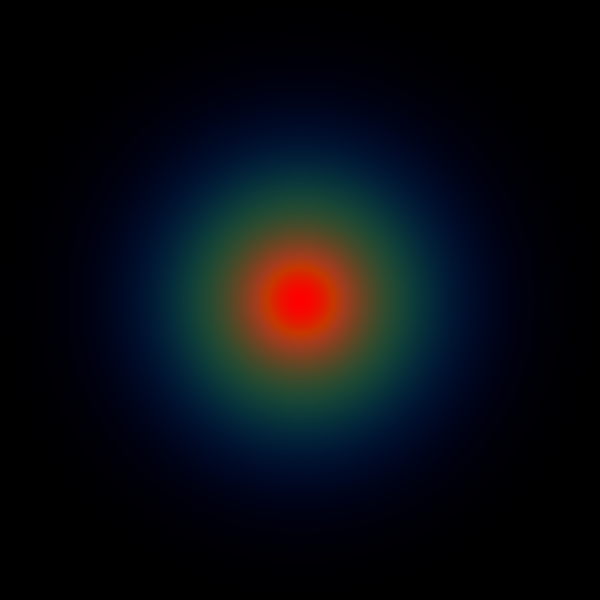} & \includegraphics[width=\cellsize,height=\cellsize]{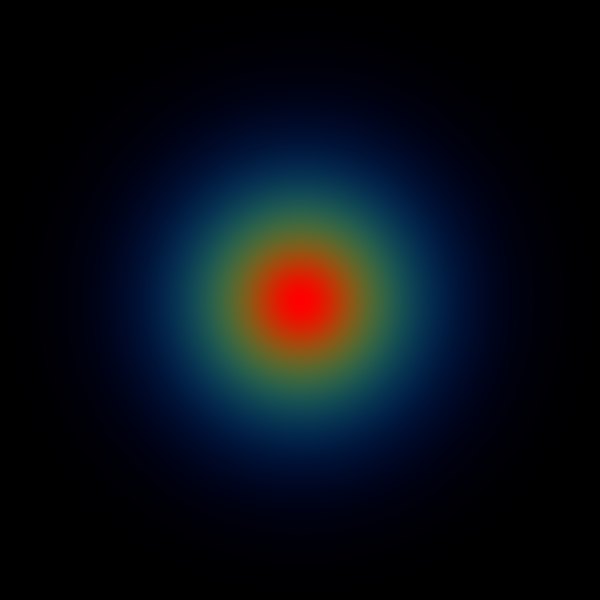} & \includegraphics[width=\cellsize,height=\cellsize]{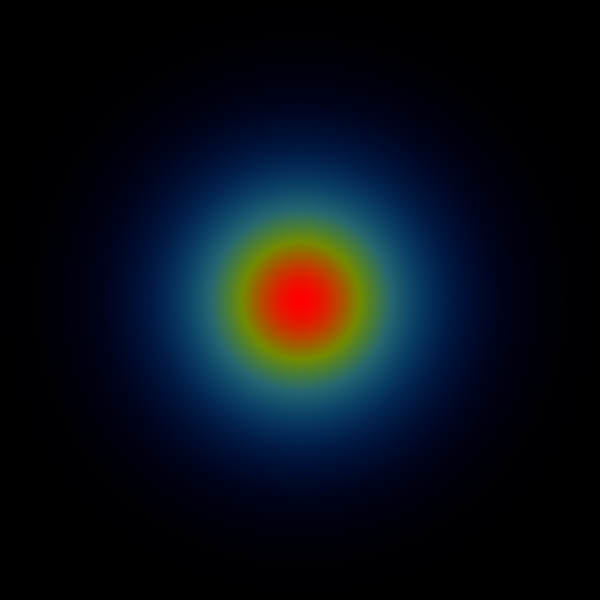} & \includegraphics[width=\cellsize,height=\cellsize]{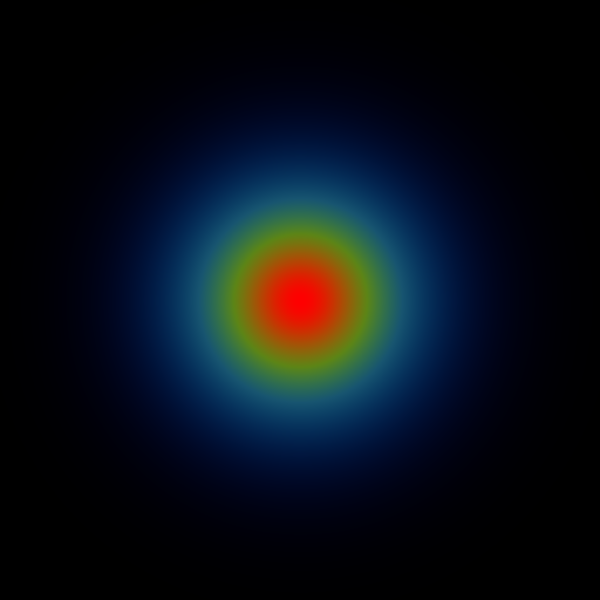} & \includegraphics[width=\cellsize,height=\cellsize]{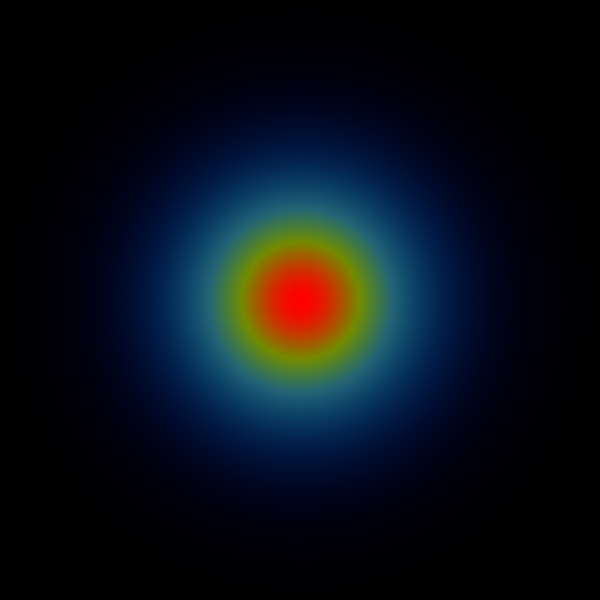} & \includegraphics[width=\cellsize,height=\cellsize]{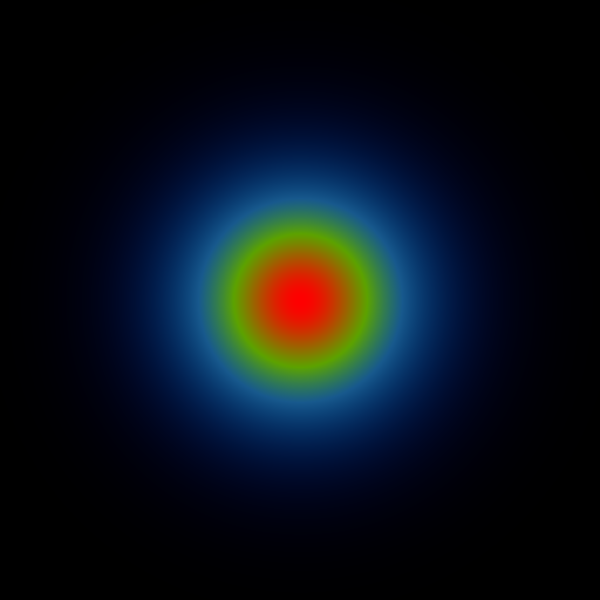} & \includegraphics[width=\cellsize,height=\cellsize]{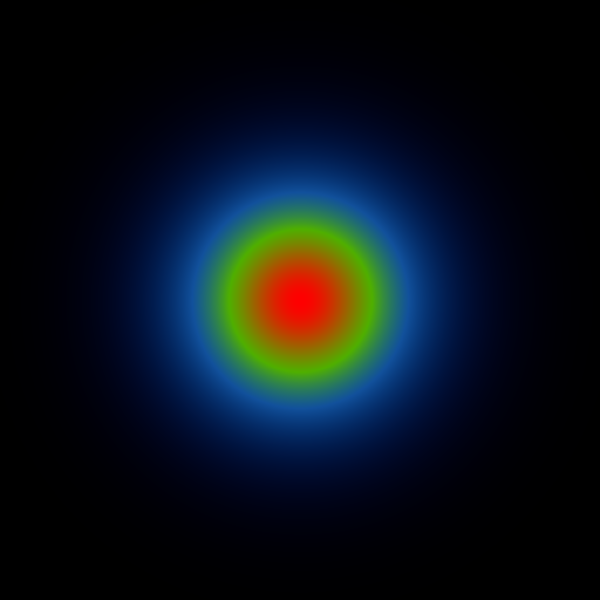} & \includegraphics[width=\cellsize,height=\cellsize]{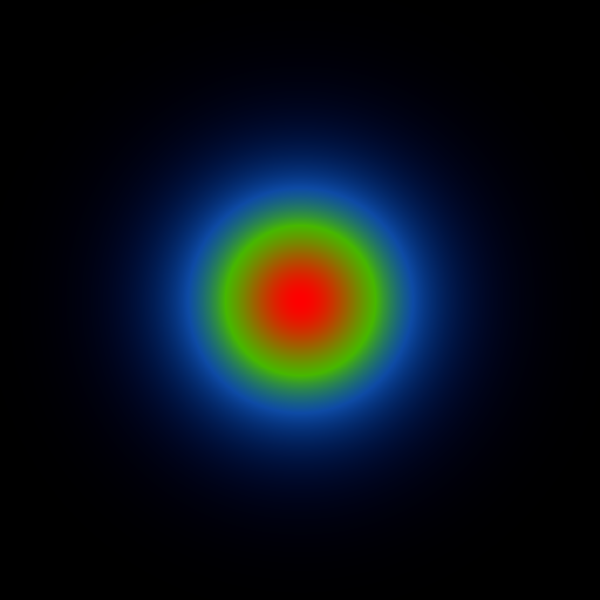} & \includegraphics[width=\cellsize,height=\cellsize]{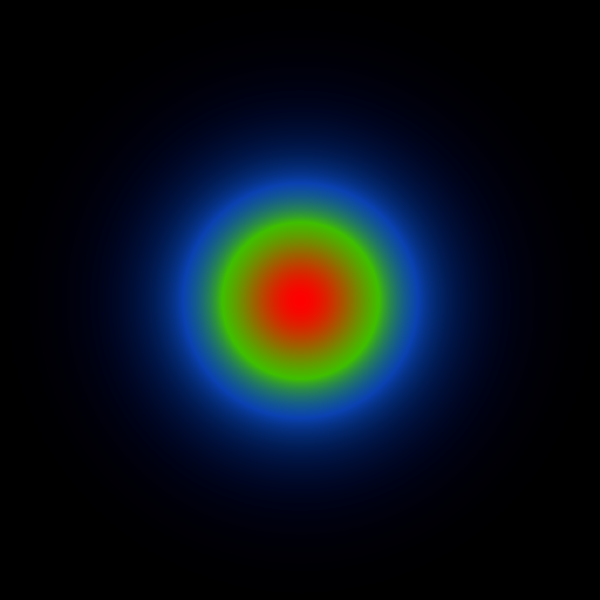} \\[0.5ex]
\raisebox{\dimexpr\cellsize/2-\height/2\relax}{\rotatebox{90}{\scriptsize\textbf{Intensity}}} & \input{\datapath n3_concentric_cross_section_PowermLaw_v2.tex} & \input{\datapath n3_concentric_cross_section_Exponential.tex} & \input{\datapath n3_concentric_cross_section_Quadratic_cm0p5.tex} & \input{\datapath n3_concentric_cross_section_citetvicini2021non_gamma0p5.tex} & \input{\datapath n3_concentric_cross_section_Blend_gamma0p5.tex} & \input{\datapath n3_concentric_cross_section_Linear.tex} & \input{\datapath n3_concentric_cross_section_Quadratic_c0p5.tex} & \input{\datapath n3_concentric_cross_section_Quadratic_c1.tex} & \input{\datapath n3_concentric_cross_section_PowermLaw_vm1.tex} \\[0.5ex]
\raisebox{\dimexpr\cellsize/2-\height/2\relax}{\rotatebox{90}{\scriptsize\textbf{$1 - \bar T_{N+1}$}}} & \includegraphics[width=\cellsize,height=\cellsize]{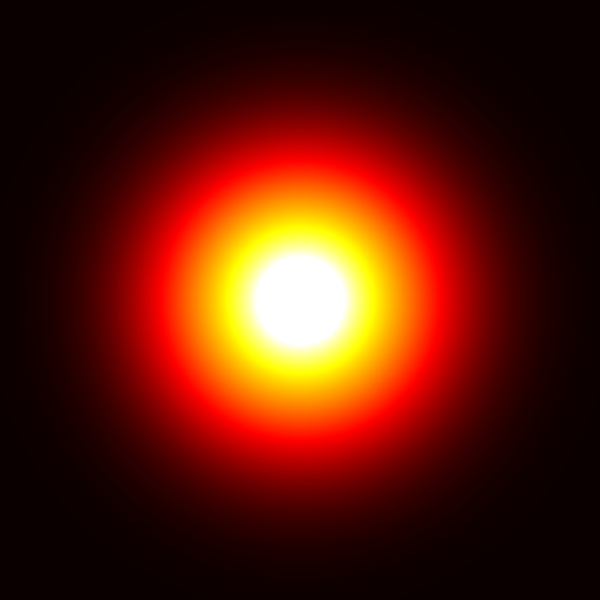} & \includegraphics[width=\cellsize,height=\cellsize]{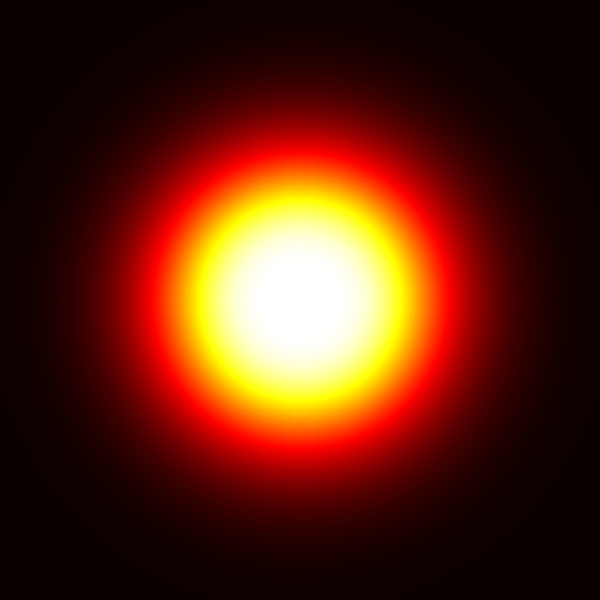} & \includegraphics[width=\cellsize,height=\cellsize]{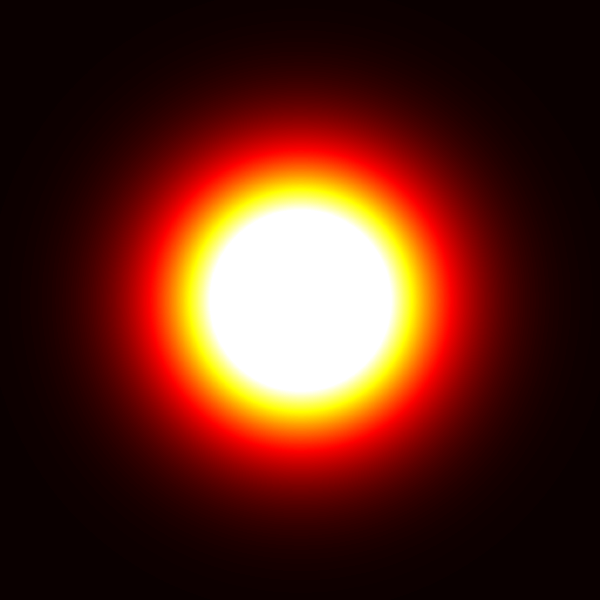} & \includegraphics[width=\cellsize,height=\cellsize]{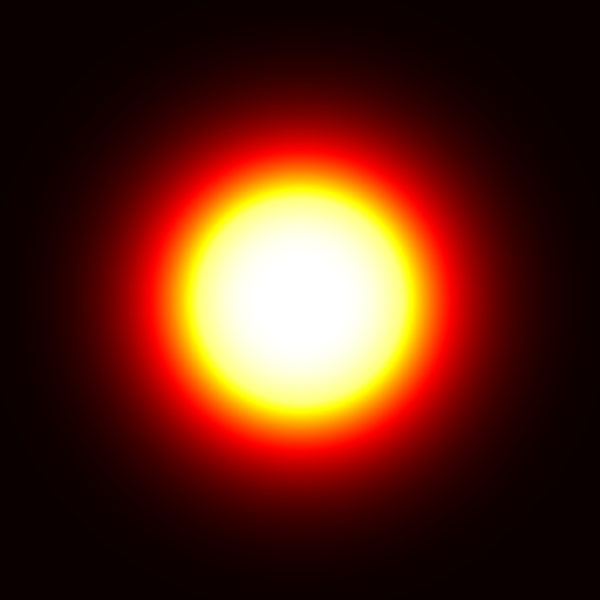} & \includegraphics[width=\cellsize,height=\cellsize]{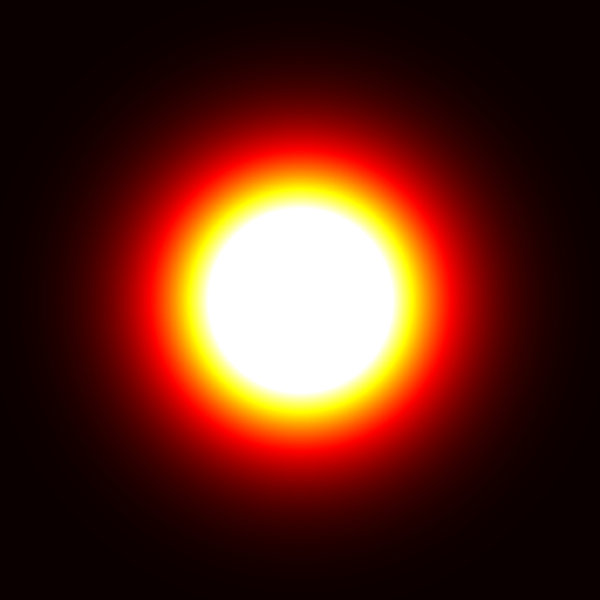} & \includegraphics[width=\cellsize,height=\cellsize]{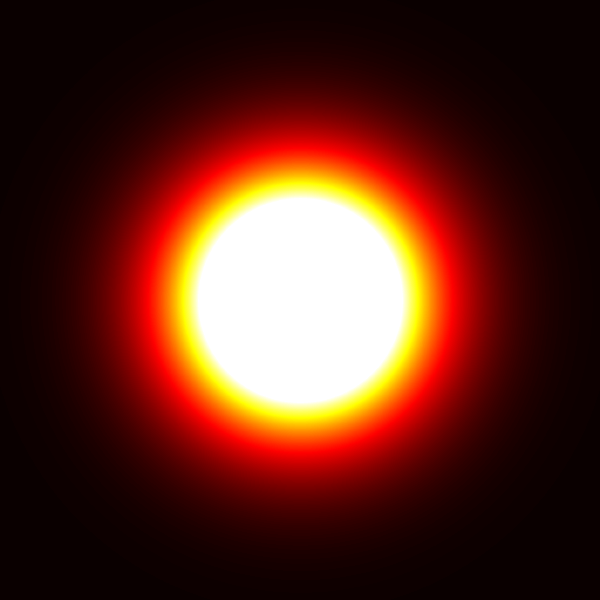} & \includegraphics[width=\cellsize,height=\cellsize]{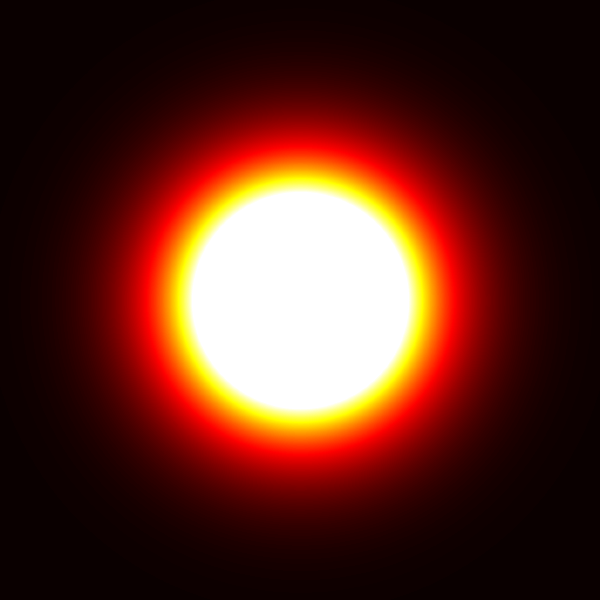} & \includegraphics[width=\cellsize,height=\cellsize]{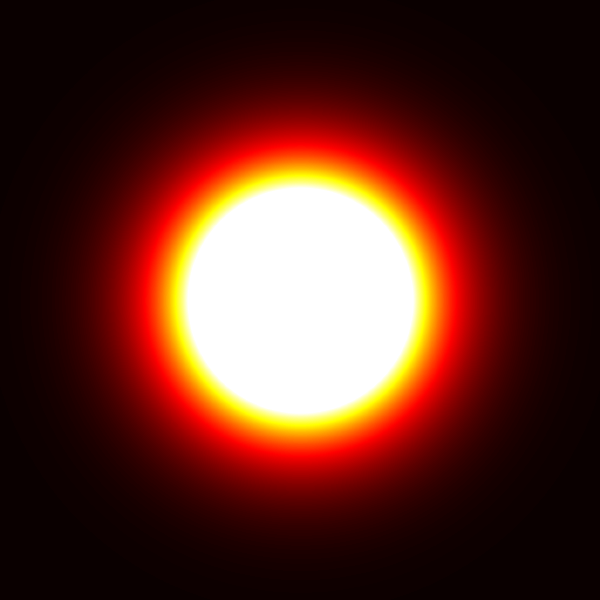} & \includegraphics[width=\cellsize,height=\cellsize]{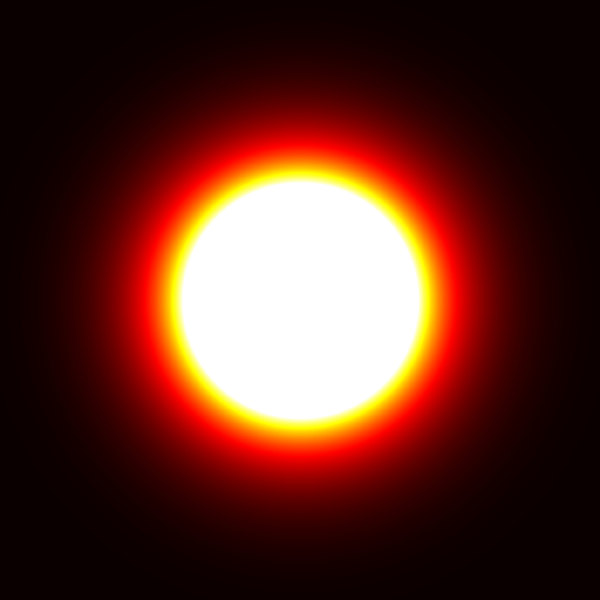} \\
\multicolumn{10}{c}{\scriptsize{(a) $N=3$ concentric Gaussians.}} \\[2ex]
\raisebox{\dimexpr\cellsize/2-\height/2\relax}{\rotatebox{90}{\scriptsize\textbf{RGB}}} & \includegraphics[width=\cellsize,height=\cellsize]{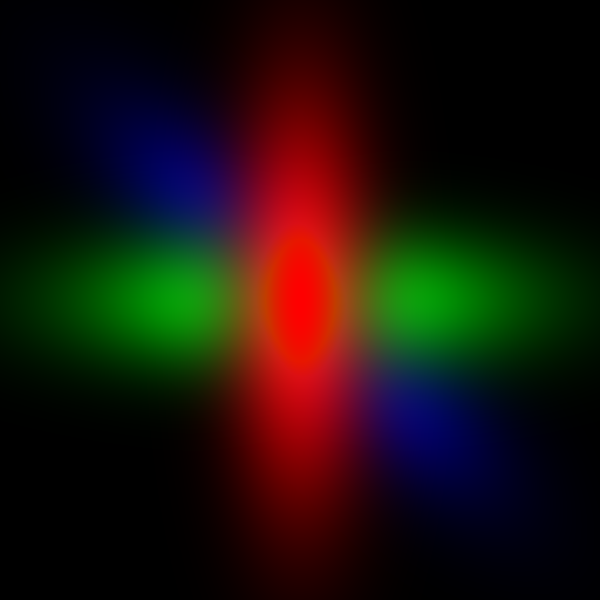} & \includegraphics[width=\cellsize,height=\cellsize]{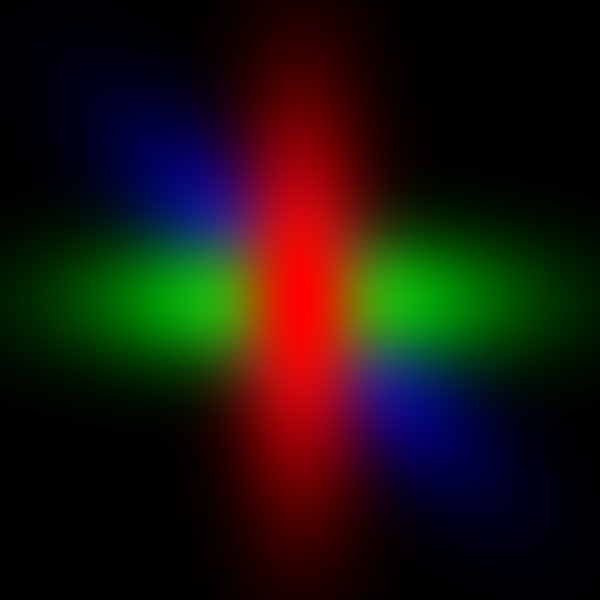} & \includegraphics[width=\cellsize,height=\cellsize]{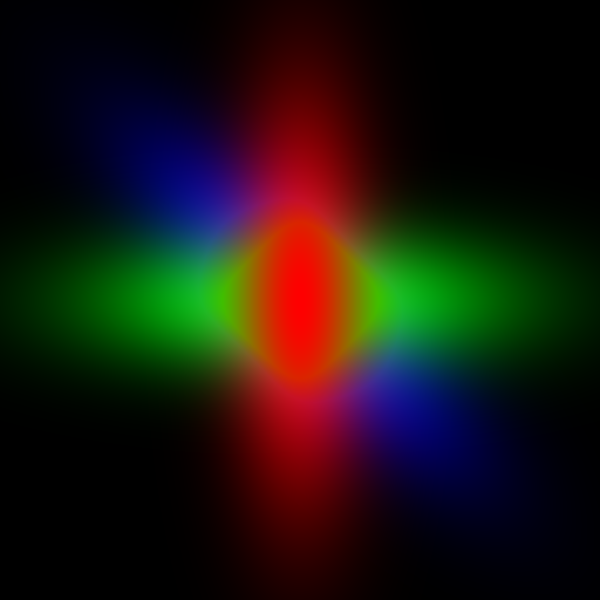} & \includegraphics[width=\cellsize,height=\cellsize]{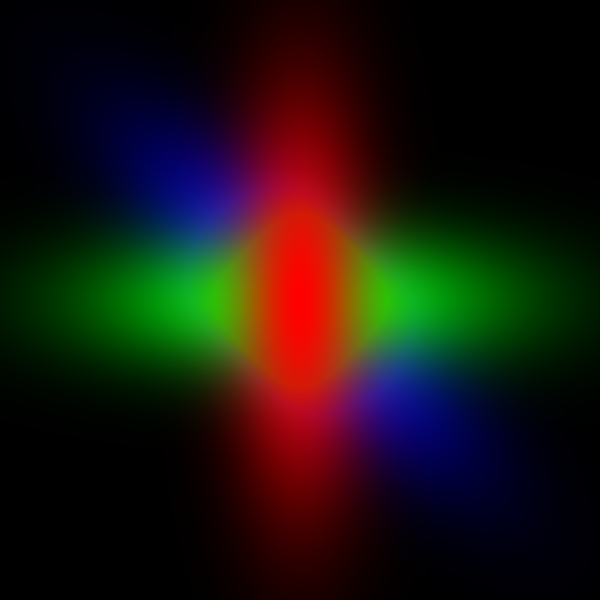} & \includegraphics[width=\cellsize,height=\cellsize]{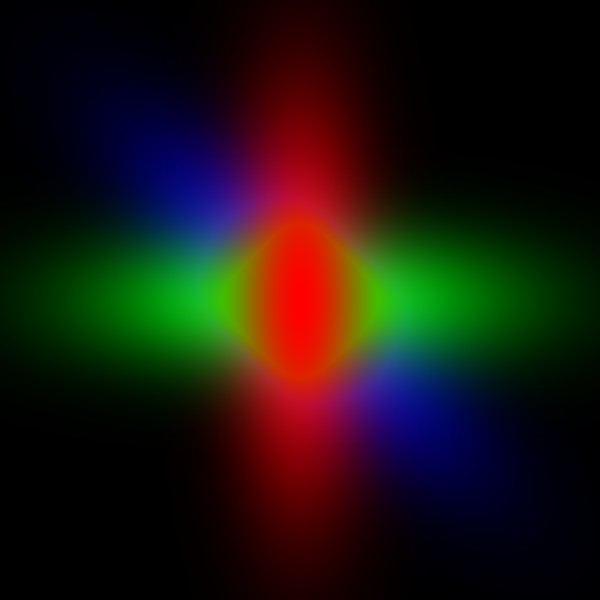} & \includegraphics[width=\cellsize,height=\cellsize]{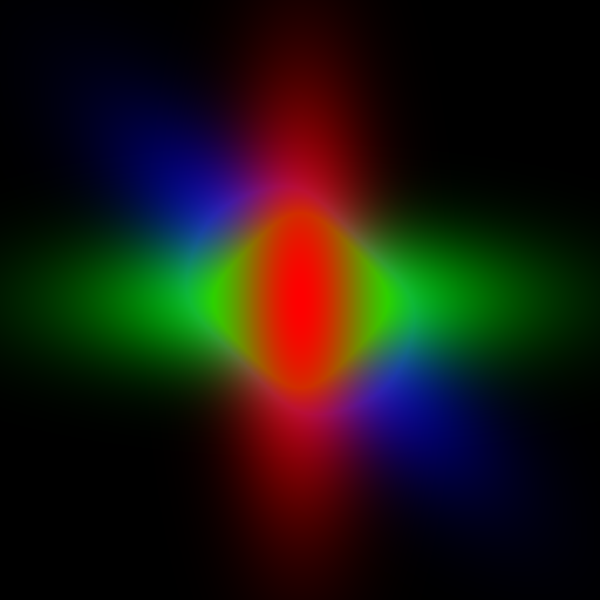} & \includegraphics[width=\cellsize,height=\cellsize]{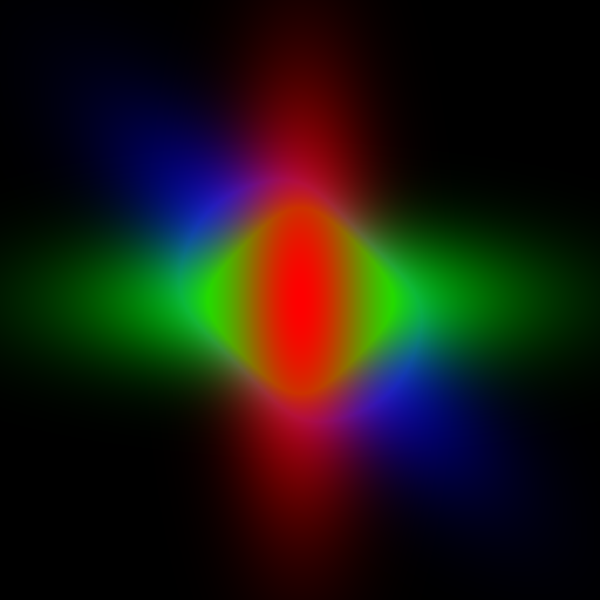} & \includegraphics[width=\cellsize,height=\cellsize]{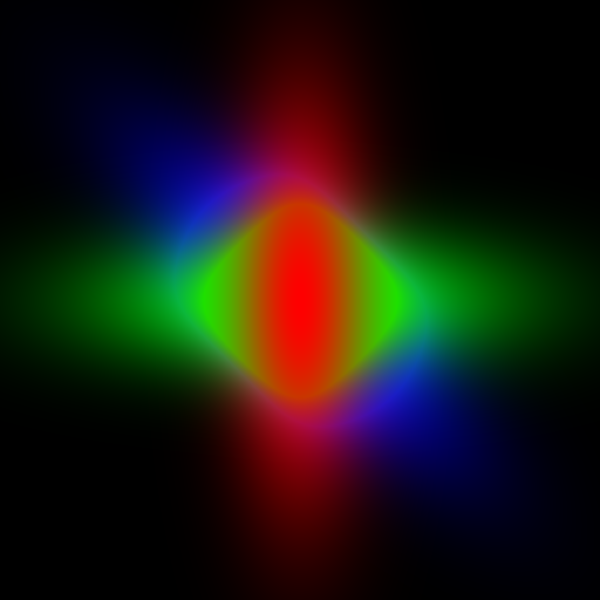} & \includegraphics[width=\cellsize,height=\cellsize]{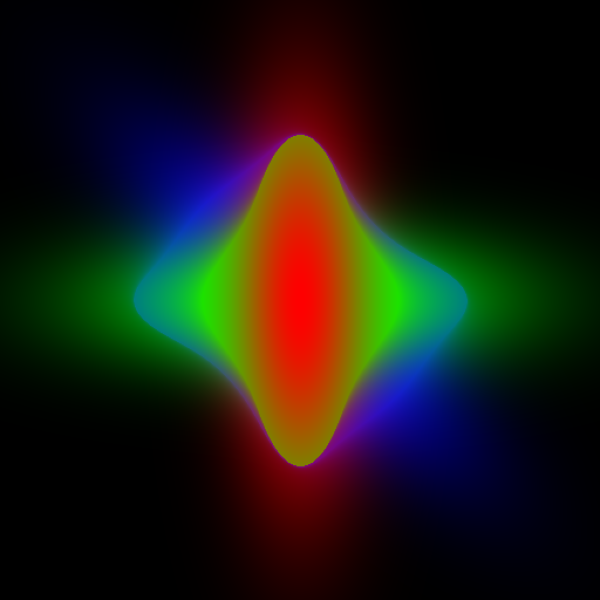} \\[0.5ex]
\raisebox{\dimexpr\cellsize/2-\height/2\relax}{\rotatebox{90}{\scriptsize\textbf{Intensity}}} & \input{\datapath n3_cyclic_cross_section_PowermLaw_v2.tex} & \input{\datapath n3_cyclic_cross_section_Exponential.tex} & \input{\datapath n3_cyclic_cross_section_Quadratic_cm0p5.tex} & \input{\datapath n3_cyclic_cross_section_citetvicini2021non_gamma0p5.tex} & \input{\datapath n3_cyclic_cross_section_Blend_gamma0p5.tex} & \input{\datapath n3_cyclic_cross_section_Linear.tex} & \input{\datapath n3_cyclic_cross_section_Quadratic_c0p5.tex} & \input{\datapath n3_cyclic_cross_section_Quadratic_c1.tex} & \input{\datapath n3_cyclic_cross_section_PowermLaw_vm1.tex} \\[0.5ex]
\raisebox{\dimexpr\cellsize/2-\height/2\relax}{\rotatebox{90}{\scriptsize\textbf{$1 - \bar T_{N+1}$}}} & \includegraphics[width=\cellsize,height=\cellsize]{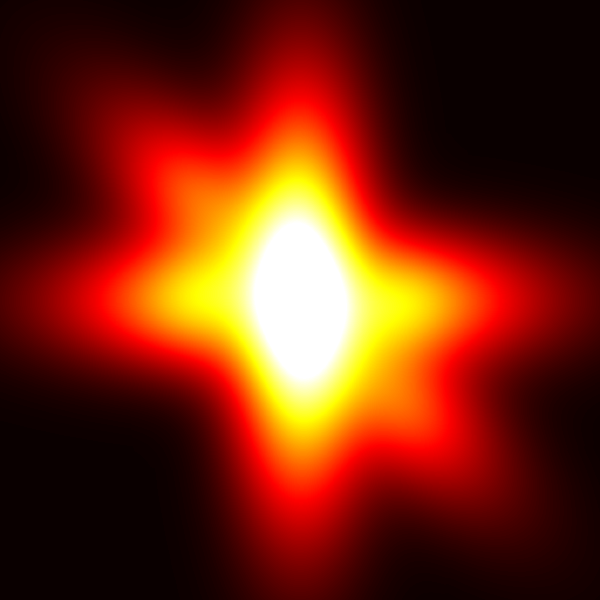} & \includegraphics[width=\cellsize,height=\cellsize]{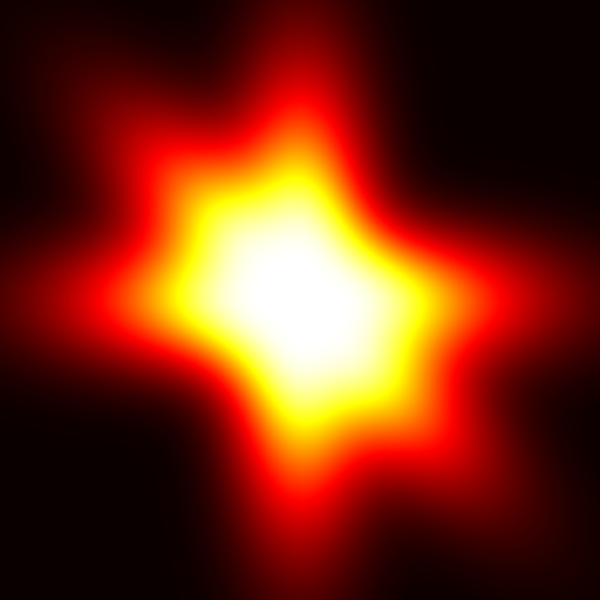} & \includegraphics[width=\cellsize,height=\cellsize]{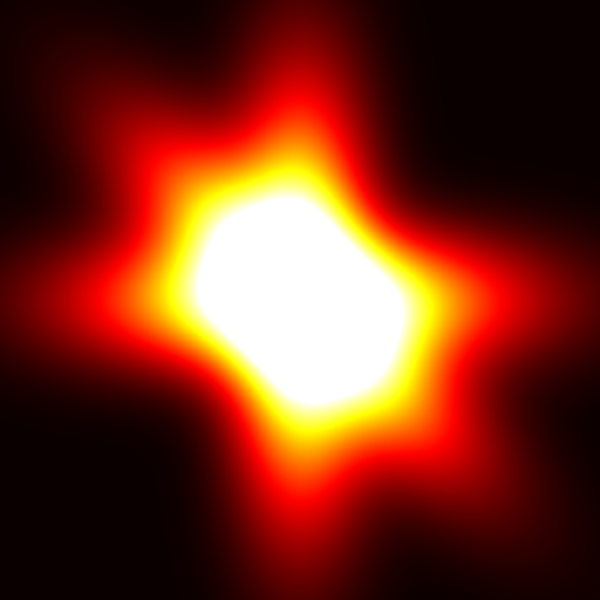} & \includegraphics[width=\cellsize,height=\cellsize]{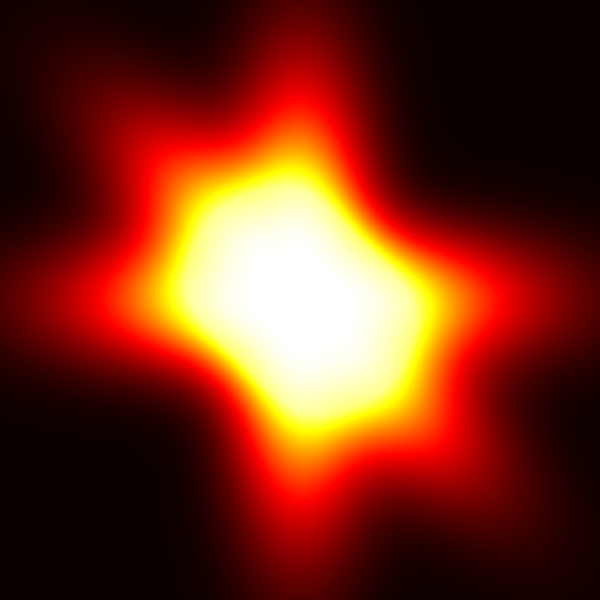} & \includegraphics[width=\cellsize,height=\cellsize]{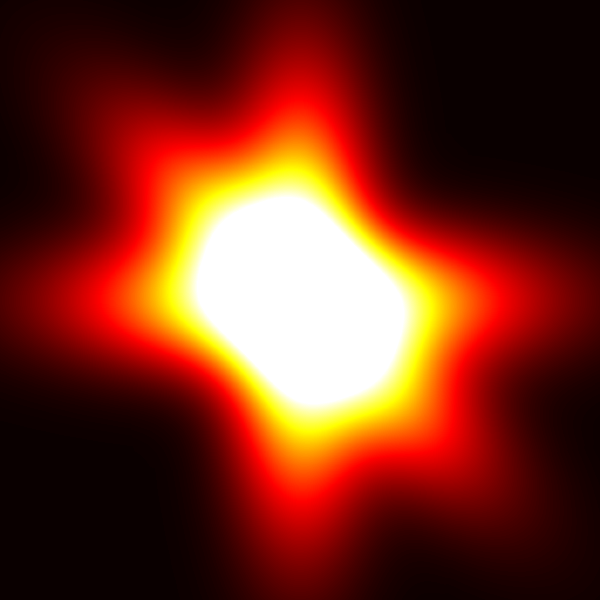} & \includegraphics[width=\cellsize,height=\cellsize]{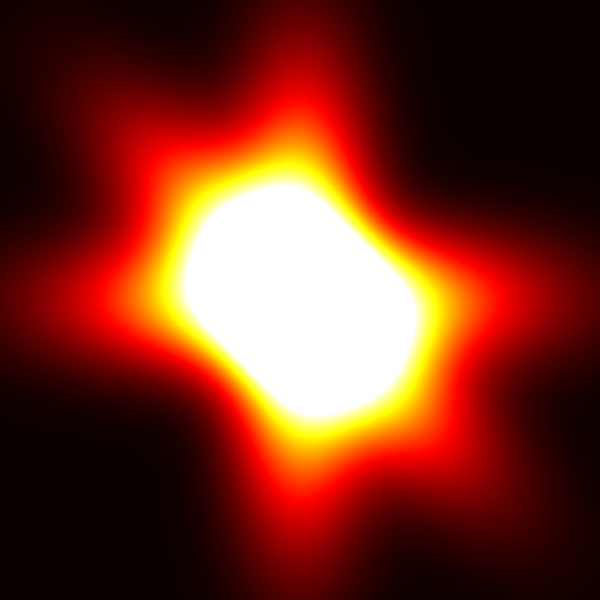} & \includegraphics[width=\cellsize,height=\cellsize]{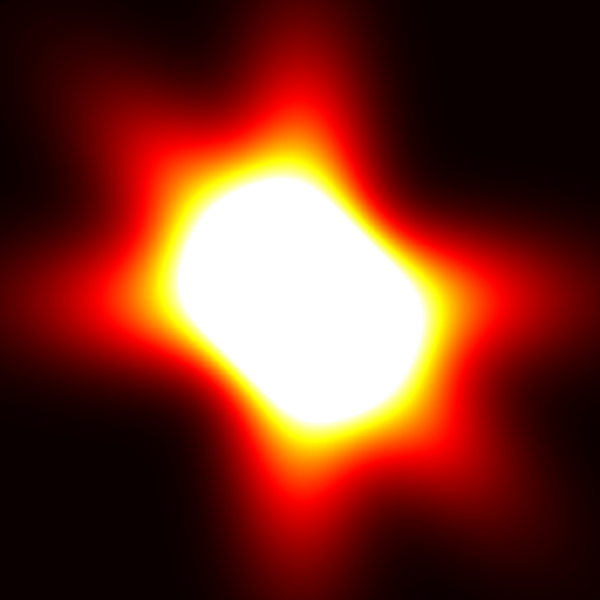} & \includegraphics[width=\cellsize,height=\cellsize]{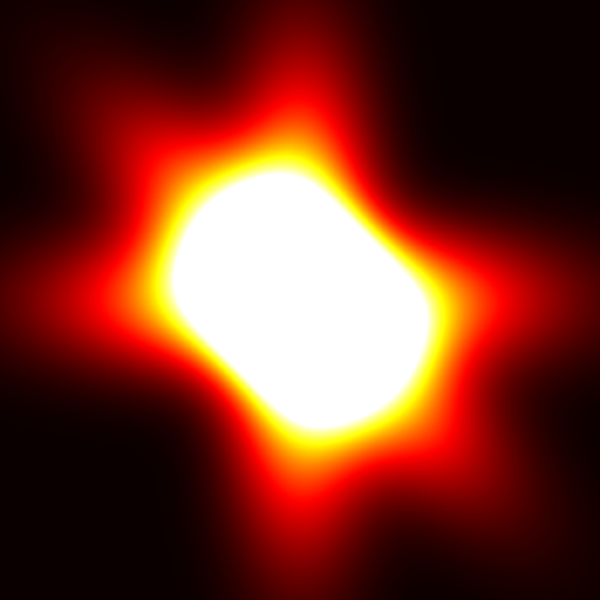} & \includegraphics[width=\cellsize,height=\cellsize]{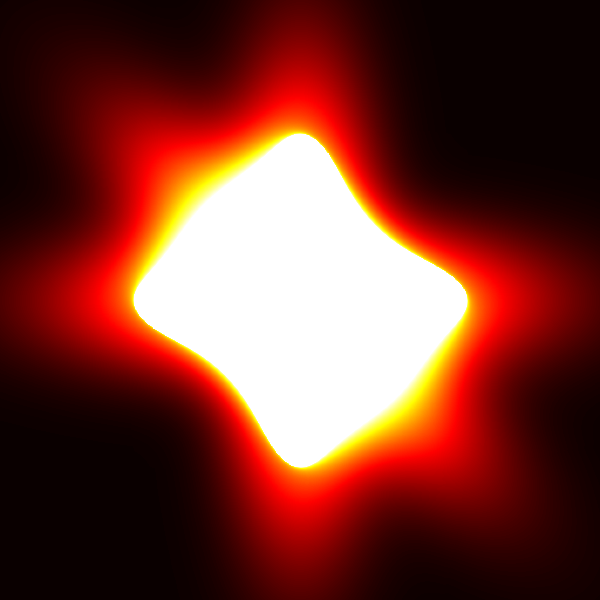} \\[0.5ex]
\multicolumn{10}{c}{\scriptsize{(b) $N=3$ anisotropic Gaussians, with different orientations.}}
\end{tabular}
\caption{\textbf{Blending comparison.} We render 3 Gaussians with different colors and depths (red is closer, then green, and blue at the back) with opacity $\aleph=1$, and (a) without anisotropy and growing in size and (b) anisotropic and with different rotations in the image plane. All splats are parallel to the image plane, and do not intersect. In both (a) and (b), the top row shows the result of blending, the middle row shows the scan-line at the center of the top image (with the black dashed line the sum of the three Gaussians), and the bottom row is the inverse discrete transmittance $1 - \bar T_{N+1}$. Faster-than-exponential transmittance (third to ninth columns) show sharper blending, which for extreme cases (\emph{power-law} with $v=-1$) results in unintuitive blending: This sharper blending behavior is in part due to transparency saturating to one (see middle row), which creates $C_1$ discontinuities on the blended primitives. Only the exponential transmittance does not show this behavior, due to the statistical uncorrelation between each splat, at the cost of more overdraws.  }
\label{fig:blending}
\end{figure*}

%% file: figures/combined_experiments/n3_concentric_cross_section_PowermLaw_v2.tex
\begin{tikzpicture}
\begin{axis}[width=\cellsize,height=\cellsize,scale only axis,yticklabels={},xmin=0,xmax=511,ymin=0,ymax=1.1,grid=major,xtick=\empty,ytick={0,0.25,0.5,0.75,1}]
\addplot[red, thick] coordinates { (0,0.000000) (4,0.000000) (8,0.000000) (12,0.000000) (16,0.000000) (20,0.000000) (24,0.000000) (28,0.000000) (32,0.000000) (36,0.000000) (40,0.000000) (44,0.000000) (48,0.000000) (52,0.000000) (56,0.000000) (60,0.000000) (64,0.000000) (68,0.000000) (72,0.000000) (76,0.000000) (80,0.000000) (84,0.000000) (88,0.000000) (92,0.000000) (96,0.000000) (100,0.000000) (104,0.000000) (108,0.001065) (112,0.001534) (116,0.002187) (120,0.003089) (124,0.004318) (128,0.005976) (132,0.008189) (136,0.011109) (140,0.014921) (144,0.019841) (148,0.026121) (152,0.034047) (156,0.043937) (160,0.056135) (164,0.071005) (168,0.088922) (172,0.110251) (176,0.135335) (180,0.164474) (184,0.197899) (188,0.235746) (192,0.278037) (196,0.324652) (200,0.375311) (204,0.429557) (208,0.486752) (212,0.546074) (216,0.606531) (220,0.666977) (224,0.726149) (228,0.782705) (232,0.835270) (236,0.882497) (240,0.923116) (244,0.955997) (248,0.980199) (252,0.995012) (256,1.000000) (260,0.995012) (264,0.980199) (268,0.955997) (272,0.923116) (276,0.882497) (280,0.835270) (284,0.782705) (288,0.726149) (292,0.666977) (296,0.606531) (300,0.546074) (304,0.486752) (308,0.429557) (312,0.375311) (316,0.324652) (320,0.278037) (324,0.235746) (328,0.197899) (332,0.164474) (336,0.135335) (340,0.110251) (344,0.088922) (348,0.071005) (352,0.056135) (356,0.043937) (360,0.034047) (364,0.026121) (368,0.019841) (372,0.014921) (376,0.011109) (380,0.008189) (384,0.005976) (388,0.004318) (392,0.003089) (396,0.002187) (400,0.001534) (404,0.001065) (408,0.000000) (412,0.000000) (416,0.000000) (420,0.000000) (424,0.000000) (428,0.000000) (432,0.000000) (436,0.000000) (440,0.000000) (444,0.000000) (448,0.000000) (452,0.000000) (456,0.000000) (460,0.000000) (464,0.000000) (468,0.000000) (472,0.000000) (476,0.000000) (480,0.000000) (484,0.000000) (488,0.000000) (492,0.000000) (496,0.000000) (500,0.000000) (504,0.000000) (508,0.000000) };
\addplot[green!70!black, thick] coordinates { (0,0.000000) (4,0.000000) (8,0.000000) (12,0.000000) (16,0.000000) (20,0.000000) (24,0.000000) (28,0.000000) (32,0.000000) (36,0.001204) (40,0.001534) (44,0.001946) (48,0.002457) (52,0.003089) (56,0.003866) (60,0.004817) (64,0.005976) (68,0.007381) (72,0.009075) (76,0.011109) (80,0.013538) (84,0.016426) (88,0.019841) (92,0.023860) (96,0.028566) (100,0.034047) (104,0.040401) (108,0.047577) (112,0.055877) (116,0.065300) (120,0.075916) (124,0.087782) (128,0.100925) (132,0.115334) (136,0.130947) (140,0.147637) (144,0.165201) (148,0.183345) (152,0.201687) (156,0.219758) (160,0.237022) (164,0.252905) (168,0.266846) (172,0.278345) (176,0.287019) (180,0.292642) (184,0.295171) (188,0.294747) (192,0.291668) (196,0.286353) (200,0.279288) (204,0.270981) (208,0.261920) (212,0.252542) (216,0.243220) (220,0.234255) (224,0.225878) (228,0.217295) (232,0.164730) (236,0.117503) (240,0.076884) (244,0.044003) (248,0.019801) (252,0.004988) (256,0.000000) (260,0.004988) (264,0.019801) (268,0.044003) (272,0.076884) (276,0.117503) (280,0.164730) (284,0.217295) (288,0.225878) (292,0.234255) (296,0.243220) (300,0.252542) (304,0.261920) (308,0.270981) (312,0.279288) (316,0.286353) (320,0.291668) (324,0.294747) (328,0.295171) (332,0.292642) (336,0.287019) (340,0.278345) (344,0.266846) (348,0.252905) (352,0.237022) (356,0.219758) (360,0.201687) (364,0.183345) (368,0.165201) (372,0.147637) (376,0.130947) (380,0.115334) (384,0.100925) (388,0.087782) (392,0.075916) (396,0.065300) (400,0.055877) (404,0.047577) (408,0.040401) (412,0.034047) (416,0.028566) (420,0.023860) (424,0.019841) (428,0.016426) (432,0.013538) (436,0.011109) (440,0.009075) (444,0.007381) (448,0.005976) (452,0.004817) (456,0.003866) (460,0.003089) (464,0.002457) (468,0.001946) (472,0.001534) (476,0.001204) (480,0.000000) (484,0.000000) (488,0.000000) (492,0.000000) (496,0.000000) (500,0.000000) (504,0.000000) (508,0.000000) };
\addplot[blue, thick] coordinates { (0,0.005976) (4,0.007004) (8,0.008189) (12,0.009550) (16,0.011109) (20,0.012891) (24,0.014921) (28,0.017227) (32,0.019841) (36,0.022712) (40,0.026002) (44,0.029686) (48,0.033798) (52,0.038370) (56,0.043432) (60,0.049015) (64,0.055143) (68,0.061838) (72,0.069115) (76,0.076980) (80,0.085429) (84,0.094445) (88,0.103999) (92,0.114044) (96,0.124514) (100,0.135326) (104,0.146379) (108,0.157092) (112,0.168009) (116,0.178655) (120,0.188836) (124,0.198347) (128,0.206984) (132,0.214543) (136,0.220840) (140,0.225711) (144,0.229032) (148,0.230716) (152,0.230731) (156,0.229095) (160,0.225881) (164,0.221213) (168,0.215259) (172,0.208221) (176,0.200323) (180,0.191801) (184,0.182886) (188,0.173798) (192,0.164736) (196,0.155872) (200,0.147346) (204,0.139271) (208,0.131730) (212,0.124779) (216,0.118454) (220,0.098768) (224,0.047973) (228,0.000000) (232,0.000000) (236,0.000000) (240,0.000000) (244,0.000000) (248,0.000000) (252,0.000000) (256,0.000000) (260,0.000000) (264,0.000000) (268,0.000000) (272,0.000000) (276,0.000000) (280,0.000000) (284,0.000000) (288,0.047973) (292,0.098768) (296,0.118454) (300,0.124779) (304,0.131730) (308,0.139271) (312,0.147346) (316,0.155872) (320,0.164736) (324,0.173798) (328,0.182886) (332,0.191801) (336,0.200323) (340,0.208221) (344,0.215259) (348,0.221213) (352,0.225881) (356,0.229095) (360,0.230731) (364,0.230716) (368,0.229032) (372,0.225711) (376,0.220840) (380,0.214543) (384,0.206984) (388,0.198347) (392,0.188836) (396,0.178655) (400,0.168009) (404,0.157092) (408,0.146379) (412,0.135326) (416,0.124514) (420,0.114044) (424,0.103999) (428,0.094445) (432,0.085429) (436,0.076980) (440,0.069115) (444,0.061838) (448,0.055143) (452,0.049015) (456,0.043432) (460,0.038370) (464,0.033798) (468,0.029686) (472,0.026002) (476,0.022712) (480,0.019841) (484,0.017227) (488,0.014921) (492,0.012891) (496,0.011109) (500,0.009550) (504,0.008189) (508,0.007004) };
\addplot[black, thick, dashed] coordinates { (0,0.005976) (4,0.007004) (8,0.008189) (12,0.009550) (16,0.011109) (20,0.012891) (24,0.014921) (28,0.017227) (32,0.019841) (36,0.023916) (40,0.027535) (44,0.031632) (48,0.036255) (52,0.041458) (56,0.047298) (60,0.053832) (64,0.061119) (68,0.069219) (72,0.078190) (76,0.088089) (80,0.098967) (84,0.110871) (88,0.123840) (92,0.137904) (96,0.153079) (100,0.169374) (104,0.186780) (108,0.205734) (112,0.225421) (116,0.246142) (120,0.267840) (124,0.290447) (128,0.313885) (132,0.338066) (136,0.362896) (140,0.388270) (144,0.414073) (148,0.440183) (152,0.466466) (156,0.492790) (160,0.519037) (164,0.545123) (168,0.571027) (172,0.596817) (176,0.622678) (180,0.648917) (184,0.675956) (188,0.704291) (192,0.734442) (196,0.766877) (200,0.801945) (204,0.839810) (208,0.880402) (212,0.923396) (216,0.968205) (220,1.000000) (224,1.000000) (228,1.000000) (232,1.000000) (236,1.000000) (240,1.000000) (244,1.000000) (248,1.000000) (252,1.000000) (256,1.000000) (260,1.000000) (264,1.000000) (268,1.000000) (272,1.000000) (276,1.000000) (280,1.000000) (284,1.000000) (288,1.000000) (292,1.000000) (296,0.968205) (300,0.923396) (304,0.880402) (308,0.839810) (312,0.801945) (316,0.766877) (320,0.734442) (324,0.704291) (328,0.675956) (332,0.648917) (336,0.622678) (340,0.596817) (344,0.571027) (348,0.545123) (352,0.519037) (356,0.492790) (360,0.466466) (364,0.440183) (368,0.414073) (372,0.388270) (376,0.362896) (380,0.338066) (384,0.313885) (388,0.290447) (392,0.267840) (396,0.246142) (400,0.225421) (404,0.205734) (408,0.186780) (412,0.169374) (416,0.153079) (420,0.137904) (424,0.123840) (428,0.110871) (432,0.098967) (436,0.088089) (440,0.078190) (444,0.069219) (448,0.061119) (452,0.053832) (456,0.047298) (460,0.041458) (464,0.036255) (468,0.031632) (472,0.027535) (476,0.023916) (480,0.019841) (484,0.017227) (488,0.014921) (492,0.012891) (496,0.011109) (500,0.009550) (504,0.008189) (508,0.007004) };
\end{axis}
\end{tikzpicture}

%% file: figures/combined_experiments/n3_concentric_cross_section_Exponential.tex
\begin{tikzpicture}
\begin{axis}[width=\cellsize,height=\cellsize,scale only axis,yticklabels={},xmin=0,xmax=511,ymin=0,ymax=1.1,grid=major,xtick=\empty,ytick={0,0.25,0.5,0.75,1}]
\addplot[red, thick] coordinates { (0,0.000000) (4,0.000000) (8,0.000000) (12,0.000000) (16,0.000000) (20,0.000000) (24,0.000000) (28,0.000000) (32,0.000000) (36,0.000000) (40,0.000000) (44,0.000000) (48,0.000000) (52,0.000000) (56,0.000000) (60,0.000000) (64,0.000000) (68,0.000000) (72,0.000000) (76,0.000000) (80,0.000000) (84,0.000000) (88,0.000000) (92,0.000000) (96,0.000000) (100,0.000000) (104,0.000000) (108,0.001065) (112,0.001534) (116,0.002187) (120,0.003089) (124,0.004318) (128,0.005976) (132,0.008189) (136,0.011109) (140,0.014921) (144,0.019841) (148,0.026121) (152,0.034047) (156,0.043937) (160,0.056135) (164,0.071005) (168,0.088922) (172,0.110251) (176,0.135335) (180,0.164474) (184,0.197899) (188,0.235746) (192,0.278037) (196,0.324652) (200,0.375311) (204,0.429557) (208,0.486752) (212,0.546074) (216,0.606531) (220,0.666977) (224,0.726149) (228,0.782705) (232,0.835270) (236,0.882497) (240,0.923116) (244,0.955997) (248,0.980199) (252,0.995012) (256,1.000000) (260,0.995012) (264,0.980199) (268,0.955997) (272,0.923116) (276,0.882497) (280,0.835270) (284,0.782705) (288,0.726149) (292,0.666977) (296,0.606531) (300,0.546074) (304,0.486752) (308,0.429557) (312,0.375311) (316,0.324652) (320,0.278037) (324,0.235746) (328,0.197899) (332,0.164474) (336,0.135335) (340,0.110251) (344,0.088922) (348,0.071005) (352,0.056135) (356,0.043937) (360,0.034047) (364,0.026121) (368,0.019841) (372,0.014921) (376,0.011109) (380,0.008189) (384,0.005976) (388,0.004318) (392,0.003089) (396,0.002187) (400,0.001534) (404,0.001065) (408,0.000000) (412,0.000000) (416,0.000000) (420,0.000000) (424,0.000000) (428,0.000000) (432,0.000000) (436,0.000000) (440,0.000000) (444,0.000000) (448,0.000000) (452,0.000000) (456,0.000000) (460,0.000000) (464,0.000000) (468,0.000000) (472,0.000000) (476,0.000000) (480,0.000000) (484,0.000000) (488,0.000000) (492,0.000000) (496,0.000000) (500,0.000000) (504,0.000000) (508,0.000000) };
\addplot[green!70!black, thick] coordinates { (0,0.000000) (4,0.000000) (8,0.000000) (12,0.000000) (16,0.000000) (20,0.000000) (24,0.000000) (28,0.000000) (32,0.000000) (36,0.001204) (40,0.001534) (44,0.001946) (48,0.002457) (52,0.003089) (56,0.003866) (60,0.004817) (64,0.005976) (68,0.007381) (72,0.009075) (76,0.011109) (80,0.013538) (84,0.016426) (88,0.019841) (92,0.023860) (96,0.028566) (100,0.034047) (104,0.040401) (108,0.047678) (112,0.056049) (116,0.065585) (120,0.076384) (124,0.088538) (128,0.102126) (132,0.117211) (136,0.133832) (140,0.151993) (144,0.171656) (148,0.192729) (152,0.215055) (156,0.238396) (160,0.262430) (164,0.286732) (168,0.310776) (172,0.333933) (176,0.355474) (180,0.374592) (184,0.390425) (188,0.402091) (192,0.408742) (196,0.409619) (200,0.404114) (204,0.391841) (208,0.372694) (212,0.346903) (216,0.315066) (220,0.278164) (224,0.237546) (228,0.194877) (232,0.152065) (236,0.111153) (240,0.074198) (244,0.043131) (248,0.019626) (252,0.004976) (256,0.000000) (260,0.004976) (264,0.019626) (268,0.043131) (272,0.074198) (276,0.111153) (280,0.152065) (284,0.194877) (288,0.237546) (292,0.278164) (296,0.315066) (300,0.346903) (304,0.372694) (308,0.391841) (312,0.404114) (316,0.409619) (320,0.408742) (324,0.402091) (328,0.390425) (332,0.374592) (336,0.355474) (340,0.333933) (344,0.310776) (348,0.286732) (352,0.262430) (356,0.238396) (360,0.215055) (364,0.192729) (368,0.171656) (372,0.151993) (376,0.133832) (380,0.117211) (384,0.102126) (388,0.088538) (392,0.076384) (396,0.065585) (400,0.056049) (404,0.047678) (408,0.040401) (412,0.034047) (416,0.028566) (420,0.023860) (424,0.019841) (428,0.016426) (432,0.013538) (436,0.011109) (440,0.009075) (444,0.007381) (448,0.005976) (452,0.004817) (456,0.003866) (460,0.003089) (464,0.002457) (468,0.001946) (472,0.001534) (476,0.001204) (480,0.000000) (484,0.000000) (488,0.000000) (492,0.000000) (496,0.000000) (500,0.000000) (504,0.000000) (508,0.000000) };
\addplot[blue, thick] coordinates { (0,0.005976) (4,0.007004) (8,0.008189) (12,0.009550) (16,0.011109) (20,0.012891) (24,0.014921) (28,0.017227) (32,0.019841) (36,0.022767) (40,0.026081) (44,0.029801) (48,0.033964) (52,0.038606) (56,0.043767) (60,0.049485) (64,0.055799) (68,0.062746) (72,0.070361) (76,0.078676) (80,0.087718) (84,0.097509) (88,0.108063) (92,0.119385) (96,0.131469) (100,0.144296) (104,0.157829) (108,0.171835) (112,0.186503) (116,0.201608) (120,0.217011) (124,0.232538) (128,0.247981) (132,0.263095) (136,0.277597) (140,0.291164) (144,0.303440) (148,0.314039) (152,0.322554) (156,0.328572) (160,0.331690) (164,0.331540) (168,0.327810) (172,0.320278) (176,0.308840) (180,0.293537) (184,0.274579) (188,0.252357) (192,0.227445) (196,0.200582) (200,0.172645) (204,0.144591) (208,0.117400) (212,0.092000) (216,0.069191) (220,0.049576) (224,0.033514) (228,0.021086) (232,0.012108) (236,0.006155) (240,0.002632) (244,0.000862) (248,0.000174) (252,0.000011) (256,0.000000) (260,0.000011) (264,0.000174) (268,0.000862) (272,0.002632) (276,0.006155) (280,0.012108) (284,0.021086) (288,0.033514) (292,0.049576) (296,0.069191) (300,0.092000) (304,0.117400) (308,0.144591) (312,0.172645) (316,0.200582) (320,0.227445) (324,0.252357) (328,0.274579) (332,0.293537) (336,0.308840) (340,0.320278) (344,0.327810) (348,0.331540) (352,0.331690) (356,0.328572) (360,0.322554) (364,0.314039) (368,0.303440) (372,0.291164) (376,0.277597) (380,0.263095) (384,0.247981) (388,0.232538) (392,0.217011) (396,0.201608) (400,0.186503) (404,0.171835) (408,0.157829) (412,0.144296) (416,0.131469) (420,0.119385) (424,0.108063) (428,0.097509) (432,0.087718) (436,0.078676) (440,0.070361) (444,0.062746) (448,0.055799) (452,0.049485) (456,0.043767) (460,0.038606) (464,0.033964) (468,0.029801) (472,0.026081) (476,0.022767) (480,0.019841) (484,0.017227) (488,0.014921) (492,0.012891) (496,0.011109) (500,0.009550) (504,0.008189) (508,0.007004) };
\addplot[black, thick, dashed] coordinates { (0,0.005976) (4,0.007004) (8,0.008189) (12,0.009550) (16,0.011109) (20,0.012891) (24,0.014921) (28,0.017227) (32,0.019841) (36,0.023971) (40,0.027615) (44,0.031747) (48,0.036421) (52,0.041695) (56,0.047633) (60,0.054303) (64,0.061775) (68,0.070127) (72,0.079436) (76,0.089785) (80,0.101256) (84,0.113935) (88,0.127904) (92,0.143245) (96,0.160035) (100,0.178343) (104,0.198231) (108,0.220578) (112,0.244086) (116,0.269381) (120,0.296483) (124,0.325393) (128,0.356083) (132,0.388495) (136,0.422538) (140,0.458078) (144,0.494937) (148,0.532889) (152,0.571656) (156,0.610905) (160,0.650255) (164,0.689277) (168,0.727508) (172,0.764461) (176,0.799649) (180,0.832604) (184,0.862902) (188,0.890194) (192,0.914224) (196,0.934854) (200,0.952070) (204,0.965989) (208,0.976847) (212,0.984977) (216,0.990787) (220,0.994717) (224,0.997209) (228,0.998668) (232,0.999443) (236,0.999805) (240,0.999947) (244,0.999990) (248,0.999999) (252,1.000000) (256,1.000000) (260,1.000000) (264,0.999999) (268,0.999990) (272,0.999947) (276,0.999805) (280,0.999443) (284,0.998668) (288,0.997209) (292,0.994717) (296,0.990787) (300,0.984977) (304,0.976847) (308,0.965989) (312,0.952070) (316,0.934854) (320,0.914224) (324,0.890194) (328,0.862902) (332,0.832604) (336,0.799649) (340,0.764461) (344,0.727508) (348,0.689277) (352,0.650255) (356,0.610905) (360,0.571656) (364,0.532889) (368,0.494937) (372,0.458078) (376,0.422538) (380,0.388495) (384,0.356083) (388,0.325393) (392,0.296483) (396,0.269381) (400,0.244086) (404,0.220578) (408,0.198231) (412,0.178343) (416,0.160035) (420,0.143245) (424,0.127904) (428,0.113935) (432,0.101256) (436,0.089785) (440,0.079436) (444,0.070127) (448,0.061775) (452,0.054303) (456,0.047633) (460,0.041695) (464,0.036421) (468,0.031747) (472,0.027615) (476,0.023971) (480,0.019841) (484,0.017227) (488,0.014921) (492,0.012891) (496,0.011109) (500,0.009550) (504,0.008189) (508,0.007004) };
\end{axis}
\end{tikzpicture}

%% file: figures/combined_experiments/n3_concentric_cross_section_Quadratic_cm0p5.tex
\begin{tikzpicture}
\begin{axis}[width=\cellsize,height=\cellsize,scale only axis,yticklabels={},xmin=0,xmax=511,ymin=0,ymax=1.1,grid=major,xtick=\empty,ytick={0,0.25,0.5,0.75,1}]
\addplot[red, thick] coordinates { (0,0.000000) (4,0.000000) (8,0.000000) (12,0.000000) (16,0.000000) (20,0.000000) (24,0.000000) (28,0.000000) (32,0.000000) (36,0.000000) (40,0.000000) (44,0.000000) (48,0.000000) (52,0.000000) (56,0.000000) (60,0.000000) (64,0.000000) (68,0.000000) (72,0.000000) (76,0.000000) (80,0.000000) (84,0.000000) (88,0.000000) (92,0.000000) (96,0.000000) (100,0.000000) (104,0.000000) (108,0.001065) (112,0.001534) (116,0.002187) (120,0.003089) (124,0.004318) (128,0.005976) (132,0.008189) (136,0.011109) (140,0.014921) (144,0.019841) (148,0.026121) (152,0.034047) (156,0.043937) (160,0.056135) (164,0.071005) (168,0.088922) (172,0.110251) (176,0.135335) (180,0.164474) (184,0.197899) (188,0.235746) (192,0.278037) (196,0.324652) (200,0.375311) (204,0.429557) (208,0.486752) (212,0.546074) (216,0.606531) (220,0.666977) (224,0.726149) (228,0.782705) (232,0.835270) (236,0.882497) (240,0.923116) (244,0.955997) (248,0.980199) (252,0.995012) (256,1.000000) (260,0.995012) (264,0.980199) (268,0.955997) (272,0.923116) (276,0.882497) (280,0.835270) (284,0.782705) (288,0.726149) (292,0.666977) (296,0.606531) (300,0.546074) (304,0.486752) (308,0.429557) (312,0.375311) (316,0.324652) (320,0.278037) (324,0.235746) (328,0.197899) (332,0.164474) (336,0.135335) (340,0.110251) (344,0.088922) (348,0.071005) (352,0.056135) (356,0.043937) (360,0.034047) (364,0.026121) (368,0.019841) (372,0.014921) (376,0.011109) (380,0.008189) (384,0.005976) (388,0.004318) (392,0.003089) (396,0.002187) (400,0.001534) (404,0.001065) (408,0.000000) (412,0.000000) (416,0.000000) (420,0.000000) (424,0.000000) (428,0.000000) (432,0.000000) (436,0.000000) (440,0.000000) (444,0.000000) (448,0.000000) (452,0.000000) (456,0.000000) (460,0.000000) (464,0.000000) (468,0.000000) (472,0.000000) (476,0.000000) (480,0.000000) (484,0.000000) (488,0.000000) (492,0.000000) (496,0.000000) (500,0.000000) (504,0.000000) (508,0.000000) };
\addplot[green!70!black, thick] coordinates { (0,0.000000) (4,0.000000) (8,0.000000) (12,0.000000) (16,0.000000) (20,0.000000) (24,0.000000) (28,0.000000) (32,0.000000) (36,0.001204) (40,0.001534) (44,0.001946) (48,0.002457) (52,0.003089) (56,0.003866) (60,0.004817) (64,0.005976) (68,0.007381) (72,0.009075) (76,0.011109) (80,0.013538) (84,0.016426) (88,0.019841) (92,0.023860) (96,0.028566) (100,0.034047) (104,0.040401) (108,0.047703) (112,0.056092) (116,0.065657) (120,0.076502) (124,0.088730) (128,0.102433) (132,0.117695) (136,0.134584) (140,0.153144) (144,0.173393) (148,0.195314) (152,0.218845) (156,0.243874) (160,0.270234) (164,0.297690) (168,0.325942) (172,0.354622) (176,0.383293) (180,0.411462) (184,0.438588) (188,0.464106) (192,0.487448) (196,0.508075) (200,0.525510) (204,0.539374) (208,0.513248) (212,0.453926) (216,0.393469) (220,0.333023) (224,0.273851) (228,0.217295) (232,0.164730) (236,0.117503) (240,0.076884) (244,0.044003) (248,0.019801) (252,0.004988) (256,0.000000) (260,0.004988) (264,0.019801) (268,0.044003) (272,0.076884) (276,0.117503) (280,0.164730) (284,0.217295) (288,0.273851) (292,0.333023) (296,0.393469) (300,0.453926) (304,0.513248) (308,0.539374) (312,0.525510) (316,0.508075) (320,0.487448) (324,0.464106) (328,0.438588) (332,0.411462) (336,0.383293) (340,0.354622) (344,0.325942) (348,0.297690) (352,0.270234) (356,0.243874) (360,0.218845) (364,0.195314) (368,0.173393) (372,0.153144) (376,0.134584) (380,0.117695) (384,0.102433) (388,0.088730) (392,0.076502) (396,0.065657) (400,0.056092) (404,0.047703) (408,0.040401) (412,0.034047) (416,0.028566) (420,0.023860) (424,0.019841) (428,0.016426) (432,0.013538) (436,0.011109) (440,0.009075) (444,0.007381) (448,0.005976) (452,0.004817) (456,0.003866) (460,0.003089) (464,0.002457) (468,0.001946) (472,0.001534) (476,0.001204) (480,0.000000) (484,0.000000) (488,0.000000) (492,0.000000) (496,0.000000) (500,0.000000) (504,0.000000) (508,0.000000) };
\addplot[blue, thick] coordinates { (0,0.005976) (4,0.007004) (8,0.008189) (12,0.009550) (16,0.011109) (20,0.012891) (24,0.014921) (28,0.017227) (32,0.019841) (36,0.022780) (40,0.026101) (44,0.029831) (48,0.034006) (52,0.038666) (56,0.043852) (60,0.049605) (64,0.055967) (68,0.062979) (72,0.070683) (76,0.079118) (80,0.088320) (84,0.098323) (88,0.109157) (92,0.120844) (96,0.133402) (100,0.146839) (104,0.161152) (108,0.176233) (112,0.192192) (116,0.208921) (120,0.226350) (124,0.244390) (128,0.262924) (132,0.281811) (136,0.300881) (140,0.319930) (144,0.338724) (148,0.356991) (152,0.374427) (156,0.390695) (160,0.405423) (164,0.418216) (168,0.428660) (172,0.436332) (176,0.440812) (180,0.424064) (184,0.363513) (188,0.300148) (192,0.234515) (196,0.167273) (200,0.099179) (204,0.031068) (208,0.000000) (212,0.000000) (216,0.000000) (220,0.000000) (224,0.000000) (228,0.000000) (232,0.000000) (236,0.000000) (240,0.000000) (244,0.000000) (248,0.000000) (252,0.000000) (256,0.000000) (260,0.000000) (264,0.000000) (268,0.000000) (272,0.000000) (276,0.000000) (280,0.000000) (284,0.000000) (288,0.000000) (292,0.000000) (296,0.000000) (300,0.000000) (304,0.000000) (308,0.031068) (312,0.099179) (316,0.167273) (320,0.234515) (324,0.300148) (328,0.363513) (332,0.424064) (336,0.440812) (340,0.436332) (344,0.428660) (348,0.418216) (352,0.405423) (356,0.390695) (360,0.374427) (364,0.356991) (368,0.338724) (372,0.319930) (376,0.300881) (380,0.281811) (384,0.262924) (388,0.244390) (392,0.226350) (396,0.208921) (400,0.192192) (404,0.176233) (408,0.161152) (412,0.146839) (416,0.133402) (420,0.120844) (424,0.109157) (428,0.098323) (432,0.088320) (436,0.079118) (440,0.070683) (444,0.062979) (448,0.055967) (452,0.049605) (456,0.043852) (460,0.038666) (464,0.034006) (468,0.029831) (472,0.026101) (476,0.022780) (480,0.019841) (484,0.017227) (488,0.014921) (492,0.012891) (496,0.011109) (500,0.009550) (504,0.008189) (508,0.007004) };
\addplot[black, thick, dashed] coordinates { (0,0.005976) (4,0.007004) (8,0.008189) (12,0.009550) (16,0.011109) (20,0.012891) (24,0.014921) (28,0.017227) (32,0.019841) (36,0.023984) (40,0.027635) (44,0.031776) (48,0.036462) (52,0.041755) (56,0.047718) (60,0.054422) (64,0.061943) (68,0.070360) (72,0.079758) (76,0.090227) (80,0.101858) (84,0.114749) (88,0.128998) (92,0.144704) (96,0.161968) (100,0.180886) (104,0.201553) (108,0.225001) (112,0.249818) (116,0.276765) (120,0.305942) (124,0.337437) (128,0.371333) (132,0.407695) (136,0.446573) (140,0.487995) (144,0.531958) (148,0.578426) (152,0.627320) (156,0.678506) (160,0.731791) (164,0.786911) (168,0.843524) (172,0.901204) (176,0.959441) (180,1.000000) (184,1.000000) (188,1.000000) (192,1.000000) (196,1.000000) (200,1.000000) (204,1.000000) (208,1.000000) (212,1.000000) (216,1.000000) (220,1.000000) (224,1.000000) (228,1.000000) (232,1.000000) (236,1.000000) (240,1.000000) (244,1.000000) (248,1.000000) (252,1.000000) (256,1.000000) (260,1.000000) (264,1.000000) (268,1.000000) (272,1.000000) (276,1.000000) (280,1.000000) (284,1.000000) (288,1.000000) (292,1.000000) (296,1.000000) (300,1.000000) (304,1.000000) (308,1.000000) (312,1.000000) (316,1.000000) (320,1.000000) (324,1.000000) (328,1.000000) (332,1.000000) (336,0.959441) (340,0.901204) (344,0.843524) (348,0.786911) (352,0.731791) (356,0.678506) (360,0.627320) (364,0.578426) (368,0.531958) (372,0.487995) (376,0.446573) (380,0.407695) (384,0.371333) (388,0.337437) (392,0.305942) (396,0.276765) (400,0.249818) (404,0.225001) (408,0.201553) (412,0.180886) (416,0.161968) (420,0.144704) (424,0.128998) (428,0.114749) (432,0.101858) (436,0.090227) (440,0.079758) (444,0.070360) (448,0.061943) (452,0.054422) (456,0.047718) (460,0.041755) (464,0.036462) (468,0.031776) (472,0.027635) (476,0.023984) (480,0.019841) (484,0.017227) (488,0.014921) (492,0.012891) (496,0.011109) (500,0.009550) (504,0.008189) (508,0.007004) };
\end{axis}
\end{tikzpicture}

%% file: figures/combined_experiments/n3_concentric_cross_section_citetvicini2021non_gamma0p5.tex
\begin{tikzpicture}
\begin{axis}[width=\cellsize,height=\cellsize,scale only axis,yticklabels={},xmin=0,xmax=511,ymin=0,ymax=1.1,grid=major,xtick=\empty,ytick={0,0.25,0.5,0.75,1}]
\addplot[red, thick] coordinates { (0,0.000000) (4,0.000000) (8,0.000000) (12,0.000000) (16,0.000000) (20,0.000000) (24,0.000000) (28,0.000000) (32,0.000000) (36,0.000000) (40,0.000000) (44,0.000000) (48,0.000000) (52,0.000000) (56,0.000000) (60,0.000000) (64,0.000000) (68,0.000000) (72,0.000000) (76,0.000000) (80,0.000000) (84,0.000000) (88,0.000000) (92,0.000000) (96,0.000000) (100,0.000000) (104,0.000000) (108,0.001065) (112,0.001534) (116,0.002187) (120,0.003089) (124,0.004318) (128,0.005976) (132,0.008189) (136,0.011109) (140,0.014921) (144,0.019841) (148,0.026121) (152,0.034047) (156,0.043937) (160,0.056135) (164,0.071005) (168,0.088922) (172,0.110251) (176,0.135335) (180,0.164474) (184,0.197899) (188,0.235746) (192,0.278037) (196,0.324652) (200,0.375311) (204,0.429557) (208,0.486752) (212,0.546074) (216,0.606531) (220,0.666977) (224,0.726149) (228,0.782705) (232,0.835270) (236,0.882497) (240,0.923116) (244,0.955997) (248,0.980199) (252,0.995012) (256,1.000000) (260,0.995012) (264,0.980199) (268,0.955997) (272,0.923116) (276,0.882497) (280,0.835270) (284,0.782705) (288,0.726149) (292,0.666977) (296,0.606531) (300,0.546074) (304,0.486752) (308,0.429557) (312,0.375311) (316,0.324652) (320,0.278037) (324,0.235746) (328,0.197899) (332,0.164474) (336,0.135335) (340,0.110251) (344,0.088922) (348,0.071005) (352,0.056135) (356,0.043937) (360,0.034047) (364,0.026121) (368,0.019841) (372,0.014921) (376,0.011109) (380,0.008189) (384,0.005976) (388,0.004318) (392,0.003089) (396,0.002187) (400,0.001534) (404,0.001065) (408,0.000000) (412,0.000000) (416,0.000000) (420,0.000000) (424,0.000000) (428,0.000000) (432,0.000000) (436,0.000000) (440,0.000000) (444,0.000000) (448,0.000000) (452,0.000000) (456,0.000000) (460,0.000000) (464,0.000000) (468,0.000000) (472,0.000000) (476,0.000000) (480,0.000000) (484,0.000000) (488,0.000000) (492,0.000000) (496,0.000000) (500,0.000000) (504,0.000000) (508,0.000000) };
\addplot[green!70!black, thick] coordinates { (0,0.000000) (4,0.000000) (8,0.000000) (12,0.000000) (16,0.000000) (20,0.000000) (24,0.000000) (28,0.000000) (32,0.000000) (36,0.001204) (40,0.001534) (44,0.001946) (48,0.002457) (52,0.003089) (56,0.003866) (60,0.004817) (64,0.005976) (68,0.007381) (72,0.009075) (76,0.011109) (80,0.013538) (84,0.016426) (88,0.019841) (92,0.023860) (96,0.028566) (100,0.034047) (104,0.040401) (108,0.047703) (112,0.056092) (116,0.065657) (120,0.076502) (124,0.088730) (128,0.102433) (132,0.117695) (136,0.134584) (140,0.153144) (144,0.173393) (148,0.195314) (152,0.218845) (156,0.243874) (160,0.270234) (164,0.297690) (168,0.325942) (172,0.354622) (176,0.383293) (180,0.411462) (184,0.438588) (188,0.464106) (192,0.487448) (196,0.508075) (200,0.514402) (204,0.481142) (208,0.442971) (212,0.400414) (216,0.354267) (220,0.305594) (224,0.255699) (228,0.206086) (232,0.158397) (236,0.114328) (240,0.075541) (244,0.043567) (248,0.019714) (252,0.004982) (256,0.000000) (260,0.004982) (264,0.019714) (268,0.043567) (272,0.075541) (276,0.114328) (280,0.158397) (284,0.206086) (288,0.255699) (292,0.305594) (296,0.354267) (300,0.400414) (304,0.442971) (308,0.481142) (312,0.514402) (316,0.508075) (320,0.487448) (324,0.464106) (328,0.438588) (332,0.411462) (336,0.383293) (340,0.354622) (344,0.325942) (348,0.297690) (352,0.270234) (356,0.243874) (360,0.218845) (364,0.195314) (368,0.173393) (372,0.153144) (376,0.134584) (380,0.117695) (384,0.102433) (388,0.088730) (392,0.076502) (396,0.065657) (400,0.056092) (404,0.047703) (408,0.040401) (412,0.034047) (416,0.028566) (420,0.023860) (424,0.019841) (428,0.016426) (432,0.013538) (436,0.011109) (440,0.009075) (444,0.007381) (448,0.005976) (452,0.004817) (456,0.003866) (460,0.003089) (464,0.002457) (468,0.001946) (472,0.001534) (476,0.001204) (480,0.000000) (484,0.000000) (488,0.000000) (492,0.000000) (496,0.000000) (500,0.000000) (504,0.000000) (508,0.000000) };
\addplot[blue, thick] coordinates { (0,0.005976) (4,0.007004) (8,0.008189) (12,0.009550) (16,0.011109) (20,0.012891) (24,0.014921) (28,0.017227) (32,0.019841) (36,0.022780) (40,0.026101) (44,0.029831) (48,0.034006) (52,0.038666) (56,0.043852) (60,0.049605) (64,0.055967) (68,0.062979) (72,0.070683) (76,0.079118) (80,0.088320) (84,0.098323) (88,0.109157) (92,0.120844) (96,0.133402) (100,0.146839) (104,0.161152) (108,0.176237) (112,0.192201) (116,0.208937) (120,0.226378) (124,0.244439) (128,0.263009) (132,0.281957) (136,0.301125) (140,0.320332) (144,0.339376) (148,0.358030) (152,0.376056) (156,0.393203) (160,0.409221) (164,0.423873) (168,0.436942) (172,0.417358) (176,0.381196) (180,0.340365) (184,0.294964) (188,0.245244) (192,0.191627) (196,0.134700) (200,0.086322) (204,0.072295) (208,0.058700) (212,0.046000) (216,0.034596) (220,0.024788) (224,0.016757) (228,0.010543) (232,0.006054) (236,0.003077) (240,0.001316) (244,0.000431) (248,0.000087) (252,0.000006) (256,0.000000) (260,0.000006) (264,0.000087) (268,0.000431) (272,0.001316) (276,0.003077) (280,0.006054) (284,0.010543) (288,0.016757) (292,0.024788) (296,0.034596) (300,0.046000) (304,0.058700) (308,0.072295) (312,0.086322) (316,0.134700) (320,0.191627) (324,0.245244) (328,0.294964) (332,0.340365) (336,0.381196) (340,0.417358) (344,0.436942) (348,0.423873) (352,0.409221) (356,0.393203) (360,0.376056) (364,0.358030) (368,0.339376) (372,0.320332) (376,0.301125) (380,0.281957) (384,0.263009) (388,0.244439) (392,0.226378) (396,0.208937) (400,0.192201) (404,0.176237) (408,0.161152) (412,0.146839) (416,0.133402) (420,0.120844) (424,0.109157) (428,0.098323) (432,0.088320) (436,0.079118) (440,0.070683) (444,0.062979) (448,0.055967) (452,0.049605) (456,0.043852) (460,0.038666) (464,0.034006) (468,0.029831) (472,0.026101) (476,0.022780) (480,0.019841) (484,0.017227) (488,0.014921) (492,0.012891) (496,0.011109) (500,0.009550) (504,0.008189) (508,0.007004) };
\addplot[black, thick, dashed] coordinates { (0,0.005976) (4,0.007004) (8,0.008189) (12,0.009550) (16,0.011109) (20,0.012891) (24,0.014921) (28,0.017227) (32,0.019841) (36,0.023984) (40,0.027635) (44,0.031776) (48,0.036462) (52,0.041755) (56,0.047718) (60,0.054422) (64,0.061943) (68,0.070360) (72,0.079758) (76,0.090227) (80,0.101858) (84,0.114749) (88,0.128998) (92,0.144704) (96,0.161968) (100,0.180886) (104,0.201553) (108,0.225005) (112,0.249826) (116,0.276781) (120,0.305969) (124,0.337486) (128,0.371418) (132,0.407840) (136,0.446817) (140,0.488397) (144,0.532610) (148,0.579465) (152,0.628948) (156,0.681014) (160,0.735590) (164,0.792568) (168,0.851806) (172,0.882231) (176,0.899825) (180,0.916302) (184,0.931451) (188,0.945097) (192,0.957112) (196,0.967427) (200,0.976035) (204,0.982995) (208,0.988423) (212,0.992489) (216,0.995394) (220,0.997359) (224,0.998604) (228,0.999334) (232,0.999721) (236,0.999902) (240,0.999973) (244,0.999995) (248,1.000000) (252,1.000000) (256,1.000000) (260,1.000000) (264,1.000000) (268,0.999995) (272,0.999973) (276,0.999902) (280,0.999721) (284,0.999334) (288,0.998604) (292,0.997359) (296,0.995394) (300,0.992489) (304,0.988423) (308,0.982995) (312,0.976035) (316,0.967427) (320,0.957112) (324,0.945097) (328,0.931451) (332,0.916302) (336,0.899825) (340,0.882231) (344,0.851806) (348,0.792568) (352,0.735590) (356,0.681014) (360,0.628948) (364,0.579465) (368,0.532610) (372,0.488397) (376,0.446817) (380,0.407840) (384,0.371418) (388,0.337486) (392,0.305969) (396,0.276781) (400,0.249826) (404,0.225005) (408,0.201553) (412,0.180886) (416,0.161968) (420,0.144704) (424,0.128998) (428,0.114749) (432,0.101858) (436,0.090227) (440,0.079758) (444,0.070360) (448,0.061943) (452,0.054422) (456,0.047718) (460,0.041755) (464,0.036462) (468,0.031776) (472,0.027635) (476,0.023984) (480,0.019841) (484,0.017227) (488,0.014921) (492,0.012891) (496,0.011109) (500,0.009550) (504,0.008189) (508,0.007004) };
\end{axis}
\end{tikzpicture}

%% file: figures/combined_experiments/n3_concentric_cross_section_Blend_gamma0p5.tex
\begin{tikzpicture}
\begin{axis}[width=\cellsize,height=\cellsize,scale only axis,yticklabels={},xmin=0,xmax=511,ymin=0,ymax=1.1,grid=major,xtick=\empty,ytick={0,0.25,0.5,0.75,1}]
\addplot[red, thick] coordinates { (0,0.000000) (4,0.000000) (8,0.000000) (12,0.000000) (16,0.000000) (20,0.000000) (24,0.000000) (28,0.000000) (32,0.000000) (36,0.000000) (40,0.000000) (44,0.000000) (48,0.000000) (52,0.000000) (56,0.000000) (60,0.000000) (64,0.000000) (68,0.000000) (72,0.000000) (76,0.000000) (80,0.000000) (84,0.000000) (88,0.000000) (92,0.000000) (96,0.000000) (100,0.000000) (104,0.000000) (108,0.001065) (112,0.001534) (116,0.002187) (120,0.003089) (124,0.004318) (128,0.005976) (132,0.008189) (136,0.011109) (140,0.014921) (144,0.019841) (148,0.026121) (152,0.034047) (156,0.043937) (160,0.056135) (164,0.071005) (168,0.088922) (172,0.110251) (176,0.135335) (180,0.164474) (184,0.197899) (188,0.235746) (192,0.278037) (196,0.324652) (200,0.375311) (204,0.429557) (208,0.486752) (212,0.546074) (216,0.606531) (220,0.666977) (224,0.726149) (228,0.782705) (232,0.835270) (236,0.882497) (240,0.923116) (244,0.955997) (248,0.980199) (252,0.995012) (256,1.000000) (260,0.995012) (264,0.980199) (268,0.955997) (272,0.923116) (276,0.882497) (280,0.835270) (284,0.782705) (288,0.726149) (292,0.666977) (296,0.606531) (300,0.546074) (304,0.486752) (308,0.429557) (312,0.375311) (316,0.324652) (320,0.278037) (324,0.235746) (328,0.197899) (332,0.164474) (336,0.135335) (340,0.110251) (344,0.088922) (348,0.071005) (352,0.056135) (356,0.043937) (360,0.034047) (364,0.026121) (368,0.019841) (372,0.014921) (376,0.011109) (380,0.008189) (384,0.005976) (388,0.004318) (392,0.003089) (396,0.002187) (400,0.001534) (404,0.001065) (408,0.000000) (412,0.000000) (416,0.000000) (420,0.000000) (424,0.000000) (428,0.000000) (432,0.000000) (436,0.000000) (440,0.000000) (444,0.000000) (448,0.000000) (452,0.000000) (456,0.000000) (460,0.000000) (464,0.000000) (468,0.000000) (472,0.000000) (476,0.000000) (480,0.000000) (484,0.000000) (488,0.000000) (492,0.000000) (496,0.000000) (500,0.000000) (504,0.000000) (508,0.000000) };
\addplot[green!70!black, thick] coordinates { (0,0.000000) (4,0.000000) (8,0.000000) (12,0.000000) (16,0.000000) (20,0.000000) (24,0.000000) (28,0.000000) (32,0.000000) (36,0.001204) (40,0.001534) (44,0.001946) (48,0.002457) (52,0.003089) (56,0.003866) (60,0.004817) (64,0.005976) (68,0.007381) (72,0.009075) (76,0.011109) (80,0.013538) (84,0.016426) (88,0.019841) (92,0.023860) (96,0.028566) (100,0.034047) (104,0.040401) (108,0.047703) (112,0.056092) (116,0.065657) (120,0.076502) (124,0.088730) (128,0.102433) (132,0.117695) (136,0.134584) (140,0.153144) (144,0.173393) (148,0.195314) (152,0.218845) (156,0.243874) (160,0.270234) (164,0.297690) (168,0.325942) (172,0.354622) (176,0.383293) (180,0.411462) (184,0.438588) (188,0.464106) (192,0.487448) (196,0.508075) (200,0.525510) (204,0.539374) (208,0.513248) (212,0.453926) (216,0.393469) (220,0.333023) (224,0.273851) (228,0.217295) (232,0.164730) (236,0.117503) (240,0.076884) (244,0.044003) (248,0.019801) (252,0.004988) (256,0.000000) (260,0.004988) (264,0.019801) (268,0.044003) (272,0.076884) (276,0.117503) (280,0.164730) (284,0.217295) (288,0.273851) (292,0.333023) (296,0.393469) (300,0.453926) (304,0.513248) (308,0.539374) (312,0.525510) (316,0.508075) (320,0.487448) (324,0.464106) (328,0.438588) (332,0.411462) (336,0.383293) (340,0.354622) (344,0.325942) (348,0.297690) (352,0.270234) (356,0.243874) (360,0.218845) (364,0.195314) (368,0.173393) (372,0.153144) (376,0.134584) (380,0.117695) (384,0.102433) (388,0.088730) (392,0.076502) (396,0.065657) (400,0.056092) (404,0.047703) (408,0.040401) (412,0.034047) (416,0.028566) (420,0.023860) (424,0.019841) (428,0.016426) (432,0.013538) (436,0.011109) (440,0.009075) (444,0.007381) (448,0.005976) (452,0.004817) (456,0.003866) (460,0.003089) (464,0.002457) (468,0.001946) (472,0.001534) (476,0.001204) (480,0.000000) (484,0.000000) (488,0.000000) (492,0.000000) (496,0.000000) (500,0.000000) (504,0.000000) (508,0.000000) };
\addplot[blue, thick] coordinates { (0,0.005976) (4,0.007004) (8,0.008189) (12,0.009550) (16,0.011109) (20,0.012891) (24,0.014921) (28,0.017227) (32,0.019841) (36,0.022780) (40,0.026101) (44,0.029831) (48,0.034006) (52,0.038666) (56,0.043852) (60,0.049605) (64,0.055967) (68,0.062979) (72,0.070683) (76,0.079118) (80,0.088320) (84,0.098323) (88,0.109157) (92,0.120844) (96,0.133402) (100,0.146839) (104,0.161152) (108,0.176237) (112,0.192201) (116,0.208937) (120,0.226378) (124,0.244439) (128,0.263009) (132,0.281957) (136,0.301125) (140,0.320332) (144,0.339376) (148,0.358030) (152,0.376056) (156,0.393203) (160,0.409221) (164,0.423873) (168,0.436942) (172,0.448253) (176,0.457685) (180,0.424064) (184,0.363513) (188,0.300148) (192,0.234515) (196,0.167273) (200,0.099179) (204,0.031068) (208,0.000000) (212,0.000000) (216,0.000000) (220,0.000000) (224,0.000000) (228,0.000000) (232,0.000000) (236,0.000000) (240,0.000000) (244,0.000000) (248,0.000000) (252,0.000000) (256,0.000000) (260,0.000000) (264,0.000000) (268,0.000000) (272,0.000000) (276,0.000000) (280,0.000000) (284,0.000000) (288,0.000000) (292,0.000000) (296,0.000000) (300,0.000000) (304,0.000000) (308,0.031068) (312,0.099179) (316,0.167273) (320,0.234515) (324,0.300148) (328,0.363513) (332,0.424064) (336,0.457685) (340,0.448253) (344,0.436942) (348,0.423873) (352,0.409221) (356,0.393203) (360,0.376056) (364,0.358030) (368,0.339376) (372,0.320332) (376,0.301125) (380,0.281957) (384,0.263009) (388,0.244439) (392,0.226378) (396,0.208937) (400,0.192201) (404,0.176237) (408,0.161152) (412,0.146839) (416,0.133402) (420,0.120844) (424,0.109157) (428,0.098323) (432,0.088320) (436,0.079118) (440,0.070683) (444,0.062979) (448,0.055967) (452,0.049605) (456,0.043852) (460,0.038666) (464,0.034006) (468,0.029831) (472,0.026101) (476,0.022780) (480,0.019841) (484,0.017227) (488,0.014921) (492,0.012891) (496,0.011109) (500,0.009550) (504,0.008189) (508,0.007004) };
\addplot[black, thick, dashed] coordinates { (0,0.005976) (4,0.007004) (8,0.008189) (12,0.009550) (16,0.011109) (20,0.012891) (24,0.014921) (28,0.017227) (32,0.019841) (36,0.023984) (40,0.027635) (44,0.031776) (48,0.036462) (52,0.041755) (56,0.047718) (60,0.054422) (64,0.061943) (68,0.070360) (72,0.079758) (76,0.090227) (80,0.101858) (84,0.114749) (88,0.128998) (92,0.144704) (96,0.161968) (100,0.180886) (104,0.201553) (108,0.225005) (112,0.249826) (116,0.276781) (120,0.305969) (124,0.337486) (128,0.371418) (132,0.407840) (136,0.446817) (140,0.488397) (144,0.532610) (148,0.579465) (152,0.628948) (156,0.681014) (160,0.735590) (164,0.792568) (168,0.851806) (172,0.913126) (176,0.976314) (180,1.000000) (184,1.000000) (188,1.000000) (192,1.000000) (196,1.000000) (200,1.000000) (204,1.000000) (208,1.000000) (212,1.000000) (216,1.000000) (220,1.000000) (224,1.000000) (228,1.000000) (232,1.000000) (236,1.000000) (240,1.000000) (244,1.000000) (248,1.000000) (252,1.000000) (256,1.000000) (260,1.000000) (264,1.000000) (268,1.000000) (272,1.000000) (276,1.000000) (280,1.000000) (284,1.000000) (288,1.000000) (292,1.000000) (296,1.000000) (300,1.000000) (304,1.000000) (308,1.000000) (312,1.000000) (316,1.000000) (320,1.000000) (324,1.000000) (328,1.000000) (332,1.000000) (336,0.976314) (340,0.913126) (344,0.851806) (348,0.792568) (352,0.735590) (356,0.681014) (360,0.628948) (364,0.579465) (368,0.532610) (372,0.488397) (376,0.446817) (380,0.407840) (384,0.371418) (388,0.337486) (392,0.305969) (396,0.276781) (400,0.249826) (404,0.225005) (408,0.201553) (412,0.180886) (416,0.161968) (420,0.144704) (424,0.128998) (428,0.114749) (432,0.101858) (436,0.090227) (440,0.079758) (444,0.070360) (448,0.061943) (452,0.054422) (456,0.047718) (460,0.041755) (464,0.036462) (468,0.031776) (472,0.027635) (476,0.023984) (480,0.019841) (484,0.017227) (488,0.014921) (492,0.012891) (496,0.011109) (500,0.009550) (504,0.008189) (508,0.007004) };
\end{axis}
\end{tikzpicture}

%% file: figures/combined_experiments/n3_concentric_cross_section_Linear.tex
\begin{tikzpicture}
\begin{axis}[width=\cellsize,height=\cellsize,scale only axis,yticklabels={},xmin=0,xmax=511,ymin=0,ymax=1.1,grid=major,xtick=\empty,ytick={0,0.25,0.5,0.75,1}]
\addplot[red, thick] coordinates { (0,0.000000) (4,0.000000) (8,0.000000) (12,0.000000) (16,0.000000) (20,0.000000) (24,0.000000) (28,0.000000) (32,0.000000) (36,0.000000) (40,0.000000) (44,0.000000) (48,0.000000) (52,0.000000) (56,0.000000) (60,0.000000) (64,0.000000) (68,0.000000) (72,0.000000) (76,0.000000) (80,0.000000) (84,0.000000) (88,0.000000) (92,0.000000) (96,0.000000) (100,0.000000) (104,0.000000) (108,0.001065) (112,0.001534) (116,0.002187) (120,0.003089) (124,0.004318) (128,0.005976) (132,0.008189) (136,0.011109) (140,0.014921) (144,0.019841) (148,0.026121) (152,0.034047) (156,0.043937) (160,0.056135) (164,0.071005) (168,0.088922) (172,0.110251) (176,0.135335) (180,0.164474) (184,0.197899) (188,0.235746) (192,0.278037) (196,0.324652) (200,0.375311) (204,0.429557) (208,0.486752) (212,0.546074) (216,0.606531) (220,0.666977) (224,0.726149) (228,0.782705) (232,0.835270) (236,0.882497) (240,0.923116) (244,0.955997) (248,0.980199) (252,0.995012) (256,1.000000) (260,0.995012) (264,0.980199) (268,0.955997) (272,0.923116) (276,0.882497) (280,0.835270) (284,0.782705) (288,0.726149) (292,0.666977) (296,0.606531) (300,0.546074) (304,0.486752) (308,0.429557) (312,0.375311) (316,0.324652) (320,0.278037) (324,0.235746) (328,0.197899) (332,0.164474) (336,0.135335) (340,0.110251) (344,0.088922) (348,0.071005) (352,0.056135) (356,0.043937) (360,0.034047) (364,0.026121) (368,0.019841) (372,0.014921) (376,0.011109) (380,0.008189) (384,0.005976) (388,0.004318) (392,0.003089) (396,0.002187) (400,0.001534) (404,0.001065) (408,0.000000) (412,0.000000) (416,0.000000) (420,0.000000) (424,0.000000) (428,0.000000) (432,0.000000) (436,0.000000) (440,0.000000) (444,0.000000) (448,0.000000) (452,0.000000) (456,0.000000) (460,0.000000) (464,0.000000) (468,0.000000) (472,0.000000) (476,0.000000) (480,0.000000) (484,0.000000) (488,0.000000) (492,0.000000) (496,0.000000) (500,0.000000) (504,0.000000) (508,0.000000) };
\addplot[green!70!black, thick] coordinates { (0,0.000000) (4,0.000000) (8,0.000000) (12,0.000000) (16,0.000000) (20,0.000000) (24,0.000000) (28,0.000000) (32,0.000000) (36,0.001204) (40,0.001534) (44,0.001946) (48,0.002457) (52,0.003089) (56,0.003866) (60,0.004817) (64,0.005976) (68,0.007381) (72,0.009075) (76,0.011109) (80,0.013538) (84,0.016426) (88,0.019841) (92,0.023860) (96,0.028566) (100,0.034047) (104,0.040401) (108,0.047729) (112,0.056135) (116,0.065729) (120,0.076621) (124,0.088922) (128,0.102740) (132,0.118179) (136,0.135335) (140,0.154295) (144,0.175131) (148,0.197899) (152,0.222635) (156,0.249352) (160,0.278037) (164,0.308647) (168,0.341108) (172,0.375311) (176,0.411112) (180,0.448332) (184,0.486752) (188,0.526122) (192,0.566154) (196,0.606531) (200,0.624689) (204,0.570443) (208,0.513248) (212,0.453926) (216,0.393469) (220,0.333023) (224,0.273851) (228,0.217295) (232,0.164730) (236,0.117503) (240,0.076884) (244,0.044003) (248,0.019801) (252,0.004988) (256,0.000000) (260,0.004988) (264,0.019801) (268,0.044003) (272,0.076884) (276,0.117503) (280,0.164730) (284,0.217295) (288,0.273851) (292,0.333023) (296,0.393469) (300,0.453926) (304,0.513248) (308,0.570443) (312,0.624689) (316,0.606531) (320,0.566154) (324,0.526122) (328,0.486752) (332,0.448332) (336,0.411112) (340,0.375311) (344,0.341108) (348,0.308647) (352,0.278037) (356,0.249352) (360,0.222635) (364,0.197899) (368,0.175131) (372,0.154295) (376,0.135335) (380,0.118179) (384,0.102740) (388,0.088922) (392,0.076621) (396,0.065729) (400,0.056135) (404,0.047729) (408,0.040401) (412,0.034047) (416,0.028566) (420,0.023860) (424,0.019841) (428,0.016426) (432,0.013538) (436,0.011109) (440,0.009075) (444,0.007381) (448,0.005976) (452,0.004817) (456,0.003866) (460,0.003089) (464,0.002457) (468,0.001946) (472,0.001534) (476,0.001204) (480,0.000000) (484,0.000000) (488,0.000000) (492,0.000000) (496,0.000000) (500,0.000000) (504,0.000000) (508,0.000000) };
\addplot[blue, thick] coordinates { (0,0.005976) (4,0.007004) (8,0.008189) (12,0.009550) (16,0.011109) (20,0.012891) (24,0.014921) (28,0.017227) (32,0.019841) (36,0.022794) (40,0.026121) (44,0.029860) (48,0.034047) (52,0.038726) (56,0.043937) (60,0.049725) (64,0.056135) (68,0.063213) (72,0.071005) (76,0.079560) (80,0.088922) (84,0.099137) (88,0.110251) (92,0.122303) (96,0.135335) (100,0.149382) (104,0.164474) (108,0.180640) (112,0.197899) (116,0.216265) (120,0.235746) (124,0.256340) (128,0.278037) (132,0.300818) (136,0.324652) (140,0.349501) (144,0.375311) (148,0.402021) (152,0.429557) (156,0.457833) (160,0.486752) (164,0.516206) (168,0.546074) (172,0.514438) (176,0.453552) (180,0.387194) (184,0.315349) (188,0.238132) (192,0.155809) (196,0.068817) (200,0.000000) (204,0.000000) (208,0.000000) (212,0.000000) (216,0.000000) (220,0.000000) (224,0.000000) (228,0.000000) (232,0.000000) (236,0.000000) (240,0.000000) (244,0.000000) (248,0.000000) (252,0.000000) (256,0.000000) (260,0.000000) (264,0.000000) (268,0.000000) (272,0.000000) (276,0.000000) (280,0.000000) (284,0.000000) (288,0.000000) (292,0.000000) (296,0.000000) (300,0.000000) (304,0.000000) (308,0.000000) (312,0.000000) (316,0.068817) (320,0.155809) (324,0.238132) (328,0.315349) (332,0.387194) (336,0.453552) (340,0.514438) (344,0.546074) (348,0.516206) (352,0.486752) (356,0.457833) (360,0.429557) (364,0.402021) (368,0.375311) (372,0.349501) (376,0.324652) (380,0.300818) (384,0.278037) (388,0.256340) (392,0.235746) (396,0.216265) (400,0.197899) (404,0.180640) (408,0.164474) (412,0.149382) (416,0.135335) (420,0.122303) (424,0.110251) (428,0.099137) (432,0.088922) (436,0.079560) (440,0.071005) (444,0.063213) (448,0.056135) (452,0.049725) (456,0.043937) (460,0.038726) (464,0.034047) (468,0.029860) (472,0.026121) (476,0.022794) (480,0.019841) (484,0.017227) (488,0.014921) (492,0.012891) (496,0.011109) (500,0.009550) (504,0.008189) (508,0.007004) };
\addplot[black, thick, dashed] coordinates { (0,0.005976) (4,0.007004) (8,0.008189) (12,0.009550) (16,0.011109) (20,0.012891) (24,0.014921) (28,0.017227) (32,0.019841) (36,0.023998) (40,0.027655) (44,0.031805) (48,0.036504) (52,0.041814) (56,0.047803) (60,0.054542) (64,0.062111) (68,0.070593) (72,0.080080) (76,0.090669) (80,0.102460) (84,0.115563) (88,0.130092) (92,0.146163) (96,0.163901) (100,0.183429) (104,0.204876) (108,0.229433) (112,0.255567) (116,0.284181) (120,0.315455) (124,0.349580) (128,0.386753) (132,0.427186) (136,0.471097) (140,0.518716) (144,0.570283) (148,0.626041) (152,0.686240) (156,0.751123) (160,0.820924) (164,0.895858) (168,0.976104) (172,1.000000) (176,1.000000) (180,1.000000) (184,1.000000) (188,1.000000) (192,1.000000) (196,1.000000) (200,1.000000) (204,1.000000) (208,1.000000) (212,1.000000) (216,1.000000) (220,1.000000) (224,1.000000) (228,1.000000) (232,1.000000) (236,1.000000) (240,1.000000) (244,1.000000) (248,1.000000) (252,1.000000) (256,1.000000) (260,1.000000) (264,1.000000) (268,1.000000) (272,1.000000) (276,1.000000) (280,1.000000) (284,1.000000) (288,1.000000) (292,1.000000) (296,1.000000) (300,1.000000) (304,1.000000) (308,1.000000) (312,1.000000) (316,1.000000) (320,1.000000) (324,1.000000) (328,1.000000) (332,1.000000) (336,1.000000) (340,1.000000) (344,0.976104) (348,0.895858) (352,0.820924) (356,0.751123) (360,0.686240) (364,0.626041) (368,0.570283) (372,0.518716) (376,0.471097) (380,0.427186) (384,0.386753) (388,0.349580) (392,0.315455) (396,0.284181) (400,0.255567) (404,0.229433) (408,0.204876) (412,0.183429) (416,0.163901) (420,0.146163) (424,0.130092) (428,0.115563) (432,0.102460) (436,0.090669) (440,0.080080) (444,0.070593) (448,0.062111) (452,0.054542) (456,0.047803) (460,0.041814) (464,0.036504) (468,0.031805) (472,0.027655) (476,0.023998) (480,0.019841) (484,0.017227) (488,0.014921) (492,0.012891) (496,0.011109) (500,0.009550) (504,0.008189) (508,0.007004) };
\end{axis}
\end{tikzpicture}

%% file: figures/combined_experiments/n3_concentric_cross_section_Quadratic_c0p5.tex
\begin{tikzpicture}
\begin{axis}[width=\cellsize,height=\cellsize,scale only axis,yticklabels={},xmin=0,xmax=511,ymin=0,ymax=1.1,grid=major,xtick=\empty,ytick={0,0.25,0.5,0.75,1}]
\addplot[red, thick] coordinates { (0,0.000000) (4,0.000000) (8,0.000000) (12,0.000000) (16,0.000000) (20,0.000000) (24,0.000000) (28,0.000000) (32,0.000000) (36,0.000000) (40,0.000000) (44,0.000000) (48,0.000000) (52,0.000000) (56,0.000000) (60,0.000000) (64,0.000000) (68,0.000000) (72,0.000000) (76,0.000000) (80,0.000000) (84,0.000000) (88,0.000000) (92,0.000000) (96,0.000000) (100,0.000000) (104,0.000000) (108,0.001065) (112,0.001534) (116,0.002187) (120,0.003089) (124,0.004318) (128,0.005976) (132,0.008189) (136,0.011109) (140,0.014921) (144,0.019841) (148,0.026121) (152,0.034047) (156,0.043937) (160,0.056135) (164,0.071005) (168,0.088922) (172,0.110251) (176,0.135335) (180,0.164474) (184,0.197899) (188,0.235746) (192,0.278037) (196,0.324652) (200,0.375311) (204,0.429557) (208,0.486752) (212,0.546074) (216,0.606531) (220,0.666977) (224,0.726149) (228,0.782705) (232,0.835270) (236,0.882497) (240,0.923116) (244,0.955997) (248,0.980199) (252,0.995012) (256,1.000000) (260,0.995012) (264,0.980199) (268,0.955997) (272,0.923116) (276,0.882497) (280,0.835270) (284,0.782705) (288,0.726149) (292,0.666977) (296,0.606531) (300,0.546074) (304,0.486752) (308,0.429557) (312,0.375311) (316,0.324652) (320,0.278037) (324,0.235746) (328,0.197899) (332,0.164474) (336,0.135335) (340,0.110251) (344,0.088922) (348,0.071005) (352,0.056135) (356,0.043937) (360,0.034047) (364,0.026121) (368,0.019841) (372,0.014921) (376,0.011109) (380,0.008189) (384,0.005976) (388,0.004318) (392,0.003089) (396,0.002187) (400,0.001534) (404,0.001065) (408,0.000000) (412,0.000000) (416,0.000000) (420,0.000000) (424,0.000000) (428,0.000000) (432,0.000000) (436,0.000000) (440,0.000000) (444,0.000000) (448,0.000000) (452,0.000000) (456,0.000000) (460,0.000000) (464,0.000000) (468,0.000000) (472,0.000000) (476,0.000000) (480,0.000000) (484,0.000000) (488,0.000000) (492,0.000000) (496,0.000000) (500,0.000000) (504,0.000000) (508,0.000000) };
\addplot[green!70!black, thick] coordinates { (0,0.000000) (4,0.000000) (8,0.000000) (12,0.000000) (16,0.000000) (20,0.000000) (24,0.000000) (28,0.000000) (32,0.000000) (36,0.001204) (40,0.001534) (44,0.001946) (48,0.002457) (52,0.003089) (56,0.003866) (60,0.004817) (64,0.005976) (68,0.007381) (72,0.009075) (76,0.011109) (80,0.013538) (84,0.016426) (88,0.019841) (92,0.023860) (96,0.028566) (100,0.034047) (104,0.040401) (108,0.047754) (112,0.056178) (116,0.065800) (120,0.076739) (124,0.089114) (128,0.103047) (132,0.118663) (136,0.136087) (140,0.155446) (144,0.176868) (148,0.200483) (152,0.226425) (156,0.254830) (160,0.285841) (164,0.319605) (168,0.356274) (172,0.396000) (176,0.438931) (180,0.485201) (184,0.534916) (188,0.588138) (192,0.644860) (196,0.675348) (200,0.624689) (204,0.570443) (208,0.513248) (212,0.453926) (216,0.393469) (220,0.333023) (224,0.273851) (228,0.217295) (232,0.164730) (236,0.117503) (240,0.076884) (244,0.044003) (248,0.019801) (252,0.004988) (256,0.000000) (260,0.004988) (264,0.019801) (268,0.044003) (272,0.076884) (276,0.117503) (280,0.164730) (284,0.217295) (288,0.273851) (292,0.333023) (296,0.393469) (300,0.453926) (304,0.513248) (308,0.570443) (312,0.624689) (316,0.675348) (320,0.644860) (324,0.588138) (328,0.534916) (332,0.485201) (336,0.438931) (340,0.396000) (344,0.356274) (348,0.319605) (352,0.285841) (356,0.254830) (360,0.226425) (364,0.200483) (368,0.176868) (372,0.155446) (376,0.136087) (380,0.118663) (384,0.103047) (388,0.089114) (392,0.076739) (396,0.065800) (400,0.056178) (404,0.047754) (408,0.040401) (412,0.034047) (416,0.028566) (420,0.023860) (424,0.019841) (428,0.016426) (432,0.013538) (436,0.011109) (440,0.009075) (444,0.007381) (448,0.005976) (452,0.004817) (456,0.003866) (460,0.003089) (464,0.002457) (468,0.001946) (472,0.001534) (476,0.001204) (480,0.000000) (484,0.000000) (488,0.000000) (492,0.000000) (496,0.000000) (500,0.000000) (504,0.000000) (508,0.000000) };
\addplot[blue, thick] coordinates { (0,0.005976) (4,0.007004) (8,0.008189) (12,0.009550) (16,0.011109) (20,0.012891) (24,0.014921) (28,0.017227) (32,0.019841) (36,0.022808) (40,0.026141) (44,0.029889) (48,0.034089) (52,0.038786) (56,0.044022) (60,0.049845) (64,0.056302) (68,0.063446) (72,0.071328) (76,0.080001) (80,0.089524) (84,0.099951) (88,0.111344) (92,0.123763) (96,0.137268) (100,0.151925) (104,0.167797) (108,0.185047) (112,0.203605) (116,0.223609) (120,0.245142) (124,0.268291) (128,0.293151) (132,0.319825) (136,0.348424) (140,0.379071) (144,0.411899) (148,0.447052) (152,0.484687) (156,0.524972) (160,0.568082) (164,0.609389) (168,0.554804) (172,0.493749) (176,0.425733) (180,0.350324) (184,0.267185) (188,0.176116) (192,0.077103) (196,0.000000) (200,0.000000) (204,0.000000) (208,0.000000) (212,0.000000) (216,0.000000) (220,0.000000) (224,0.000000) (228,0.000000) (232,0.000000) (236,0.000000) (240,0.000000) (244,0.000000) (248,0.000000) (252,0.000000) (256,0.000000) (260,0.000000) (264,0.000000) (268,0.000000) (272,0.000000) (276,0.000000) (280,0.000000) (284,0.000000) (288,0.000000) (292,0.000000) (296,0.000000) (300,0.000000) (304,0.000000) (308,0.000000) (312,0.000000) (316,0.000000) (320,0.077103) (324,0.176116) (328,0.267185) (332,0.350324) (336,0.425733) (340,0.493749) (344,0.554804) (348,0.609389) (352,0.568082) (356,0.524972) (360,0.484687) (364,0.447052) (368,0.411899) (372,0.379071) (376,0.348424) (380,0.319825) (384,0.293151) (388,0.268291) (392,0.245142) (396,0.223609) (400,0.203605) (404,0.185047) (408,0.167797) (412,0.151925) (416,0.137268) (420,0.123763) (424,0.111344) (428,0.099951) (432,0.089524) (436,0.080001) (440,0.071328) (444,0.063446) (448,0.056302) (452,0.049845) (456,0.044022) (460,0.038786) (464,0.034089) (468,0.029889) (472,0.026141) (476,0.022808) (480,0.019841) (484,0.017227) (488,0.014921) (492,0.012891) (496,0.011109) (500,0.009550) (504,0.008189) (508,0.007004) };
\addplot[black, thick, dashed] coordinates { (0,0.005976) (4,0.007004) (8,0.008189) (12,0.009550) (16,0.011109) (20,0.012891) (24,0.014921) (28,0.017227) (32,0.019841) (36,0.024012) (40,0.027675) (44,0.031834) (48,0.036546) (52,0.041874) (56,0.047888) (60,0.054662) (64,0.062279) (68,0.070827) (72,0.080403) (76,0.091110) (80,0.103062) (84,0.116377) (88,0.131185) (92,0.147622) (96,0.165834) (100,0.185972) (104,0.208198) (108,0.233866) (112,0.261317) (116,0.291597) (120,0.324969) (124,0.361722) (128,0.402174) (132,0.446676) (136,0.495620) (140,0.549438) (144,0.608608) (148,0.673657) (152,0.745160) (156,0.823739) (160,0.910058) (164,1.000000) (168,1.000000) (172,1.000000) (176,1.000000) (180,1.000000) (184,1.000000) (188,1.000000) (192,1.000000) (196,1.000000) (200,1.000000) (204,1.000000) (208,1.000000) (212,1.000000) (216,1.000000) (220,1.000000) (224,1.000000) (228,1.000000) (232,1.000000) (236,1.000000) (240,1.000000) (244,1.000000) (248,1.000000) (252,1.000000) (256,1.000000) (260,1.000000) (264,1.000000) (268,1.000000) (272,1.000000) (276,1.000000) (280,1.000000) (284,1.000000) (288,1.000000) (292,1.000000) (296,1.000000) (300,1.000000) (304,1.000000) (308,1.000000) (312,1.000000) (316,1.000000) (320,1.000000) (324,1.000000) (328,1.000000) (332,1.000000) (336,1.000000) (340,1.000000) (344,1.000000) (348,1.000000) (352,0.910058) (356,0.823739) (360,0.745160) (364,0.673657) (368,0.608608) (372,0.549438) (376,0.495620) (380,0.446676) (384,0.402174) (388,0.361722) (392,0.324969) (396,0.291597) (400,0.261317) (404,0.233866) (408,0.208198) (412,0.185972) (416,0.165834) (420,0.147622) (424,0.131185) (428,0.116377) (432,0.103062) (436,0.091110) (440,0.080403) (444,0.070827) (448,0.062279) (452,0.054662) (456,0.047888) (460,0.041874) (464,0.036546) (468,0.031834) (472,0.027675) (476,0.024012) (480,0.019841) (484,0.017227) (488,0.014921) (492,0.012891) (496,0.011109) (500,0.009550) (504,0.008189) (508,0.007004) };
\end{axis}
\end{tikzpicture}

%% file: figures/combined_experiments/n3_concentric_cross_section_Quadratic_c1.tex
\begin{tikzpicture}
\begin{axis}[width=\cellsize,height=\cellsize,scale only axis,yticklabels={},xmin=0,xmax=511,ymin=0,ymax=1.1,grid=major,xtick=\empty,ytick={0,0.25,0.5,0.75,1}]
\addplot[red, thick] coordinates { (0,0.000000) (4,0.000000) (8,0.000000) (12,0.000000) (16,0.000000) (20,0.000000) (24,0.000000) (28,0.000000) (32,0.000000) (36,0.000000) (40,0.000000) (44,0.000000) (48,0.000000) (52,0.000000) (56,0.000000) (60,0.000000) (64,0.000000) (68,0.000000) (72,0.000000) (76,0.000000) (80,0.000000) (84,0.000000) (88,0.000000) (92,0.000000) (96,0.000000) (100,0.000000) (104,0.000000) (108,0.001065) (112,0.001534) (116,0.002187) (120,0.003089) (124,0.004318) (128,0.005976) (132,0.008189) (136,0.011109) (140,0.014921) (144,0.019841) (148,0.026121) (152,0.034047) (156,0.043937) (160,0.056135) (164,0.071005) (168,0.088922) (172,0.110251) (176,0.135335) (180,0.164474) (184,0.197899) (188,0.235746) (192,0.278037) (196,0.324652) (200,0.375311) (204,0.429557) (208,0.486752) (212,0.546074) (216,0.606531) (220,0.666977) (224,0.726149) (228,0.782705) (232,0.835270) (236,0.882497) (240,0.923116) (244,0.955997) (248,0.980199) (252,0.995012) (256,1.000000) (260,0.995012) (264,0.980199) (268,0.955997) (272,0.923116) (276,0.882497) (280,0.835270) (284,0.782705) (288,0.726149) (292,0.666977) (296,0.606531) (300,0.546074) (304,0.486752) (308,0.429557) (312,0.375311) (316,0.324652) (320,0.278037) (324,0.235746) (328,0.197899) (332,0.164474) (336,0.135335) (340,0.110251) (344,0.088922) (348,0.071005) (352,0.056135) (356,0.043937) (360,0.034047) (364,0.026121) (368,0.019841) (372,0.014921) (376,0.011109) (380,0.008189) (384,0.005976) (388,0.004318) (392,0.003089) (396,0.002187) (400,0.001534) (404,0.001065) (408,0.000000) (412,0.000000) (416,0.000000) (420,0.000000) (424,0.000000) (428,0.000000) (432,0.000000) (436,0.000000) (440,0.000000) (444,0.000000) (448,0.000000) (452,0.000000) (456,0.000000) (460,0.000000) (464,0.000000) (468,0.000000) (472,0.000000) (476,0.000000) (480,0.000000) (484,0.000000) (488,0.000000) (492,0.000000) (496,0.000000) (500,0.000000) (504,0.000000) (508,0.000000) };
\addplot[green!70!black, thick] coordinates { (0,0.000000) (4,0.000000) (8,0.000000) (12,0.000000) (16,0.000000) (20,0.000000) (24,0.000000) (28,0.000000) (32,0.000000) (36,0.001204) (40,0.001534) (44,0.001946) (48,0.002457) (52,0.003089) (56,0.003866) (60,0.004817) (64,0.005976) (68,0.007381) (72,0.009075) (76,0.011109) (80,0.013538) (84,0.016426) (88,0.019841) (92,0.023860) (96,0.028566) (100,0.034047) (104,0.040401) (108,0.047780) (112,0.056221) (116,0.065872) (120,0.076857) (124,0.089306) (128,0.103354) (132,0.119147) (136,0.136839) (140,0.156597) (144,0.178606) (148,0.203068) (152,0.230215) (156,0.260308) (160,0.293645) (164,0.330563) (168,0.371440) (172,0.416689) (176,0.466750) (180,0.522071) (184,0.583080) (188,0.650153) (192,0.721963) (196,0.675348) (200,0.624689) (204,0.570443) (208,0.513248) (212,0.453926) (216,0.393469) (220,0.333023) (224,0.273851) (228,0.217295) (232,0.164730) (236,0.117503) (240,0.076884) (244,0.044003) (248,0.019801) (252,0.004988) (256,0.000000) (260,0.004988) (264,0.019801) (268,0.044003) (272,0.076884) (276,0.117503) (280,0.164730) (284,0.217295) (288,0.273851) (292,0.333023) (296,0.393469) (300,0.453926) (304,0.513248) (308,0.570443) (312,0.624689) (316,0.675348) (320,0.721963) (324,0.650153) (328,0.583080) (332,0.522071) (336,0.466750) (340,0.416689) (344,0.371440) (348,0.330563) (352,0.293645) (356,0.260308) (360,0.230215) (364,0.203068) (368,0.178606) (372,0.156597) (376,0.136839) (380,0.119147) (384,0.103354) (388,0.089306) (392,0.076857) (396,0.065872) (400,0.056221) (404,0.047780) (408,0.040401) (412,0.034047) (416,0.028566) (420,0.023860) (424,0.019841) (428,0.016426) (432,0.013538) (436,0.011109) (440,0.009075) (444,0.007381) (448,0.005976) (452,0.004817) (456,0.003866) (460,0.003089) (464,0.002457) (468,0.001946) (472,0.001534) (476,0.001204) (480,0.000000) (484,0.000000) (488,0.000000) (492,0.000000) (496,0.000000) (500,0.000000) (504,0.000000) (508,0.000000) };
\addplot[blue, thick] coordinates { (0,0.005976) (4,0.007004) (8,0.008189) (12,0.009550) (16,0.011109) (20,0.012891) (24,0.014921) (28,0.017227) (32,0.019841) (36,0.022822) (40,0.026161) (44,0.029918) (48,0.034131) (52,0.038845) (56,0.044107) (60,0.049964) (64,0.056470) (68,0.063679) (72,0.071650) (76,0.080443) (80,0.090125) (84,0.100766) (88,0.112438) (92,0.125222) (96,0.139201) (100,0.154468) (104,0.171119) (108,0.189454) (112,0.209311) (116,0.230953) (120,0.254537) (124,0.280241) (128,0.308264) (132,0.338832) (136,0.372196) (140,0.408642) (144,0.448486) (148,0.492082) (152,0.539817) (156,0.592111) (160,0.649411) (164,0.598432) (168,0.539638) (172,0.473060) (176,0.397914) (180,0.313455) (184,0.219021) (188,0.114101) (192,0.000000) (196,0.000000) (200,0.000000) (204,0.000000) (208,0.000000) (212,0.000000) (216,0.000000) (220,0.000000) (224,0.000000) (228,0.000000) (232,0.000000) (236,0.000000) (240,0.000000) (244,0.000000) (248,0.000000) (252,0.000000) (256,0.000000) (260,0.000000) (264,0.000000) (268,0.000000) (272,0.000000) (276,0.000000) (280,0.000000) (284,0.000000) (288,0.000000) (292,0.000000) (296,0.000000) (300,0.000000) (304,0.000000) (308,0.000000) (312,0.000000) (316,0.000000) (320,0.000000) (324,0.114101) (328,0.219021) (332,0.313455) (336,0.397914) (340,0.473060) (344,0.539638) (348,0.598432) (352,0.649411) (356,0.592111) (360,0.539817) (364,0.492082) (368,0.448486) (372,0.408642) (376,0.372196) (380,0.338832) (384,0.308264) (388,0.280241) (392,0.254537) (396,0.230953) (400,0.209311) (404,0.189454) (408,0.171119) (412,0.154468) (416,0.139201) (420,0.125222) (424,0.112438) (428,0.100766) (432,0.090125) (436,0.080443) (440,0.071650) (444,0.063679) (448,0.056470) (452,0.049964) (456,0.044107) (460,0.038845) (464,0.034131) (468,0.029918) (472,0.026161) (476,0.022822) (480,0.019841) (484,0.017227) (488,0.014921) (492,0.012891) (496,0.011109) (500,0.009550) (504,0.008189) (508,0.007004) };
\addplot[black, thick, dashed] coordinates { (0,0.005976) (4,0.007004) (8,0.008189) (12,0.009550) (16,0.011109) (20,0.012891) (24,0.014921) (28,0.017227) (32,0.019841) (36,0.024025) (40,0.027695) (44,0.031863) (48,0.036588) (52,0.041934) (56,0.047973) (60,0.054782) (64,0.062446) (68,0.071060) (72,0.080725) (76,0.091552) (80,0.103664) (84,0.117192) (88,0.132279) (92,0.149082) (96,0.167767) (100,0.188515) (104,0.211521) (108,0.238298) (112,0.267066) (116,0.299013) (120,0.334483) (124,0.373865) (128,0.417594) (132,0.466167) (136,0.520144) (140,0.580160) (144,0.646933) (148,0.721272) (152,0.804080) (156,0.896356) (160,0.999191) (164,1.000000) (168,1.000000) (172,1.000000) (176,1.000000) (180,1.000000) (184,1.000000) (188,1.000000) (192,1.000000) (196,1.000000) (200,1.000000) (204,1.000000) (208,1.000000) (212,1.000000) (216,1.000000) (220,1.000000) (224,1.000000) (228,1.000000) (232,1.000000) (236,1.000000) (240,1.000000) (244,1.000000) (248,1.000000) (252,1.000000) (256,1.000000) (260,1.000000) (264,1.000000) (268,1.000000) (272,1.000000) (276,1.000000) (280,1.000000) (284,1.000000) (288,1.000000) (292,1.000000) (296,1.000000) (300,1.000000) (304,1.000000) (308,1.000000) (312,1.000000) (316,1.000000) (320,1.000000) (324,1.000000) (328,1.000000) (332,1.000000) (336,1.000000) (340,1.000000) (344,1.000000) (348,1.000000) (352,0.999191) (356,0.896356) (360,0.804080) (364,0.721272) (368,0.646933) (372,0.580160) (376,0.520144) (380,0.466167) (384,0.417594) (388,0.373865) (392,0.334483) (396,0.299013) (400,0.267066) (404,0.238298) (408,0.211521) (412,0.188515) (416,0.167767) (420,0.149082) (424,0.132279) (428,0.117192) (432,0.103664) (436,0.091552) (440,0.080725) (444,0.071060) (448,0.062446) (452,0.054782) (456,0.047973) (460,0.041934) (464,0.036588) (468,0.031863) (472,0.027695) (476,0.024025) (480,0.019841) (484,0.017227) (488,0.014921) (492,0.012891) (496,0.011109) (500,0.009550) (504,0.008189) (508,0.007004) };
\end{axis}
\end{tikzpicture}

%% file: figures/combined_experiments/n3_concentric_cross_section_PowermLaw_vm1.tex
\begin{tikzpicture}
\begin{axis}[width=\cellsize,height=\cellsize,scale only axis,yticklabels={},xmin=0,xmax=511,ymin=0,ymax=1.1,grid=major,xtick=\empty,ytick={0,0.25,0.5,0.75,1}]
\addplot[red, thick] coordinates { (0,0.000000) (4,0.000000) (8,0.000000) (12,0.000000) (16,0.000000) (20,0.000000) (24,0.000000) (28,0.000000) (32,0.000000) (36,0.000000) (40,0.000000) (44,0.000000) (48,0.000000) (52,0.000000) (56,0.000000) (60,0.000000) (64,0.000000) (68,0.000000) (72,0.000000) (76,0.000000) (80,0.000000) (84,0.000000) (88,0.000000) (92,0.000000) (96,0.000000) (100,0.000000) (104,0.000000) (108,0.001065) (112,0.001534) (116,0.002187) (120,0.003089) (124,0.004318) (128,0.005976) (132,0.008189) (136,0.011109) (140,0.014921) (144,0.019841) (148,0.026121) (152,0.034047) (156,0.043937) (160,0.056135) (164,0.071005) (168,0.088922) (172,0.110251) (176,0.135335) (180,0.164474) (184,0.197899) (188,0.235746) (192,0.278037) (196,0.324652) (200,0.375311) (204,0.429557) (208,0.486752) (212,0.546074) (216,0.606531) (220,0.666977) (224,0.726149) (228,0.782705) (232,0.835270) (236,0.882497) (240,0.923116) (244,0.955997) (248,0.980199) (252,0.995012) (256,1.000000) (260,0.995012) (264,0.980199) (268,0.955997) (272,0.923116) (276,0.882497) (280,0.835270) (284,0.782705) (288,0.726149) (292,0.666977) (296,0.606531) (300,0.546074) (304,0.486752) (308,0.429557) (312,0.375311) (316,0.324652) (320,0.278037) (324,0.235746) (328,0.197899) (332,0.164474) (336,0.135335) (340,0.110251) (344,0.088922) (348,0.071005) (352,0.056135) (356,0.043937) (360,0.034047) (364,0.026121) (368,0.019841) (372,0.014921) (376,0.011109) (380,0.008189) (384,0.005976) (388,0.004318) (392,0.003089) (396,0.002187) (400,0.001534) (404,0.001065) (408,0.000000) (412,0.000000) (416,0.000000) (420,0.000000) (424,0.000000) (428,0.000000) (432,0.000000) (436,0.000000) (440,0.000000) (444,0.000000) (448,0.000000) (452,0.000000) (456,0.000000) (460,0.000000) (464,0.000000) (468,0.000000) (472,0.000000) (476,0.000000) (480,0.000000) (484,0.000000) (488,0.000000) (492,0.000000) (496,0.000000) (500,0.000000) (504,0.000000) (508,0.000000) };
\addplot[green!70!black, thick] coordinates { (0,0.000000) (4,0.000000) (8,0.000000) (12,0.000000) (16,0.000000) (20,0.000000) (24,0.000000) (28,0.000000) (32,0.000000) (36,0.001204) (40,0.001534) (44,0.001946) (48,0.002457) (52,0.003089) (56,0.003866) (60,0.004817) (64,0.005976) (68,0.007381) (72,0.009075) (76,0.011109) (80,0.013538) (84,0.016426) (88,0.019841) (92,0.023860) (96,0.028566) (100,0.034047) (104,0.040401) (108,0.047780) (112,0.056221) (116,0.065873) (120,0.076858) (124,0.089308) (128,0.103359) (132,0.119159) (136,0.136864) (140,0.156650) (144,0.178713) (148,0.203280) (152,0.230626) (156,0.261087) (160,0.295095) (164,0.333213) (168,0.376197) (172,0.425093) (176,0.481392) (180,0.547295) (184,0.626205) (188,0.723704) (192,0.721963) (196,0.675348) (200,0.624689) (204,0.570443) (208,0.513248) (212,0.453926) (216,0.393469) (220,0.333023) (224,0.273851) (228,0.217295) (232,0.164730) (236,0.117503) (240,0.076884) (244,0.044003) (248,0.019801) (252,0.004988) (256,0.000000) (260,0.004988) (264,0.019801) (268,0.044003) (272,0.076884) (276,0.117503) (280,0.164730) (284,0.217295) (288,0.273851) (292,0.333023) (296,0.393469) (300,0.453926) (304,0.513248) (308,0.570443) (312,0.624689) (316,0.675348) (320,0.721963) (324,0.723704) (328,0.626205) (332,0.547295) (336,0.481392) (340,0.425093) (344,0.376197) (348,0.333213) (352,0.295095) (356,0.261087) (360,0.230626) (364,0.203280) (368,0.178713) (372,0.156650) (376,0.136864) (380,0.119159) (384,0.103359) (388,0.089308) (392,0.076858) (396,0.065873) (400,0.056221) (404,0.047780) (408,0.040401) (412,0.034047) (416,0.028566) (420,0.023860) (424,0.019841) (428,0.016426) (432,0.013538) (436,0.011109) (440,0.009075) (444,0.007381) (448,0.005976) (452,0.004817) (456,0.003866) (460,0.003089) (464,0.002457) (468,0.001946) (472,0.001534) (476,0.001204) (480,0.000000) (484,0.000000) (488,0.000000) (492,0.000000) (496,0.000000) (500,0.000000) (504,0.000000) (508,0.000000) };
\addplot[blue, thick] coordinates { (0,0.005976) (4,0.007004) (8,0.008189) (12,0.009550) (16,0.011109) (20,0.012891) (24,0.014921) (28,0.017227) (32,0.019841) (36,0.022822) (40,0.026162) (44,0.029918) (48,0.034131) (52,0.038846) (56,0.044108) (60,0.049966) (64,0.056473) (68,0.063684) (72,0.071659) (76,0.080458) (80,0.090150) (84,0.100807) (88,0.112505) (92,0.125330) (96,0.139375) (100,0.154743) (104,0.171551) (108,0.190156) (112,0.210404) (116,0.232642) (120,0.257131) (124,0.284205) (128,0.314298) (132,0.347990) (136,0.386078) (140,0.429695) (144,0.480514) (148,0.541122) (152,0.615771) (156,0.694976) (160,0.648770) (164,0.595782) (168,0.534882) (172,0.464657) (176,0.383273) (180,0.288230) (184,0.175896) (188,0.040550) (192,0.000000) (196,0.000000) (200,0.000000) (204,0.000000) (208,0.000000) (212,0.000000) (216,0.000000) (220,0.000000) (224,0.000000) (228,0.000000) (232,0.000000) (236,0.000000) (240,0.000000) (244,0.000000) (248,0.000000) (252,0.000000) (256,0.000000) (260,0.000000) (264,0.000000) (268,0.000000) (272,0.000000) (276,0.000000) (280,0.000000) (284,0.000000) (288,0.000000) (292,0.000000) (296,0.000000) (300,0.000000) (304,0.000000) (308,0.000000) (312,0.000000) (316,0.000000) (320,0.000000) (324,0.040550) (328,0.175896) (332,0.288230) (336,0.383273) (340,0.464657) (344,0.534882) (348,0.595782) (352,0.648770) (356,0.694976) (360,0.615771) (364,0.541122) (368,0.480514) (372,0.429695) (376,0.386078) (380,0.347990) (384,0.314298) (388,0.284205) (392,0.257131) (396,0.232642) (400,0.210404) (404,0.190156) (408,0.171551) (412,0.154743) (416,0.139375) (420,0.125330) (424,0.112505) (428,0.100807) (432,0.090150) (436,0.080458) (440,0.071659) (444,0.063684) (448,0.056473) (452,0.049966) (456,0.044108) (460,0.038846) (464,0.034131) (468,0.029918) (472,0.026162) (476,0.022822) (480,0.019841) (484,0.017227) (488,0.014921) (492,0.012891) (496,0.011109) (500,0.009550) (504,0.008189) (508,0.007004) };
\addplot[black, thick, dashed] coordinates { (0,0.005976) (4,0.007004) (8,0.008189) (12,0.009550) (16,0.011109) (20,0.012891) (24,0.014921) (28,0.017227) (32,0.019841) (36,0.024026) (40,0.027695) (44,0.031863) (48,0.036588) (52,0.041935) (56,0.047974) (60,0.054783) (64,0.062449) (68,0.071065) (72,0.080734) (76,0.091567) (80,0.103689) (84,0.117233) (88,0.132346) (92,0.149190) (96,0.167941) (100,0.188791) (104,0.211953) (108,0.239001) (112,0.268159) (116,0.300702) (120,0.337078) (124,0.377831) (128,0.423634) (132,0.475337) (136,0.534051) (140,0.601266) (144,0.679068) (148,0.770524) (152,0.880444) (156,1.000000) (160,1.000000) (164,1.000000) (168,1.000000) (172,1.000000) (176,1.000000) (180,1.000000) (184,1.000000) (188,1.000000) (192,1.000000) (196,1.000000) (200,1.000000) (204,1.000000) (208,1.000000) (212,1.000000) (216,1.000000) (220,1.000000) (224,1.000000) (228,1.000000) (232,1.000000) (236,1.000000) (240,1.000000) (244,1.000000) (248,1.000000) (252,1.000000) (256,1.000000) (260,1.000000) (264,1.000000) (268,1.000000) (272,1.000000) (276,1.000000) (280,1.000000) (284,1.000000) (288,1.000000) (292,1.000000) (296,1.000000) (300,1.000000) (304,1.000000) (308,1.000000) (312,1.000000) (316,1.000000) (320,1.000000) (324,1.000000) (328,1.000000) (332,1.000000) (336,1.000000) (340,1.000000) (344,1.000000) (348,1.000000) (352,1.000000) (356,1.000000) (360,0.880444) (364,0.770524) (368,0.679068) (372,0.601266) (376,0.534051) (380,0.475337) (384,0.423634) (388,0.377831) (392,0.337078) (396,0.300702) (400,0.268159) (404,0.239001) (408,0.211953) (412,0.188791) (416,0.167941) (420,0.149190) (424,0.132346) (428,0.117233) (432,0.103689) (436,0.091567) (440,0.080734) (444,0.071065) (448,0.062449) (452,0.054783) (456,0.047974) (460,0.041935) (464,0.036588) (468,0.031863) (472,0.027695) (476,0.024026) (480,0.019841) (484,0.017227) (488,0.014921) (492,0.012891) (496,0.011109) (500,0.009550) (504,0.008189) (508,0.007004) };
\end{axis}
\end{tikzpicture}

%% file: figures/combined_experiments/n3_cyclic_cross_section_PowermLaw_v2.tex
\begin{tikzpicture}
\begin{axis}[width=\cellsize,height=\cellsize,scale only axis,yticklabels={},xmin=0,xmax=511,ymin=0,ymax=1.1,grid=major,xtick=\empty,ytick={0,0.25,0.5,0.75,1}]
\addplot[red, thick] coordinates { (0,0.000000) (4,0.000000) (8,0.000000) (12,0.000000) (16,0.000000) (20,0.000000) (24,0.000000) (28,0.000000) (32,0.000000) (36,0.000000) (40,0.000000) (44,0.000000) (48,0.000000) (52,0.000000) (56,0.000000) (60,0.000000) (64,0.000000) (68,0.000000) (72,0.000000) (76,0.000000) (80,0.000000) (84,0.000000) (88,0.000000) (92,0.000000) (96,0.000000) (100,0.000000) (104,0.000000) (108,0.001065) (112,0.001534) (116,0.002187) (120,0.003089) (124,0.004318) (128,0.005976) (132,0.008189) (136,0.011109) (140,0.014921) (144,0.019841) (148,0.026121) (152,0.034047) (156,0.043937) (160,0.056135) (164,0.071005) (168,0.088922) (172,0.110251) (176,0.135335) (180,0.164474) (184,0.197899) (188,0.235746) (192,0.278037) (196,0.324652) (200,0.375311) (204,0.429557) (208,0.486752) (212,0.546074) (216,0.606531) (220,0.666977) (224,0.726149) (228,0.782705) (232,0.835270) (236,0.882497) (240,0.923116) (244,0.955997) (248,0.980199) (252,0.995012) (256,1.000000) (260,0.995012) (264,0.980199) (268,0.955997) (272,0.923116) (276,0.882497) (280,0.835270) (284,0.782705) (288,0.726149) (292,0.666977) (296,0.606531) (300,0.546074) (304,0.486752) (308,0.429557) (312,0.375311) (316,0.324652) (320,0.278037) (324,0.235746) (328,0.197899) (332,0.164474) (336,0.135335) (340,0.110251) (344,0.088922) (348,0.071005) (352,0.056135) (356,0.043937) (360,0.034047) (364,0.026121) (368,0.019841) (372,0.014921) (376,0.011109) (380,0.008189) (384,0.005976) (388,0.004318) (392,0.003089) (396,0.002187) (400,0.001534) (404,0.001065) (408,0.000000) (412,0.000000) (416,0.000000) (420,0.000000) (424,0.000000) (428,0.000000) (432,0.000000) (436,0.000000) (440,0.000000) (444,0.000000) (448,0.000000) (452,0.000000) (456,0.000000) (460,0.000000) (464,0.000000) (468,0.000000) (472,0.000000) (476,0.000000) (480,0.000000) (484,0.000000) (488,0.000000) (492,0.000000) (496,0.000000) (500,0.000000) (504,0.000000) (508,0.000000) };
\addplot[green!70!black, thick] coordinates { (0,0.102740) (4,0.110251) (8,0.118179) (12,0.126537) (16,0.135335) (20,0.144585) (24,0.154295) (28,0.164474) (32,0.175131) (36,0.186270) (40,0.197899) (44,0.210019) (48,0.222635) (52,0.235746) (56,0.249352) (60,0.263451) (64,0.278037) (68,0.293106) (72,0.308647) (76,0.324652) (80,0.341108) (84,0.358000) (88,0.375311) (92,0.393022) (96,0.411112) (100,0.429557) (104,0.448332) (108,0.465918) (112,0.484521) (116,0.503031) (120,0.521284) (124,0.539076) (128,0.556154) (132,0.572206) (136,0.586864) (140,0.599698) (144,0.610225) (148,0.617926) (152,0.622277) (156,0.622781) (160,0.619029) (164,0.610746) (168,0.597849) (172,0.580484) (176,0.559037) (180,0.534118) (184,0.506516) (188,0.477128) (192,0.446877) (196,0.416641) (200,0.387188) (204,0.359141) (208,0.332966) (212,0.308970) (216,0.287331) (220,0.268113) (224,0.251303) (228,0.217295) (232,0.164730) (236,0.117503) (240,0.076884) (244,0.044003) (248,0.019801) (252,0.004988) (256,0.000000) (260,0.004988) (264,0.019801) (268,0.044003) (272,0.076884) (276,0.117503) (280,0.164730) (284,0.217295) (288,0.251303) (292,0.268113) (296,0.287331) (300,0.308970) (304,0.332966) (308,0.359141) (312,0.387188) (316,0.416641) (320,0.446877) (324,0.477128) (328,0.506516) (332,0.534118) (336,0.559037) (340,0.580484) (344,0.597849) (348,0.610746) (352,0.619029) (356,0.622781) (360,0.622277) (364,0.617926) (368,0.610225) (372,0.599698) (376,0.586864) (380,0.572206) (384,0.556154) (388,0.539076) (392,0.521284) (396,0.503031) (400,0.484521) (404,0.465918) (408,0.448332) (412,0.429557) (416,0.411112) (420,0.393022) (424,0.375311) (428,0.358000) (432,0.341108) (436,0.324652) (440,0.308647) (444,0.293106) (448,0.278037) (452,0.263451) (456,0.249352) (460,0.235746) (464,0.222635) (468,0.210019) (472,0.197899) (476,0.186270) (480,0.175131) (484,0.164474) (488,0.154295) (492,0.144585) (496,0.135335) (500,0.126537) (504,0.118179) (508,0.110251) };
\addplot[blue, thick] coordinates { (0,0.000000) (4,0.000000) (8,0.000000) (12,0.000000) (16,0.000000) (20,0.000000) (24,0.000000) (28,0.000000) (32,0.000000) (36,0.000000) (40,0.000000) (44,0.000000) (48,0.000000) (52,0.000000) (56,0.000000) (60,0.000673) (64,0.000856) (68,0.001083) (72,0.001362) (76,0.001703) (80,0.002117) (84,0.002616) (88,0.003215) (92,0.003929) (96,0.004774) (100,0.005770) (104,0.006934) (108,0.008276) (112,0.009833) (116,0.011618) (120,0.013650) (124,0.015947) (128,0.018525) (132,0.021394) (136,0.024562) (140,0.028028) (144,0.031783) (148,0.035809) (152,0.040077) (156,0.044545) (160,0.049160) (164,0.053854) (168,0.058553) (172,0.063174) (176,0.067630) (180,0.071838) (184,0.075719) (188,0.079209) (192,0.082260) (196,0.084843) (200,0.086948) (204,0.088587) (208,0.089791) (212,0.090602) (216,0.091074) (220,0.064910) (224,0.022547) (228,0.000000) (232,0.000000) (236,0.000000) (240,0.000000) (244,0.000000) (248,0.000000) (252,0.000000) (256,0.000000) (260,0.000000) (264,0.000000) (268,0.000000) (272,0.000000) (276,0.000000) (280,0.000000) (284,0.000000) (288,0.022547) (292,0.064910) (296,0.091074) (300,0.090602) (304,0.089791) (308,0.088587) (312,0.086948) (316,0.084843) (320,0.082260) (324,0.079209) (328,0.075719) (332,0.071838) (336,0.067630) (340,0.063174) (344,0.058553) (348,0.053854) (352,0.049160) (356,0.044545) (360,0.040077) (364,0.035809) (368,0.031783) (372,0.028028) (376,0.024562) (380,0.021394) (384,0.018525) (388,0.015947) (392,0.013650) (396,0.011618) (400,0.009833) (404,0.008276) (408,0.006934) (412,0.005770) (416,0.004774) (420,0.003929) (424,0.003215) (428,0.002616) (432,0.002117) (436,0.001703) (440,0.001362) (444,0.001083) (448,0.000856) (452,0.000673) (456,0.000000) (460,0.000000) (464,0.000000) (468,0.000000) (472,0.000000) (476,0.000000) (480,0.000000) (484,0.000000) (488,0.000000) (492,0.000000) (496,0.000000) (500,0.000000) (504,0.000000) (508,0.000000) };
\addplot[black, thick, dashed] coordinates { (0,0.102740) (4,0.110251) (8,0.118179) (12,0.126537) (16,0.135335) (20,0.144585) (24,0.154295) (28,0.164474) (32,0.175131) (36,0.186270) (40,0.197899) (44,0.210019) (48,0.222635) (52,0.235746) (56,0.249352) (60,0.264123) (64,0.278893) (68,0.294189) (72,0.310009) (76,0.326355) (80,0.343225) (84,0.360616) (88,0.378526) (92,0.396951) (96,0.415886) (100,0.435327) (104,0.455266) (108,0.475258) (112,0.495888) (116,0.516836) (120,0.538023) (124,0.559342) (128,0.580655) (132,0.601789) (136,0.622535) (140,0.642646) (144,0.661849) (148,0.679857) (152,0.696401) (156,0.711263) (160,0.724323) (164,0.735605) (168,0.745324) (172,0.753909) (176,0.762003) (180,0.770430) (184,0.780134) (188,0.792083) (192,0.807175) (196,0.826136) (200,0.849447) (204,0.877286) (208,0.909509) (212,0.945647) (216,0.984936) (220,1.000000) (224,1.000000) (228,1.000000) (232,1.000000) (236,1.000000) (240,1.000000) (244,1.000000) (248,1.000000) (252,1.000000) (256,1.000000) (260,1.000000) (264,1.000000) (268,1.000000) (272,1.000000) (276,1.000000) (280,1.000000) (284,1.000000) (288,1.000000) (292,1.000000) (296,0.984936) (300,0.945647) (304,0.909509) (308,0.877286) (312,0.849447) (316,0.826136) (320,0.807175) (324,0.792083) (328,0.780134) (332,0.770430) (336,0.762003) (340,0.753909) (344,0.745324) (348,0.735605) (352,0.724323) (356,0.711263) (360,0.696401) (364,0.679857) (368,0.661849) (372,0.642646) (376,0.622535) (380,0.601789) (384,0.580655) (388,0.559342) (392,0.538023) (396,0.516836) (400,0.495888) (404,0.475258) (408,0.455266) (412,0.435327) (416,0.415886) (420,0.396951) (424,0.378526) (428,0.360616) (432,0.343225) (436,0.326355) (440,0.310009) (444,0.294189) (448,0.278893) (452,0.264123) (456,0.249352) (460,0.235746) (464,0.222635) (468,0.210019) (472,0.197899) (476,0.186270) (480,0.175131) (484,0.164474) (488,0.154295) (492,0.144585) (496,0.135335) (500,0.126537) (504,0.118179) (508,0.110251) };
\end{axis}
\end{tikzpicture}

%% file: figures/combined_experiments/n3_cyclic_cross_section_Exponential.tex
\begin{tikzpicture}
\begin{axis}[width=\cellsize,height=\cellsize,scale only axis,yticklabels={},xmin=0,xmax=511,ymin=0,ymax=1.1,grid=major,xtick=\empty,ytick={0,0.25,0.5,0.75,1}]
\addplot[red, thick] coordinates { (0,0.000000) (4,0.000000) (8,0.000000) (12,0.000000) (16,0.000000) (20,0.000000) (24,0.000000) (28,0.000000) (32,0.000000) (36,0.000000) (40,0.000000) (44,0.000000) (48,0.000000) (52,0.000000) (56,0.000000) (60,0.000000) (64,0.000000) (68,0.000000) (72,0.000000) (76,0.000000) (80,0.000000) (84,0.000000) (88,0.000000) (92,0.000000) (96,0.000000) (100,0.000000) (104,0.000000) (108,0.001065) (112,0.001534) (116,0.002187) (120,0.003089) (124,0.004318) (128,0.005976) (132,0.008189) (136,0.011109) (140,0.014921) (144,0.019841) (148,0.026121) (152,0.034047) (156,0.043937) (160,0.056135) (164,0.071005) (168,0.088922) (172,0.110251) (176,0.135335) (180,0.164474) (184,0.197899) (188,0.235746) (192,0.278037) (196,0.324652) (200,0.375311) (204,0.429557) (208,0.486752) (212,0.546074) (216,0.606531) (220,0.666977) (224,0.726149) (228,0.782705) (232,0.835270) (236,0.882497) (240,0.923116) (244,0.955997) (248,0.980199) (252,0.995012) (256,1.000000) (260,0.995012) (264,0.980199) (268,0.955997) (272,0.923116) (276,0.882497) (280,0.835270) (284,0.782705) (288,0.726149) (292,0.666977) (296,0.606531) (300,0.546074) (304,0.486752) (308,0.429557) (312,0.375311) (316,0.324652) (320,0.278037) (324,0.235746) (328,0.197899) (332,0.164474) (336,0.135335) (340,0.110251) (344,0.088922) (348,0.071005) (352,0.056135) (356,0.043937) (360,0.034047) (364,0.026121) (368,0.019841) (372,0.014921) (376,0.011109) (380,0.008189) (384,0.005976) (388,0.004318) (392,0.003089) (396,0.002187) (400,0.001534) (404,0.001065) (408,0.000000) (412,0.000000) (416,0.000000) (420,0.000000) (424,0.000000) (428,0.000000) (432,0.000000) (436,0.000000) (440,0.000000) (444,0.000000) (448,0.000000) (452,0.000000) (456,0.000000) (460,0.000000) (464,0.000000) (468,0.000000) (472,0.000000) (476,0.000000) (480,0.000000) (484,0.000000) (488,0.000000) (492,0.000000) (496,0.000000) (500,0.000000) (504,0.000000) (508,0.000000) };
\addplot[green!70!black, thick] coordinates { (0,0.102740) (4,0.110251) (8,0.118179) (12,0.126537) (16,0.135335) (20,0.144585) (24,0.154295) (28,0.164474) (32,0.175131) (36,0.186270) (40,0.197899) (44,0.210019) (48,0.222635) (52,0.235746) (56,0.249352) (60,0.263451) (64,0.278037) (68,0.293106) (72,0.308647) (76,0.324652) (80,0.341108) (84,0.358000) (88,0.375311) (92,0.393022) (96,0.411112) (100,0.429557) (104,0.448332) (108,0.466909) (112,0.486006) (116,0.505228) (120,0.524497) (124,0.543717) (128,0.562771) (132,0.581519) (136,0.599793) (140,0.617389) (144,0.634070) (148,0.649554) (152,0.663520) (156,0.675600) (160,0.685387) (164,0.692435) (168,0.696272) (172,0.696411) (176,0.692369) (180,0.683691) (184,0.669971) (188,0.650893) (192,0.626251) (196,0.595992) (200,0.560240) (204,0.519322) (208,0.473787) (212,0.424415) (216,0.372206) (220,0.318369) (224,0.264285) (228,0.211460) (232,0.161468) (236,0.115882) (240,0.076203) (244,0.043783) (248,0.019757) (252,0.004985) (256,0.000000) (260,0.004985) (264,0.019757) (268,0.043783) (272,0.076203) (276,0.115882) (280,0.161468) (284,0.211460) (288,0.264285) (292,0.318369) (296,0.372206) (300,0.424415) (304,0.473787) (308,0.519322) (312,0.560240) (316,0.595992) (320,0.626251) (324,0.650893) (328,0.669971) (332,0.683691) (336,0.692369) (340,0.696411) (344,0.696272) (348,0.692435) (352,0.685387) (356,0.675600) (360,0.663520) (364,0.649554) (368,0.634070) (372,0.617389) (376,0.599793) (380,0.581519) (384,0.562771) (388,0.543717) (392,0.524497) (396,0.505228) (400,0.486006) (404,0.466909) (408,0.448332) (412,0.429557) (416,0.411112) (420,0.393022) (424,0.375311) (428,0.358000) (432,0.341108) (436,0.324652) (440,0.308647) (444,0.293106) (448,0.278037) (452,0.263451) (456,0.249352) (460,0.235746) (464,0.222635) (468,0.210019) (472,0.197899) (476,0.186270) (480,0.175131) (484,0.164474) (488,0.154295) (492,0.144585) (496,0.135335) (500,0.126537) (504,0.118179) (508,0.110251) };
\addplot[blue, thick] coordinates { (0,0.000000) (4,0.000000) (8,0.000000) (12,0.000000) (16,0.000000) (20,0.000000) (24,0.000000) (28,0.000000) (32,0.000000) (36,0.000000) (40,0.000000) (44,0.000000) (48,0.000000) (52,0.000000) (56,0.000000) (60,0.000935) (64,0.001200) (68,0.001529) (72,0.001936) (76,0.002436) (80,0.003043) (84,0.003775) (88,0.004652) (92,0.005692) (96,0.006916) (100,0.008343) (104,0.009993) (108,0.011869) (112,0.014002) (116,0.016394) (120,0.019044) (124,0.021946) (128,0.025085) (132,0.028429) (136,0.031939) (140,0.035557) (144,0.039209) (148,0.042809) (152,0.046251) (156,0.049419) (160,0.052186) (164,0.054421) (168,0.055997) (172,0.056795) (176,0.056718) (180,0.055702) (184,0.053720) (188,0.050795) (192,0.047004) (196,0.042476) (200,0.037391) (204,0.031968) (208,0.026451) (212,0.021087) (216,0.016106) (220,0.011701) (224,0.008008) (228,0.005093) (232,0.002951) (236,0.001512) (240,0.000651) (244,0.000214) (248,0.000043) (252,0.000003) (256,0.000000) (260,0.000003) (264,0.000043) (268,0.000214) (272,0.000651) (276,0.001512) (280,0.002951) (284,0.005093) (288,0.008008) (292,0.011701) (296,0.016106) (300,0.021087) (304,0.026451) (308,0.031968) (312,0.037391) (316,0.042476) (320,0.047004) (324,0.050795) (328,0.053720) (332,0.055702) (336,0.056718) (340,0.056795) (344,0.055997) (348,0.054421) (352,0.052186) (356,0.049419) (360,0.046251) (364,0.042809) (368,0.039209) (372,0.035557) (376,0.031939) (380,0.028429) (384,0.025085) (388,0.021946) (392,0.019044) (396,0.016394) (400,0.014002) (404,0.011869) (408,0.009993) (412,0.008343) (416,0.006916) (420,0.005692) (424,0.004652) (428,0.003775) (432,0.003043) (436,0.002436) (440,0.001936) (444,0.001529) (448,0.001200) (452,0.000935) (456,0.000000) (460,0.000000) (464,0.000000) (468,0.000000) (472,0.000000) (476,0.000000) (480,0.000000) (484,0.000000) (488,0.000000) (492,0.000000) (496,0.000000) (500,0.000000) (504,0.000000) (508,0.000000) };
\addplot[black, thick, dashed] coordinates { (0,0.102740) (4,0.110251) (8,0.118179) (12,0.126537) (16,0.135335) (20,0.144585) (24,0.154295) (28,0.164474) (32,0.175131) (36,0.186270) (40,0.197899) (44,0.210019) (48,0.222635) (52,0.235746) (56,0.249352) (60,0.264385) (64,0.279237) (68,0.294635) (72,0.310584) (76,0.327088) (80,0.344151) (84,0.361775) (88,0.379963) (92,0.398714) (96,0.418028) (100,0.437900) (104,0.458324) (108,0.479843) (112,0.501542) (116,0.523809) (120,0.546630) (124,0.569981) (128,0.593831) (132,0.618137) (136,0.642841) (140,0.667867) (144,0.693120) (148,0.718485) (152,0.743818) (156,0.768956) (160,0.793707) (164,0.817862) (168,0.841190) (172,0.863456) (176,0.884423) (180,0.903867) (184,0.921590) (188,0.937434) (192,0.951292) (196,0.963120) (200,0.972942) (204,0.980848) (208,0.986991) (212,0.991576) (216,0.994843) (220,0.997047) (224,0.998442) (228,0.999257) (232,0.999690) (236,0.999891) (240,0.999970) (244,0.999995) (248,1.000000) (252,1.000000) (256,1.000000) (260,1.000000) (264,1.000000) (268,0.999995) (272,0.999970) (276,0.999891) (280,0.999690) (284,0.999257) (288,0.998442) (292,0.997047) (296,0.994843) (300,0.991576) (304,0.986991) (308,0.980848) (312,0.972942) (316,0.963120) (320,0.951292) (324,0.937434) (328,0.921590) (332,0.903867) (336,0.884423) (340,0.863456) (344,0.841190) (348,0.817862) (352,0.793707) (356,0.768956) (360,0.743818) (364,0.718485) (368,0.693120) (372,0.667867) (376,0.642841) (380,0.618137) (384,0.593831) (388,0.569981) (392,0.546630) (396,0.523809) (400,0.501542) (404,0.479843) (408,0.458324) (412,0.437900) (416,0.418028) (420,0.398714) (424,0.379963) (428,0.361775) (432,0.344151) (436,0.327088) (440,0.310584) (444,0.294635) (448,0.279237) (452,0.264385) (456,0.249352) (460,0.235746) (464,0.222635) (468,0.210019) (472,0.197899) (476,0.186270) (480,0.175131) (484,0.164474) (488,0.154295) (492,0.144585) (496,0.135335) (500,0.126537) (504,0.118179) (508,0.110251) };
\end{axis}
\end{tikzpicture}

%% file: figures/combined_experiments/n3_cyclic_cross_section_Quadratic_cm0p5.tex
\begin{tikzpicture}
\begin{axis}[width=\cellsize,height=\cellsize,scale only axis,yticklabels={},xmin=0,xmax=511,ymin=0,ymax=1.1,grid=major,xtick=\empty,ytick={0,0.25,0.5,0.75,1}]
\addplot[red, thick] coordinates { (0,0.000000) (4,0.000000) (8,0.000000) (12,0.000000) (16,0.000000) (20,0.000000) (24,0.000000) (28,0.000000) (32,0.000000) (36,0.000000) (40,0.000000) (44,0.000000) (48,0.000000) (52,0.000000) (56,0.000000) (60,0.000000) (64,0.000000) (68,0.000000) (72,0.000000) (76,0.000000) (80,0.000000) (84,0.000000) (88,0.000000) (92,0.000000) (96,0.000000) (100,0.000000) (104,0.000000) (108,0.001065) (112,0.001534) (116,0.002187) (120,0.003089) (124,0.004318) (128,0.005976) (132,0.008189) (136,0.011109) (140,0.014921) (144,0.019841) (148,0.026121) (152,0.034047) (156,0.043937) (160,0.056135) (164,0.071005) (168,0.088922) (172,0.110251) (176,0.135335) (180,0.164474) (184,0.197899) (188,0.235746) (192,0.278037) (196,0.324652) (200,0.375311) (204,0.429557) (208,0.486752) (212,0.546074) (216,0.606531) (220,0.666977) (224,0.726149) (228,0.782705) (232,0.835270) (236,0.882497) (240,0.923116) (244,0.955997) (248,0.980199) (252,0.995012) (256,1.000000) (260,0.995012) (264,0.980199) (268,0.955997) (272,0.923116) (276,0.882497) (280,0.835270) (284,0.782705) (288,0.726149) (292,0.666977) (296,0.606531) (300,0.546074) (304,0.486752) (308,0.429557) (312,0.375311) (316,0.324652) (320,0.278037) (324,0.235746) (328,0.197899) (332,0.164474) (336,0.135335) (340,0.110251) (344,0.088922) (348,0.071005) (352,0.056135) (356,0.043937) (360,0.034047) (364,0.026121) (368,0.019841) (372,0.014921) (376,0.011109) (380,0.008189) (384,0.005976) (388,0.004318) (392,0.003089) (396,0.002187) (400,0.001534) (404,0.001065) (408,0.000000) (412,0.000000) (416,0.000000) (420,0.000000) (424,0.000000) (428,0.000000) (432,0.000000) (436,0.000000) (440,0.000000) (444,0.000000) (448,0.000000) (452,0.000000) (456,0.000000) (460,0.000000) (464,0.000000) (468,0.000000) (472,0.000000) (476,0.000000) (480,0.000000) (484,0.000000) (488,0.000000) (492,0.000000) (496,0.000000) (500,0.000000) (504,0.000000) (508,0.000000) };
\addplot[green!70!black, thick] coordinates { (0,0.102740) (4,0.110251) (8,0.118179) (12,0.126537) (16,0.135335) (20,0.144585) (24,0.154295) (28,0.164474) (32,0.175131) (36,0.186270) (40,0.197899) (44,0.210019) (48,0.222635) (52,0.235746) (56,0.249352) (60,0.263451) (64,0.278037) (68,0.293106) (72,0.308647) (76,0.324652) (80,0.341108) (84,0.358000) (88,0.375311) (92,0.393022) (96,0.411112) (100,0.429557) (104,0.448332) (108,0.467158) (112,0.486379) (116,0.505782) (120,0.525309) (124,0.544895) (128,0.564462) (132,0.583920) (136,0.603162) (140,0.622065) (144,0.640488) (148,0.658266) (152,0.675214) (156,0.691124) (160,0.705768) (164,0.718897) (168,0.730250) (172,0.739558) (176,0.746553) (180,0.750983) (184,0.752621) (188,0.751282) (192,0.721963) (196,0.675348) (200,0.624689) (204,0.570443) (208,0.513248) (212,0.453926) (216,0.393469) (220,0.333023) (224,0.273851) (228,0.217295) (232,0.164730) (236,0.117503) (240,0.076884) (244,0.044003) (248,0.019801) (252,0.004988) (256,0.000000) (260,0.004988) (264,0.019801) (268,0.044003) (272,0.076884) (276,0.117503) (280,0.164730) (284,0.217295) (288,0.273851) (292,0.333023) (296,0.393469) (300,0.453926) (304,0.513248) (308,0.570443) (312,0.624689) (316,0.675348) (320,0.721963) (324,0.751282) (328,0.752621) (332,0.750983) (336,0.746553) (340,0.739558) (344,0.730250) (348,0.718897) (352,0.705768) (356,0.691124) (360,0.675214) (364,0.658266) (368,0.640488) (372,0.622065) (376,0.603162) (380,0.583920) (384,0.564462) (388,0.544895) (392,0.525309) (396,0.505782) (400,0.486379) (404,0.467158) (408,0.448332) (412,0.429557) (416,0.411112) (420,0.393022) (424,0.375311) (428,0.358000) (432,0.341108) (436,0.324652) (440,0.308647) (444,0.293106) (448,0.278037) (452,0.263451) (456,0.249352) (460,0.235746) (464,0.222635) (468,0.210019) (472,0.197899) (476,0.186270) (480,0.175131) (484,0.164474) (488,0.154295) (492,0.144585) (496,0.135335) (500,0.126537) (504,0.118179) (508,0.110251) };
\addplot[blue, thick] coordinates { (0,0.000000) (4,0.000000) (8,0.000000) (12,0.000000) (16,0.000000) (20,0.000000) (24,0.000000) (28,0.000000) (32,0.000000) (36,0.000000) (40,0.000000) (44,0.000000) (48,0.000000) (52,0.000000) (56,0.000000) (60,0.001102) (64,0.001431) (68,0.001846) (72,0.002369) (76,0.003021) (80,0.003830) (84,0.004828) (88,0.006049) (92,0.007535) (96,0.009330) (100,0.011484) (104,0.014053) (108,0.017083) (112,0.020653) (116,0.024819) (120,0.029645) (124,0.035195) (128,0.041527) (132,0.048694) (136,0.056736) (140,0.065678) (144,0.075524) (148,0.086251) (152,0.097802) (156,0.110076) (160,0.122926) (164,0.136150) (168,0.149483) (172,0.150192) (176,0.118111) (180,0.084542) (184,0.049481) (188,0.012972) (192,0.000000) (196,0.000000) (200,0.000000) (204,0.000000) (208,0.000000) (212,0.000000) (216,0.000000) (220,0.000000) (224,0.000000) (228,0.000000) (232,0.000000) (236,0.000000) (240,0.000000) (244,0.000000) (248,0.000000) (252,0.000000) (256,0.000000) (260,0.000000) (264,0.000000) (268,0.000000) (272,0.000000) (276,0.000000) (280,0.000000) (284,0.000000) (288,0.000000) (292,0.000000) (296,0.000000) (300,0.000000) (304,0.000000) (308,0.000000) (312,0.000000) (316,0.000000) (320,0.000000) (324,0.012972) (328,0.049481) (332,0.084542) (336,0.118111) (340,0.150192) (344,0.149483) (348,0.136150) (352,0.122926) (356,0.110076) (360,0.097802) (364,0.086251) (368,0.075524) (372,0.065678) (376,0.056736) (380,0.048694) (384,0.041527) (388,0.035195) (392,0.029645) (396,0.024819) (400,0.020653) (404,0.017083) (408,0.014053) (412,0.011484) (416,0.009330) (420,0.007535) (424,0.006049) (428,0.004828) (432,0.003830) (436,0.003021) (440,0.002369) (444,0.001846) (448,0.001431) (452,0.001102) (456,0.000000) (460,0.000000) (464,0.000000) (468,0.000000) (472,0.000000) (476,0.000000) (480,0.000000) (484,0.000000) (488,0.000000) (492,0.000000) (496,0.000000) (500,0.000000) (504,0.000000) (508,0.000000) };
\addplot[black, thick, dashed] coordinates { (0,0.102740) (4,0.110251) (8,0.118179) (12,0.126537) (16,0.135335) (20,0.144585) (24,0.154295) (28,0.164474) (32,0.175131) (36,0.186270) (40,0.197899) (44,0.210019) (48,0.222635) (52,0.235746) (56,0.249352) (60,0.264553) (64,0.279468) (68,0.294952) (72,0.311016) (76,0.327674) (80,0.344939) (84,0.362828) (88,0.381360) (92,0.400557) (96,0.420442) (100,0.441041) (104,0.462384) (108,0.485306) (112,0.508566) (116,0.532788) (120,0.558043) (124,0.584408) (128,0.611966) (132,0.640802) (136,0.671006) (140,0.702664) (144,0.735853) (148,0.770639) (152,0.807063) (156,0.845137) (160,0.884829) (164,0.926052) (168,0.968655) (172,1.000000) (176,1.000000) (180,1.000000) (184,1.000000) (188,1.000000) (192,1.000000) (196,1.000000) (200,1.000000) (204,1.000000) (208,1.000000) (212,1.000000) (216,1.000000) (220,1.000000) (224,1.000000) (228,1.000000) (232,1.000000) (236,1.000000) (240,1.000000) (244,1.000000) (248,1.000000) (252,1.000000) (256,1.000000) (260,1.000000) (264,1.000000) (268,1.000000) (272,1.000000) (276,1.000000) (280,1.000000) (284,1.000000) (288,1.000000) (292,1.000000) (296,1.000000) (300,1.000000) (304,1.000000) (308,1.000000) (312,1.000000) (316,1.000000) (320,1.000000) (324,1.000000) (328,1.000000) (332,1.000000) (336,1.000000) (340,1.000000) (344,0.968655) (348,0.926052) (352,0.884829) (356,0.845137) (360,0.807063) (364,0.770639) (368,0.735853) (372,0.702664) (376,0.671006) (380,0.640802) (384,0.611966) (388,0.584408) (392,0.558043) (396,0.532788) (400,0.508566) (404,0.485306) (408,0.462384) (412,0.441041) (416,0.420442) (420,0.400557) (424,0.381360) (428,0.362828) (432,0.344939) (436,0.327674) (440,0.311016) (444,0.294952) (448,0.279468) (452,0.264553) (456,0.249352) (460,0.235746) (464,0.222635) (468,0.210019) (472,0.197899) (476,0.186270) (480,0.175131) (484,0.164474) (488,0.154295) (492,0.144585) (496,0.135335) (500,0.126537) (504,0.118179) (508,0.110251) };
\end{axis}
\end{tikzpicture}

%% file: figures/combined_experiments/n3_cyclic_cross_section_citetvicini2021non_gamma0p5.tex
\begin{tikzpicture}
\begin{axis}[width=\cellsize,height=\cellsize,scale only axis,yticklabels={},xmin=0,xmax=511,ymin=0,ymax=1.1,grid=major,xtick=\empty,ytick={0,0.25,0.5,0.75,1}]
\addplot[red, thick] coordinates { (0,0.000000) (4,0.000000) (8,0.000000) (12,0.000000) (16,0.000000) (20,0.000000) (24,0.000000) (28,0.000000) (32,0.000000) (36,0.000000) (40,0.000000) (44,0.000000) (48,0.000000) (52,0.000000) (56,0.000000) (60,0.000000) (64,0.000000) (68,0.000000) (72,0.000000) (76,0.000000) (80,0.000000) (84,0.000000) (88,0.000000) (92,0.000000) (96,0.000000) (100,0.000000) (104,0.000000) (108,0.001065) (112,0.001534) (116,0.002187) (120,0.003089) (124,0.004318) (128,0.005976) (132,0.008189) (136,0.011109) (140,0.014921) (144,0.019841) (148,0.026121) (152,0.034047) (156,0.043937) (160,0.056135) (164,0.071005) (168,0.088922) (172,0.110251) (176,0.135335) (180,0.164474) (184,0.197899) (188,0.235746) (192,0.278037) (196,0.324652) (200,0.375311) (204,0.429557) (208,0.486752) (212,0.546074) (216,0.606531) (220,0.666977) (224,0.726149) (228,0.782705) (232,0.835270) (236,0.882497) (240,0.923116) (244,0.955997) (248,0.980199) (252,0.995012) (256,1.000000) (260,0.995012) (264,0.980199) (268,0.955997) (272,0.923116) (276,0.882497) (280,0.835270) (284,0.782705) (288,0.726149) (292,0.666977) (296,0.606531) (300,0.546074) (304,0.486752) (308,0.429557) (312,0.375311) (316,0.324652) (320,0.278037) (324,0.235746) (328,0.197899) (332,0.164474) (336,0.135335) (340,0.110251) (344,0.088922) (348,0.071005) (352,0.056135) (356,0.043937) (360,0.034047) (364,0.026121) (368,0.019841) (372,0.014921) (376,0.011109) (380,0.008189) (384,0.005976) (388,0.004318) (392,0.003089) (396,0.002187) (400,0.001534) (404,0.001065) (408,0.000000) (412,0.000000) (416,0.000000) (420,0.000000) (424,0.000000) (428,0.000000) (432,0.000000) (436,0.000000) (440,0.000000) (444,0.000000) (448,0.000000) (452,0.000000) (456,0.000000) (460,0.000000) (464,0.000000) (468,0.000000) (472,0.000000) (476,0.000000) (480,0.000000) (484,0.000000) (488,0.000000) (492,0.000000) (496,0.000000) (500,0.000000) (504,0.000000) (508,0.000000) };
\addplot[green!70!black, thick] coordinates { (0,0.102740) (4,0.110251) (8,0.118179) (12,0.126537) (16,0.135335) (20,0.144585) (24,0.154295) (28,0.164474) (32,0.175131) (36,0.186270) (40,0.197899) (44,0.210019) (48,0.222635) (52,0.235746) (56,0.249352) (60,0.263451) (64,0.278037) (68,0.293106) (72,0.308647) (76,0.324652) (80,0.341108) (84,0.358000) (88,0.375311) (92,0.393022) (96,0.411112) (100,0.429557) (104,0.448332) (108,0.467158) (112,0.486379) (116,0.505782) (120,0.525309) (124,0.544895) (128,0.564462) (132,0.583920) (136,0.603162) (140,0.622065) (144,0.640488) (148,0.658266) (152,0.675214) (156,0.691124) (160,0.705768) (164,0.718897) (168,0.730250) (172,0.739558) (176,0.746553) (180,0.750983) (184,0.736036) (188,0.707573) (192,0.674107) (196,0.635670) (200,0.592464) (204,0.544882) (208,0.493518) (212,0.439170) (216,0.382838) (220,0.325696) (224,0.269068) (228,0.214378) (232,0.163099) (236,0.116693) (240,0.076543) (244,0.043893) (248,0.019779) (252,0.004986) (256,0.000000) (260,0.004986) (264,0.019779) (268,0.043893) (272,0.076543) (276,0.116693) (280,0.163099) (284,0.214378) (288,0.269068) (292,0.325696) (296,0.382838) (300,0.439170) (304,0.493518) (308,0.544882) (312,0.592464) (316,0.635670) (320,0.674107) (324,0.707573) (328,0.736036) (332,0.750983) (336,0.746553) (340,0.739558) (344,0.730250) (348,0.718897) (352,0.705768) (356,0.691124) (360,0.675214) (364,0.658266) (368,0.640488) (372,0.622065) (376,0.603162) (380,0.583920) (384,0.564462) (388,0.544895) (392,0.525309) (396,0.505782) (400,0.486379) (404,0.467158) (408,0.448332) (412,0.429557) (416,0.411112) (420,0.393022) (424,0.375311) (428,0.358000) (432,0.341108) (436,0.324652) (440,0.308647) (444,0.293106) (448,0.278037) (452,0.263451) (456,0.249352) (460,0.235746) (464,0.222635) (468,0.210019) (472,0.197899) (476,0.186270) (480,0.175131) (484,0.164474) (488,0.154295) (492,0.144585) (496,0.135335) (500,0.126537) (504,0.118179) (508,0.110251) };
\addplot[blue, thick] coordinates { (0,0.000000) (4,0.000000) (8,0.000000) (12,0.000000) (16,0.000000) (20,0.000000) (24,0.000000) (28,0.000000) (32,0.000000) (36,0.000000) (40,0.000000) (44,0.000000) (48,0.000000) (52,0.000000) (56,0.000000) (60,0.001102) (64,0.001431) (68,0.001846) (72,0.002369) (76,0.003021) (80,0.003830) (84,0.004828) (88,0.006049) (92,0.007535) (96,0.009330) (100,0.011484) (104,0.014053) (108,0.017089) (112,0.020663) (116,0.024837) (120,0.029678) (124,0.035252) (128,0.041626) (132,0.048860) (136,0.057012) (140,0.066130) (144,0.076251) (148,0.087401) (152,0.099590) (156,0.112812) (160,0.127041) (164,0.119028) (168,0.101423) (172,0.081920) (176,0.060323) (180,0.036476) (184,0.026860) (188,0.025398) (192,0.023502) (196,0.021238) (200,0.018696) (204,0.015984) (208,0.013226) (212,0.010543) (216,0.008053) (220,0.005851) (224,0.004004) (228,0.002546) (232,0.001476) (236,0.000756) (240,0.000325) (244,0.000107) (248,0.000022) (252,0.000001) (256,0.000000) (260,0.000001) (264,0.000022) (268,0.000107) (272,0.000325) (276,0.000756) (280,0.001476) (284,0.002546) (288,0.004004) (292,0.005851) (296,0.008053) (300,0.010543) (304,0.013226) (308,0.015984) (312,0.018696) (316,0.021238) (320,0.023502) (324,0.025398) (328,0.026860) (332,0.036476) (336,0.060323) (340,0.081920) (344,0.101423) (348,0.119028) (352,0.127041) (356,0.112812) (360,0.099590) (364,0.087401) (368,0.076251) (372,0.066130) (376,0.057012) (380,0.048860) (384,0.041626) (388,0.035252) (392,0.029678) (396,0.024837) (400,0.020663) (404,0.017089) (408,0.014053) (412,0.011484) (416,0.009330) (420,0.007535) (424,0.006049) (428,0.004828) (432,0.003830) (436,0.003021) (440,0.002369) (444,0.001846) (448,0.001431) (452,0.001102) (456,0.000000) (460,0.000000) (464,0.000000) (468,0.000000) (472,0.000000) (476,0.000000) (480,0.000000) (484,0.000000) (488,0.000000) (492,0.000000) (496,0.000000) (500,0.000000) (504,0.000000) (508,0.000000) };
\addplot[black, thick, dashed] coordinates { (0,0.102740) (4,0.110251) (8,0.118179) (12,0.126537) (16,0.135335) (20,0.144585) (24,0.154295) (28,0.164474) (32,0.175131) (36,0.186270) (40,0.197899) (44,0.210019) (48,0.222635) (52,0.235746) (56,0.249352) (60,0.264553) (64,0.279468) (68,0.294952) (72,0.311016) (76,0.327674) (80,0.344939) (84,0.362828) (88,0.381360) (92,0.400557) (96,0.420442) (100,0.441041) (104,0.462384) (108,0.485311) (112,0.508576) (116,0.532806) (120,0.558076) (124,0.584465) (128,0.612064) (132,0.640969) (136,0.671283) (140,0.703116) (144,0.736580) (148,0.771788) (152,0.808852) (156,0.847873) (160,0.888944) (164,0.908931) (168,0.920595) (172,0.931728) (176,0.942212) (180,0.951934) (184,0.960795) (188,0.968717) (192,0.975646) (196,0.981560) (200,0.986471) (204,0.990424) (208,0.993495) (212,0.995788) (216,0.997421) (220,0.998524) (224,0.999221) (228,0.999629) (232,0.999845) (236,0.999946) (240,0.999985) (244,0.999997) (248,1.000000) (252,1.000000) (256,1.000000) (260,1.000000) (264,1.000000) (268,0.999997) (272,0.999985) (276,0.999946) (280,0.999845) (284,0.999629) (288,0.999221) (292,0.998524) (296,0.997421) (300,0.995788) (304,0.993495) (308,0.990424) (312,0.986471) (316,0.981560) (320,0.975646) (324,0.968717) (328,0.960795) (332,0.951934) (336,0.942212) (340,0.931728) (344,0.920595) (348,0.908931) (352,0.888944) (356,0.847873) (360,0.808852) (364,0.771788) (368,0.736580) (372,0.703116) (376,0.671283) (380,0.640969) (384,0.612064) (388,0.584465) (392,0.558076) (396,0.532806) (400,0.508576) (404,0.485311) (408,0.462384) (412,0.441041) (416,0.420442) (420,0.400557) (424,0.381360) (428,0.362828) (432,0.344939) (436,0.327674) (440,0.311016) (444,0.294952) (448,0.279468) (452,0.264553) (456,0.249352) (460,0.235746) (464,0.222635) (468,0.210019) (472,0.197899) (476,0.186270) (480,0.175131) (484,0.164474) (488,0.154295) (492,0.144585) (496,0.135335) (500,0.126537) (504,0.118179) (508,0.110251) };
\end{axis}
\end{tikzpicture}

%% file: figures/combined_experiments/n3_cyclic_cross_section_Blend_gamma0p5.tex
\begin{tikzpicture}
\begin{axis}[width=\cellsize,height=\cellsize,scale only axis,yticklabels={},xmin=0,xmax=511,ymin=0,ymax=1.1,grid=major,xtick=\empty,ytick={0,0.25,0.5,0.75,1}]
\addplot[red, thick] coordinates { (0,0.000000) (4,0.000000) (8,0.000000) (12,0.000000) (16,0.000000) (20,0.000000) (24,0.000000) (28,0.000000) (32,0.000000) (36,0.000000) (40,0.000000) (44,0.000000) (48,0.000000) (52,0.000000) (56,0.000000) (60,0.000000) (64,0.000000) (68,0.000000) (72,0.000000) (76,0.000000) (80,0.000000) (84,0.000000) (88,0.000000) (92,0.000000) (96,0.000000) (100,0.000000) (104,0.000000) (108,0.001065) (112,0.001534) (116,0.002187) (120,0.003089) (124,0.004318) (128,0.005976) (132,0.008189) (136,0.011109) (140,0.014921) (144,0.019841) (148,0.026121) (152,0.034047) (156,0.043937) (160,0.056135) (164,0.071005) (168,0.088922) (172,0.110251) (176,0.135335) (180,0.164474) (184,0.197899) (188,0.235746) (192,0.278037) (196,0.324652) (200,0.375311) (204,0.429557) (208,0.486752) (212,0.546074) (216,0.606531) (220,0.666977) (224,0.726149) (228,0.782705) (232,0.835270) (236,0.882497) (240,0.923116) (244,0.955997) (248,0.980199) (252,0.995012) (256,1.000000) (260,0.995012) (264,0.980199) (268,0.955997) (272,0.923116) (276,0.882497) (280,0.835270) (284,0.782705) (288,0.726149) (292,0.666977) (296,0.606531) (300,0.546074) (304,0.486752) (308,0.429557) (312,0.375311) (316,0.324652) (320,0.278037) (324,0.235746) (328,0.197899) (332,0.164474) (336,0.135335) (340,0.110251) (344,0.088922) (348,0.071005) (352,0.056135) (356,0.043937) (360,0.034047) (364,0.026121) (368,0.019841) (372,0.014921) (376,0.011109) (380,0.008189) (384,0.005976) (388,0.004318) (392,0.003089) (396,0.002187) (400,0.001534) (404,0.001065) (408,0.000000) (412,0.000000) (416,0.000000) (420,0.000000) (424,0.000000) (428,0.000000) (432,0.000000) (436,0.000000) (440,0.000000) (444,0.000000) (448,0.000000) (452,0.000000) (456,0.000000) (460,0.000000) (464,0.000000) (468,0.000000) (472,0.000000) (476,0.000000) (480,0.000000) (484,0.000000) (488,0.000000) (492,0.000000) (496,0.000000) (500,0.000000) (504,0.000000) (508,0.000000) };
\addplot[green!70!black, thick] coordinates { (0,0.102740) (4,0.110251) (8,0.118179) (12,0.126537) (16,0.135335) (20,0.144585) (24,0.154295) (28,0.164474) (32,0.175131) (36,0.186270) (40,0.197899) (44,0.210019) (48,0.222635) (52,0.235746) (56,0.249352) (60,0.263451) (64,0.278037) (68,0.293106) (72,0.308647) (76,0.324652) (80,0.341108) (84,0.358000) (88,0.375311) (92,0.393022) (96,0.411112) (100,0.429557) (104,0.448332) (108,0.467158) (112,0.486379) (116,0.505782) (120,0.525309) (124,0.544895) (128,0.564462) (132,0.583920) (136,0.603162) (140,0.622065) (144,0.640488) (148,0.658266) (152,0.675214) (156,0.691124) (160,0.705768) (164,0.718897) (168,0.730250) (172,0.739558) (176,0.746553) (180,0.750983) (184,0.752621) (188,0.751282) (192,0.721963) (196,0.675348) (200,0.624689) (204,0.570443) (208,0.513248) (212,0.453926) (216,0.393469) (220,0.333023) (224,0.273851) (228,0.217295) (232,0.164730) (236,0.117503) (240,0.076884) (244,0.044003) (248,0.019801) (252,0.004988) (256,0.000000) (260,0.004988) (264,0.019801) (268,0.044003) (272,0.076884) (276,0.117503) (280,0.164730) (284,0.217295) (288,0.273851) (292,0.333023) (296,0.393469) (300,0.453926) (304,0.513248) (308,0.570443) (312,0.624689) (316,0.675348) (320,0.721963) (324,0.751282) (328,0.752621) (332,0.750983) (336,0.746553) (340,0.739558) (344,0.730250) (348,0.718897) (352,0.705768) (356,0.691124) (360,0.675214) (364,0.658266) (368,0.640488) (372,0.622065) (376,0.603162) (380,0.583920) (384,0.564462) (388,0.544895) (392,0.525309) (396,0.505782) (400,0.486379) (404,0.467158) (408,0.448332) (412,0.429557) (416,0.411112) (420,0.393022) (424,0.375311) (428,0.358000) (432,0.341108) (436,0.324652) (440,0.308647) (444,0.293106) (448,0.278037) (452,0.263451) (456,0.249352) (460,0.235746) (464,0.222635) (468,0.210019) (472,0.197899) (476,0.186270) (480,0.175131) (484,0.164474) (488,0.154295) (492,0.144585) (496,0.135335) (500,0.126537) (504,0.118179) (508,0.110251) };
\addplot[blue, thick] coordinates { (0,0.000000) (4,0.000000) (8,0.000000) (12,0.000000) (16,0.000000) (20,0.000000) (24,0.000000) (28,0.000000) (32,0.000000) (36,0.000000) (40,0.000000) (44,0.000000) (48,0.000000) (52,0.000000) (56,0.000000) (60,0.001102) (64,0.001431) (68,0.001846) (72,0.002369) (76,0.003021) (80,0.003830) (84,0.004828) (88,0.006049) (92,0.007535) (96,0.009330) (100,0.011484) (104,0.014053) (108,0.017089) (112,0.020663) (116,0.024837) (120,0.029678) (124,0.035252) (128,0.041626) (132,0.048860) (136,0.057012) (140,0.066130) (144,0.076251) (148,0.087401) (152,0.099590) (156,0.112812) (160,0.127041) (164,0.142237) (168,0.158341) (172,0.150192) (176,0.118111) (180,0.084542) (184,0.049481) (188,0.012972) (192,0.000000) (196,0.000000) (200,0.000000) (204,0.000000) (208,0.000000) (212,0.000000) (216,0.000000) (220,0.000000) (224,0.000000) (228,0.000000) (232,0.000000) (236,0.000000) (240,0.000000) (244,0.000000) (248,0.000000) (252,0.000000) (256,0.000000) (260,0.000000) (264,0.000000) (268,0.000000) (272,0.000000) (276,0.000000) (280,0.000000) (284,0.000000) (288,0.000000) (292,0.000000) (296,0.000000) (300,0.000000) (304,0.000000) (308,0.000000) (312,0.000000) (316,0.000000) (320,0.000000) (324,0.012972) (328,0.049481) (332,0.084542) (336,0.118111) (340,0.150192) (344,0.158341) (348,0.142237) (352,0.127041) (356,0.112812) (360,0.099590) (364,0.087401) (368,0.076251) (372,0.066130) (376,0.057012) (380,0.048860) (384,0.041626) (388,0.035252) (392,0.029678) (396,0.024837) (400,0.020663) (404,0.017089) (408,0.014053) (412,0.011484) (416,0.009330) (420,0.007535) (424,0.006049) (428,0.004828) (432,0.003830) (436,0.003021) (440,0.002369) (444,0.001846) (448,0.001431) (452,0.001102) (456,0.000000) (460,0.000000) (464,0.000000) (468,0.000000) (472,0.000000) (476,0.000000) (480,0.000000) (484,0.000000) (488,0.000000) (492,0.000000) (496,0.000000) (500,0.000000) (504,0.000000) (508,0.000000) };
\addplot[black, thick, dashed] coordinates { (0,0.102740) (4,0.110251) (8,0.118179) (12,0.126537) (16,0.135335) (20,0.144585) (24,0.154295) (28,0.164474) (32,0.175131) (36,0.186270) (40,0.197899) (44,0.210019) (48,0.222635) (52,0.235746) (56,0.249352) (60,0.264553) (64,0.279468) (68,0.294952) (72,0.311016) (76,0.327674) (80,0.344939) (84,0.362828) (88,0.381360) (92,0.400557) (96,0.420442) (100,0.441041) (104,0.462384) (108,0.485311) (112,0.508576) (116,0.532806) (120,0.558076) (124,0.584465) (128,0.612064) (132,0.640969) (136,0.671283) (140,0.703116) (144,0.736580) (148,0.771788) (152,0.808852) (156,0.847873) (160,0.888944) (164,0.932140) (168,0.977512) (172,1.000000) (176,1.000000) (180,1.000000) (184,1.000000) (188,1.000000) (192,1.000000) (196,1.000000) (200,1.000000) (204,1.000000) (208,1.000000) (212,1.000000) (216,1.000000) (220,1.000000) (224,1.000000) (228,1.000000) (232,1.000000) (236,1.000000) (240,1.000000) (244,1.000000) (248,1.000000) (252,1.000000) (256,1.000000) (260,1.000000) (264,1.000000) (268,1.000000) (272,1.000000) (276,1.000000) (280,1.000000) (284,1.000000) (288,1.000000) (292,1.000000) (296,1.000000) (300,1.000000) (304,1.000000) (308,1.000000) (312,1.000000) (316,1.000000) (320,1.000000) (324,1.000000) (328,1.000000) (332,1.000000) (336,1.000000) (340,1.000000) (344,0.977512) (348,0.932140) (352,0.888944) (356,0.847873) (360,0.808852) (364,0.771788) (368,0.736580) (372,0.703116) (376,0.671283) (380,0.640969) (384,0.612064) (388,0.584465) (392,0.558076) (396,0.532806) (400,0.508576) (404,0.485311) (408,0.462384) (412,0.441041) (416,0.420442) (420,0.400557) (424,0.381360) (428,0.362828) (432,0.344939) (436,0.327674) (440,0.311016) (444,0.294952) (448,0.279468) (452,0.264553) (456,0.249352) (460,0.235746) (464,0.222635) (468,0.210019) (472,0.197899) (476,0.186270) (480,0.175131) (484,0.164474) (488,0.154295) (492,0.144585) (496,0.135335) (500,0.126537) (504,0.118179) (508,0.110251) };
\end{axis}
\end{tikzpicture}

%% file: figures/combined_experiments/n3_cyclic_cross_section_Linear.tex
\begin{tikzpicture}
\begin{axis}[width=\cellsize,height=\cellsize,scale only axis,yticklabels={},xmin=0,xmax=511,ymin=0,ymax=1.1,grid=major,xtick=\empty,ytick={0,0.25,0.5,0.75,1}]
\addplot[red, thick] coordinates { (0,0.000000) (4,0.000000) (8,0.000000) (12,0.000000) (16,0.000000) (20,0.000000) (24,0.000000) (28,0.000000) (32,0.000000) (36,0.000000) (40,0.000000) (44,0.000000) (48,0.000000) (52,0.000000) (56,0.000000) (60,0.000000) (64,0.000000) (68,0.000000) (72,0.000000) (76,0.000000) (80,0.000000) (84,0.000000) (88,0.000000) (92,0.000000) (96,0.000000) (100,0.000000) (104,0.000000) (108,0.001065) (112,0.001534) (116,0.002187) (120,0.003089) (124,0.004318) (128,0.005976) (132,0.008189) (136,0.011109) (140,0.014921) (144,0.019841) (148,0.026121) (152,0.034047) (156,0.043937) (160,0.056135) (164,0.071005) (168,0.088922) (172,0.110251) (176,0.135335) (180,0.164474) (184,0.197899) (188,0.235746) (192,0.278037) (196,0.324652) (200,0.375311) (204,0.429557) (208,0.486752) (212,0.546074) (216,0.606531) (220,0.666977) (224,0.726149) (228,0.782705) (232,0.835270) (236,0.882497) (240,0.923116) (244,0.955997) (248,0.980199) (252,0.995012) (256,1.000000) (260,0.995012) (264,0.980199) (268,0.955997) (272,0.923116) (276,0.882497) (280,0.835270) (284,0.782705) (288,0.726149) (292,0.666977) (296,0.606531) (300,0.546074) (304,0.486752) (308,0.429557) (312,0.375311) (316,0.324652) (320,0.278037) (324,0.235746) (328,0.197899) (332,0.164474) (336,0.135335) (340,0.110251) (344,0.088922) (348,0.071005) (352,0.056135) (356,0.043937) (360,0.034047) (364,0.026121) (368,0.019841) (372,0.014921) (376,0.011109) (380,0.008189) (384,0.005976) (388,0.004318) (392,0.003089) (396,0.002187) (400,0.001534) (404,0.001065) (408,0.000000) (412,0.000000) (416,0.000000) (420,0.000000) (424,0.000000) (428,0.000000) (432,0.000000) (436,0.000000) (440,0.000000) (444,0.000000) (448,0.000000) (452,0.000000) (456,0.000000) (460,0.000000) (464,0.000000) (468,0.000000) (472,0.000000) (476,0.000000) (480,0.000000) (484,0.000000) (488,0.000000) (492,0.000000) (496,0.000000) (500,0.000000) (504,0.000000) (508,0.000000) };
\addplot[green!70!black, thick] coordinates { (0,0.102740) (4,0.110251) (8,0.118179) (12,0.126537) (16,0.135335) (20,0.144585) (24,0.154295) (28,0.164474) (32,0.175131) (36,0.186270) (40,0.197899) (44,0.210019) (48,0.222635) (52,0.235746) (56,0.249352) (60,0.263451) (64,0.278037) (68,0.293106) (72,0.308647) (76,0.324652) (80,0.341108) (84,0.358000) (88,0.375311) (92,0.393022) (96,0.411112) (100,0.429557) (104,0.448332) (108,0.467407) (112,0.486752) (116,0.506336) (120,0.526122) (124,0.546074) (128,0.566154) (132,0.586320) (136,0.606531) (140,0.626741) (144,0.646905) (148,0.666977) (152,0.686908) (156,0.706648) (160,0.726149) (164,0.745359) (168,0.764228) (172,0.782705) (176,0.800737) (180,0.818276) (184,0.802101) (188,0.764254) (192,0.721963) (196,0.675348) (200,0.624689) (204,0.570443) (208,0.513248) (212,0.453926) (216,0.393469) (220,0.333023) (224,0.273851) (228,0.217295) (232,0.164730) (236,0.117503) (240,0.076884) (244,0.044003) (248,0.019801) (252,0.004988) (256,0.000000) (260,0.004988) (264,0.019801) (268,0.044003) (272,0.076884) (276,0.117503) (280,0.164730) (284,0.217295) (288,0.273851) (292,0.333023) (296,0.393469) (300,0.453926) (304,0.513248) (308,0.570443) (312,0.624689) (316,0.675348) (320,0.721963) (324,0.764254) (328,0.802101) (332,0.818276) (336,0.800737) (340,0.782705) (344,0.764228) (348,0.745359) (352,0.726149) (356,0.706648) (360,0.686908) (364,0.666977) (368,0.646905) (372,0.626741) (376,0.606531) (380,0.586320) (384,0.566154) (388,0.546074) (392,0.526122) (396,0.506336) (400,0.486752) (404,0.467407) (408,0.448332) (412,0.429557) (416,0.411112) (420,0.393022) (424,0.375311) (428,0.358000) (432,0.341108) (436,0.324652) (440,0.308647) (444,0.293106) (448,0.278037) (452,0.263451) (456,0.249352) (460,0.235746) (464,0.222635) (468,0.210019) (472,0.197899) (476,0.186270) (480,0.175131) (484,0.164474) (488,0.154295) (492,0.144585) (496,0.135335) (500,0.126537) (504,0.118179) (508,0.110251) };
\addplot[blue, thick] coordinates { (0,0.000000) (4,0.000000) (8,0.000000) (12,0.000000) (16,0.000000) (20,0.000000) (24,0.000000) (28,0.000000) (32,0.000000) (36,0.000000) (40,0.000000) (44,0.000000) (48,0.000000) (52,0.000000) (56,0.000000) (60,0.001269) (64,0.001662) (68,0.002163) (72,0.002801) (76,0.003607) (80,0.004618) (84,0.005881) (88,0.007447) (92,0.009377) (96,0.011744) (100,0.014625) (104,0.018113) (108,0.022309) (112,0.027324) (116,0.033281) (120,0.040312) (124,0.048558) (128,0.058167) (132,0.069291) (136,0.082085) (140,0.096703) (144,0.113293) (148,0.131994) (152,0.152930) (156,0.176204) (160,0.201897) (164,0.183635) (168,0.146850) (172,0.107045) (176,0.063927) (180,0.017250) (184,0.000000) (188,0.000000) (192,0.000000) (196,0.000000) (200,0.000000) (204,0.000000) (208,0.000000) (212,0.000000) (216,0.000000) (220,0.000000) (224,0.000000) (228,0.000000) (232,0.000000) (236,0.000000) (240,0.000000) (244,0.000000) (248,0.000000) (252,0.000000) (256,0.000000) (260,0.000000) (264,0.000000) (268,0.000000) (272,0.000000) (276,0.000000) (280,0.000000) (284,0.000000) (288,0.000000) (292,0.000000) (296,0.000000) (300,0.000000) (304,0.000000) (308,0.000000) (312,0.000000) (316,0.000000) (320,0.000000) (324,0.000000) (328,0.000000) (332,0.017250) (336,0.063927) (340,0.107045) (344,0.146850) (348,0.183635) (352,0.201897) (356,0.176204) (360,0.152930) (364,0.131994) (368,0.113293) (372,0.096703) (376,0.082085) (380,0.069291) (384,0.058167) (388,0.048558) (392,0.040312) (396,0.033281) (400,0.027324) (404,0.022309) (408,0.018113) (412,0.014625) (416,0.011744) (420,0.009377) (424,0.007447) (428,0.005881) (432,0.004618) (436,0.003607) (440,0.002801) (444,0.002163) (448,0.001662) (452,0.001269) (456,0.000000) (460,0.000000) (464,0.000000) (468,0.000000) (472,0.000000) (476,0.000000) (480,0.000000) (484,0.000000) (488,0.000000) (492,0.000000) (496,0.000000) (500,0.000000) (504,0.000000) (508,0.000000) };
\addplot[black, thick, dashed] coordinates { (0,0.102740) (4,0.110251) (8,0.118179) (12,0.126537) (16,0.135335) (20,0.144585) (24,0.154295) (28,0.164474) (32,0.175131) (36,0.186270) (40,0.197899) (44,0.210019) (48,0.222635) (52,0.235746) (56,0.249352) (60,0.264720) (64,0.279699) (68,0.295269) (72,0.311448) (76,0.328259) (80,0.345726) (84,0.363881) (88,0.382758) (92,0.402400) (96,0.422856) (100,0.444183) (104,0.466445) (108,0.490780) (112,0.515610) (116,0.541804) (120,0.569522) (124,0.598950) (128,0.630297) (132,0.663800) (136,0.699725) (140,0.738365) (144,0.780039) (148,0.825092) (152,0.873885) (156,0.926790) (160,0.984180) (164,1.000000) (168,1.000000) (172,1.000000) (176,1.000000) (180,1.000000) (184,1.000000) (188,1.000000) (192,1.000000) (196,1.000000) (200,1.000000) (204,1.000000) (208,1.000000) (212,1.000000) (216,1.000000) (220,1.000000) (224,1.000000) (228,1.000000) (232,1.000000) (236,1.000000) (240,1.000000) (244,1.000000) (248,1.000000) (252,1.000000) (256,1.000000) (260,1.000000) (264,1.000000) (268,1.000000) (272,1.000000) (276,1.000000) (280,1.000000) (284,1.000000) (288,1.000000) (292,1.000000) (296,1.000000) (300,1.000000) (304,1.000000) (308,1.000000) (312,1.000000) (316,1.000000) (320,1.000000) (324,1.000000) (328,1.000000) (332,1.000000) (336,1.000000) (340,1.000000) (344,1.000000) (348,1.000000) (352,0.984180) (356,0.926790) (360,0.873885) (364,0.825092) (368,0.780039) (372,0.738365) (376,0.699725) (380,0.663800) (384,0.630297) (388,0.598950) (392,0.569522) (396,0.541804) (400,0.515610) (404,0.490780) (408,0.466445) (412,0.444183) (416,0.422856) (420,0.402400) (424,0.382758) (428,0.363881) (432,0.345726) (436,0.328259) (440,0.311448) (444,0.295269) (448,0.279699) (452,0.264720) (456,0.249352) (460,0.235746) (464,0.222635) (468,0.210019) (472,0.197899) (476,0.186270) (480,0.175131) (484,0.164474) (488,0.154295) (492,0.144585) (496,0.135335) (500,0.126537) (504,0.118179) (508,0.110251) };
\end{axis}
\end{tikzpicture}

%% file: figures/combined_experiments/n3_cyclic_cross_section_Quadratic_c0p5.tex
\begin{tikzpicture}
\begin{axis}[width=\cellsize,height=\cellsize,scale only axis,yticklabels={},xmin=0,xmax=511,ymin=0,ymax=1.1,grid=major,xtick=\empty,ytick={0,0.25,0.5,0.75,1}]
\addplot[red, thick] coordinates { (0,0.000000) (4,0.000000) (8,0.000000) (12,0.000000) (16,0.000000) (20,0.000000) (24,0.000000) (28,0.000000) (32,0.000000) (36,0.000000) (40,0.000000) (44,0.000000) (48,0.000000) (52,0.000000) (56,0.000000) (60,0.000000) (64,0.000000) (68,0.000000) (72,0.000000) (76,0.000000) (80,0.000000) (84,0.000000) (88,0.000000) (92,0.000000) (96,0.000000) (100,0.000000) (104,0.000000) (108,0.001065) (112,0.001534) (116,0.002187) (120,0.003089) (124,0.004318) (128,0.005976) (132,0.008189) (136,0.011109) (140,0.014921) (144,0.019841) (148,0.026121) (152,0.034047) (156,0.043937) (160,0.056135) (164,0.071005) (168,0.088922) (172,0.110251) (176,0.135335) (180,0.164474) (184,0.197899) (188,0.235746) (192,0.278037) (196,0.324652) (200,0.375311) (204,0.429557) (208,0.486752) (212,0.546074) (216,0.606531) (220,0.666977) (224,0.726149) (228,0.782705) (232,0.835270) (236,0.882497) (240,0.923116) (244,0.955997) (248,0.980199) (252,0.995012) (256,1.000000) (260,0.995012) (264,0.980199) (268,0.955997) (272,0.923116) (276,0.882497) (280,0.835270) (284,0.782705) (288,0.726149) (292,0.666977) (296,0.606531) (300,0.546074) (304,0.486752) (308,0.429557) (312,0.375311) (316,0.324652) (320,0.278037) (324,0.235746) (328,0.197899) (332,0.164474) (336,0.135335) (340,0.110251) (344,0.088922) (348,0.071005) (352,0.056135) (356,0.043937) (360,0.034047) (364,0.026121) (368,0.019841) (372,0.014921) (376,0.011109) (380,0.008189) (384,0.005976) (388,0.004318) (392,0.003089) (396,0.002187) (400,0.001534) (404,0.001065) (408,0.000000) (412,0.000000) (416,0.000000) (420,0.000000) (424,0.000000) (428,0.000000) (432,0.000000) (436,0.000000) (440,0.000000) (444,0.000000) (448,0.000000) (452,0.000000) (456,0.000000) (460,0.000000) (464,0.000000) (468,0.000000) (472,0.000000) (476,0.000000) (480,0.000000) (484,0.000000) (488,0.000000) (492,0.000000) (496,0.000000) (500,0.000000) (504,0.000000) (508,0.000000) };
\addplot[green!70!black, thick] coordinates { (0,0.102740) (4,0.110251) (8,0.118179) (12,0.126537) (16,0.135335) (20,0.144585) (24,0.154295) (28,0.164474) (32,0.175131) (36,0.186270) (40,0.197899) (44,0.210019) (48,0.222635) (52,0.235746) (56,0.249352) (60,0.263451) (64,0.278037) (68,0.293106) (72,0.308647) (76,0.324652) (80,0.341108) (84,0.358000) (88,0.375311) (92,0.393022) (96,0.411112) (100,0.429557) (104,0.448332) (108,0.467656) (112,0.487126) (116,0.506889) (120,0.526934) (124,0.547253) (128,0.567846) (132,0.588721) (136,0.609900) (140,0.631417) (144,0.653323) (148,0.675688) (152,0.698601) (156,0.722172) (160,0.746530) (164,0.771822) (168,0.798206) (172,0.825851) (176,0.854921) (180,0.835526) (184,0.802101) (188,0.764254) (192,0.721963) (196,0.675348) (200,0.624689) (204,0.570443) (208,0.513248) (212,0.453926) (216,0.393469) (220,0.333023) (224,0.273851) (228,0.217295) (232,0.164730) (236,0.117503) (240,0.076884) (244,0.044003) (248,0.019801) (252,0.004988) (256,0.000000) (260,0.004988) (264,0.019801) (268,0.044003) (272,0.076884) (276,0.117503) (280,0.164730) (284,0.217295) (288,0.273851) (292,0.333023) (296,0.393469) (300,0.453926) (304,0.513248) (308,0.570443) (312,0.624689) (316,0.675348) (320,0.721963) (324,0.764254) (328,0.802101) (332,0.835526) (336,0.854921) (340,0.825851) (344,0.798206) (348,0.771822) (352,0.746530) (356,0.722172) (360,0.698601) (364,0.675688) (368,0.653323) (372,0.631417) (376,0.609900) (380,0.588721) (384,0.567846) (388,0.547253) (392,0.526934) (396,0.506889) (400,0.487126) (404,0.467656) (408,0.448332) (412,0.429557) (416,0.411112) (420,0.393022) (424,0.375311) (428,0.358000) (432,0.341108) (436,0.324652) (440,0.308647) (444,0.293106) (448,0.278037) (452,0.263451) (456,0.249352) (460,0.235746) (464,0.222635) (468,0.210019) (472,0.197899) (476,0.186270) (480,0.175131) (484,0.164474) (488,0.154295) (492,0.144585) (496,0.135335) (500,0.126537) (504,0.118179) (508,0.110251) };
\addplot[blue, thick] coordinates { (0,0.000000) (4,0.000000) (8,0.000000) (12,0.000000) (16,0.000000) (20,0.000000) (24,0.000000) (28,0.000000) (32,0.000000) (36,0.000000) (40,0.000000) (44,0.000000) (48,0.000000) (52,0.000000) (56,0.000000) (60,0.001436) (64,0.001893) (68,0.002480) (72,0.003233) (76,0.004192) (80,0.005406) (84,0.006933) (88,0.008844) (92,0.011220) (96,0.014158) (100,0.017767) (104,0.022174) (108,0.027534) (112,0.033995) (116,0.041743) (120,0.050979) (124,0.061921) (128,0.074806) (132,0.089888) (136,0.107434) (140,0.127728) (144,0.151062) (148,0.177736) (152,0.208057) (156,0.233891) (160,0.197335) (164,0.157173) (168,0.112872) (172,0.063898) (176,0.009743) (180,0.000000) (184,0.000000) (188,0.000000) (192,0.000000) (196,0.000000) (200,0.000000) (204,0.000000) (208,0.000000) (212,0.000000) (216,0.000000) (220,0.000000) (224,0.000000) (228,0.000000) (232,0.000000) (236,0.000000) (240,0.000000) (244,0.000000) (248,0.000000) (252,0.000000) (256,0.000000) (260,0.000000) (264,0.000000) (268,0.000000) (272,0.000000) (276,0.000000) (280,0.000000) (284,0.000000) (288,0.000000) (292,0.000000) (296,0.000000) (300,0.000000) (304,0.000000) (308,0.000000) (312,0.000000) (316,0.000000) (320,0.000000) (324,0.000000) (328,0.000000) (332,0.000000) (336,0.009743) (340,0.063898) (344,0.112872) (348,0.157173) (352,0.197335) (356,0.233891) (360,0.208057) (364,0.177736) (368,0.151062) (372,0.127728) (376,0.107434) (380,0.089888) (384,0.074806) (388,0.061921) (392,0.050979) (396,0.041743) (400,0.033995) (404,0.027534) (408,0.022174) (412,0.017767) (416,0.014158) (420,0.011220) (424,0.008844) (428,0.006933) (432,0.005406) (436,0.004192) (440,0.003233) (444,0.002480) (448,0.001893) (452,0.001436) (456,0.000000) (460,0.000000) (464,0.000000) (468,0.000000) (472,0.000000) (476,0.000000) (480,0.000000) (484,0.000000) (488,0.000000) (492,0.000000) (496,0.000000) (500,0.000000) (504,0.000000) (508,0.000000) };
\addplot[black, thick, dashed] coordinates { (0,0.102740) (4,0.110251) (8,0.118179) (12,0.126537) (16,0.135335) (20,0.144585) (24,0.154295) (28,0.164474) (32,0.175131) (36,0.186270) (40,0.197899) (44,0.210019) (48,0.222635) (52,0.235746) (56,0.249352) (60,0.264887) (64,0.279930) (68,0.295586) (72,0.311881) (76,0.328844) (80,0.346514) (84,0.364933) (88,0.384155) (92,0.404243) (96,0.425270) (100,0.447324) (104,0.470505) (108,0.496255) (112,0.522654) (116,0.550820) (120,0.581002) (124,0.613492) (128,0.648628) (132,0.686797) (136,0.728443) (140,0.774066) (144,0.824226) (148,0.879546) (152,0.940706) (156,1.000000) (160,1.000000) (164,1.000000) (168,1.000000) (172,1.000000) (176,1.000000) (180,1.000000) (184,1.000000) (188,1.000000) (192,1.000000) (196,1.000000) (200,1.000000) (204,1.000000) (208,1.000000) (212,1.000000) (216,1.000000) (220,1.000000) (224,1.000000) (228,1.000000) (232,1.000000) (236,1.000000) (240,1.000000) (244,1.000000) (248,1.000000) (252,1.000000) (256,1.000000) (260,1.000000) (264,1.000000) (268,1.000000) (272,1.000000) (276,1.000000) (280,1.000000) (284,1.000000) (288,1.000000) (292,1.000000) (296,1.000000) (300,1.000000) (304,1.000000) (308,1.000000) (312,1.000000) (316,1.000000) (320,1.000000) (324,1.000000) (328,1.000000) (332,1.000000) (336,1.000000) (340,1.000000) (344,1.000000) (348,1.000000) (352,1.000000) (356,1.000000) (360,0.940706) (364,0.879546) (368,0.824226) (372,0.774066) (376,0.728443) (380,0.686797) (384,0.648628) (388,0.613492) (392,0.581002) (396,0.550820) (400,0.522654) (404,0.496255) (408,0.470505) (412,0.447324) (416,0.425270) (420,0.404243) (424,0.384155) (428,0.364933) (432,0.346514) (436,0.328844) (440,0.311881) (444,0.295586) (448,0.279930) (452,0.264887) (456,0.249352) (460,0.235746) (464,0.222635) (468,0.210019) (472,0.197899) (476,0.186270) (480,0.175131) (484,0.164474) (488,0.154295) (492,0.144585) (496,0.135335) (500,0.126537) (504,0.118179) (508,0.110251) };
\end{axis}
\end{tikzpicture}

%% file: figures/combined_experiments/n3_cyclic_cross_section_Quadratic_c1.tex
\begin{tikzpicture}
\begin{axis}[width=\cellsize,height=\cellsize,scale only axis,yticklabels={},xmin=0,xmax=511,ymin=0,ymax=1.1,grid=major,xtick=\empty,ytick={0,0.25,0.5,0.75,1}]
\addplot[red, thick] coordinates { (0,0.000000) (4,0.000000) (8,0.000000) (12,0.000000) (16,0.000000) (20,0.000000) (24,0.000000) (28,0.000000) (32,0.000000) (36,0.000000) (40,0.000000) (44,0.000000) (48,0.000000) (52,0.000000) (56,0.000000) (60,0.000000) (64,0.000000) (68,0.000000) (72,0.000000) (76,0.000000) (80,0.000000) (84,0.000000) (88,0.000000) (92,0.000000) (96,0.000000) (100,0.000000) (104,0.000000) (108,0.001065) (112,0.001534) (116,0.002187) (120,0.003089) (124,0.004318) (128,0.005976) (132,0.008189) (136,0.011109) (140,0.014921) (144,0.019841) (148,0.026121) (152,0.034047) (156,0.043937) (160,0.056135) (164,0.071005) (168,0.088922) (172,0.110251) (176,0.135335) (180,0.164474) (184,0.197899) (188,0.235746) (192,0.278037) (196,0.324652) (200,0.375311) (204,0.429557) (208,0.486752) (212,0.546074) (216,0.606531) (220,0.666977) (224,0.726149) (228,0.782705) (232,0.835270) (236,0.882497) (240,0.923116) (244,0.955997) (248,0.980199) (252,0.995012) (256,1.000000) (260,0.995012) (264,0.980199) (268,0.955997) (272,0.923116) (276,0.882497) (280,0.835270) (284,0.782705) (288,0.726149) (292,0.666977) (296,0.606531) (300,0.546074) (304,0.486752) (308,0.429557) (312,0.375311) (316,0.324652) (320,0.278037) (324,0.235746) (328,0.197899) (332,0.164474) (336,0.135335) (340,0.110251) (344,0.088922) (348,0.071005) (352,0.056135) (356,0.043937) (360,0.034047) (364,0.026121) (368,0.019841) (372,0.014921) (376,0.011109) (380,0.008189) (384,0.005976) (388,0.004318) (392,0.003089) (396,0.002187) (400,0.001534) (404,0.001065) (408,0.000000) (412,0.000000) (416,0.000000) (420,0.000000) (424,0.000000) (428,0.000000) (432,0.000000) (436,0.000000) (440,0.000000) (444,0.000000) (448,0.000000) (452,0.000000) (456,0.000000) (460,0.000000) (464,0.000000) (468,0.000000) (472,0.000000) (476,0.000000) (480,0.000000) (484,0.000000) (488,0.000000) (492,0.000000) (496,0.000000) (500,0.000000) (504,0.000000) (508,0.000000) };
\addplot[green!70!black, thick] coordinates { (0,0.102740) (4,0.110251) (8,0.118179) (12,0.126537) (16,0.135335) (20,0.144585) (24,0.154295) (28,0.164474) (32,0.175131) (36,0.186270) (40,0.197899) (44,0.210019) (48,0.222635) (52,0.235746) (56,0.249352) (60,0.263451) (64,0.278037) (68,0.293106) (72,0.308647) (76,0.324652) (80,0.341108) (84,0.358000) (88,0.375311) (92,0.393022) (96,0.411112) (100,0.429557) (104,0.448332) (108,0.467904) (112,0.487499) (116,0.507443) (120,0.527747) (124,0.548432) (128,0.569537) (132,0.591122) (136,0.613269) (140,0.636092) (144,0.659740) (148,0.684399) (152,0.710295) (156,0.737696) (160,0.766911) (164,0.798284) (168,0.832185) (172,0.868998) (176,0.864665) (180,0.835526) (184,0.802101) (188,0.764254) (192,0.721963) (196,0.675348) (200,0.624689) (204,0.570443) (208,0.513248) (212,0.453926) (216,0.393469) (220,0.333023) (224,0.273851) (228,0.217295) (232,0.164730) (236,0.117503) (240,0.076884) (244,0.044003) (248,0.019801) (252,0.004988) (256,0.000000) (260,0.004988) (264,0.019801) (268,0.044003) (272,0.076884) (276,0.117503) (280,0.164730) (284,0.217295) (288,0.273851) (292,0.333023) (296,0.393469) (300,0.453926) (304,0.513248) (308,0.570443) (312,0.624689) (316,0.675348) (320,0.721963) (324,0.764254) (328,0.802101) (332,0.835526) (336,0.864665) (340,0.868998) (344,0.832185) (348,0.798284) (352,0.766911) (356,0.737696) (360,0.710295) (364,0.684399) (368,0.659740) (372,0.636092) (376,0.613269) (380,0.591122) (384,0.569537) (388,0.548432) (392,0.527747) (396,0.507443) (400,0.487499) (404,0.467904) (408,0.448332) (412,0.429557) (416,0.411112) (420,0.393022) (424,0.375311) (428,0.358000) (432,0.341108) (436,0.324652) (440,0.308647) (444,0.293106) (448,0.278037) (452,0.263451) (456,0.249352) (460,0.235746) (464,0.222635) (468,0.210019) (472,0.197899) (476,0.186270) (480,0.175131) (484,0.164474) (488,0.154295) (492,0.144585) (496,0.135335) (500,0.126537) (504,0.118179) (508,0.110251) };
\addplot[blue, thick] coordinates { (0,0.000000) (4,0.000000) (8,0.000000) (12,0.000000) (16,0.000000) (20,0.000000) (24,0.000000) (28,0.000000) (32,0.000000) (36,0.000000) (40,0.000000) (44,0.000000) (48,0.000000) (52,0.000000) (56,0.000000) (60,0.001603) (64,0.002124) (68,0.002797) (72,0.003666) (76,0.004777) (80,0.006193) (84,0.007986) (88,0.010241) (92,0.013063) (96,0.016572) (100,0.020908) (104,0.026234) (108,0.032760) (112,0.040666) (116,0.050205) (120,0.061645) (124,0.075284) (128,0.091445) (132,0.110485) (136,0.132784) (140,0.158754) (144,0.188831) (148,0.223479) (152,0.255658) (156,0.218367) (160,0.176954) (164,0.130711) (168,0.078894) (172,0.020751) (176,0.000000) (180,0.000000) (184,0.000000) (188,0.000000) (192,0.000000) (196,0.000000) (200,0.000000) (204,0.000000) (208,0.000000) (212,0.000000) (216,0.000000) (220,0.000000) (224,0.000000) (228,0.000000) (232,0.000000) (236,0.000000) (240,0.000000) (244,0.000000) (248,0.000000) (252,0.000000) (256,0.000000) (260,0.000000) (264,0.000000) (268,0.000000) (272,0.000000) (276,0.000000) (280,0.000000) (284,0.000000) (288,0.000000) (292,0.000000) (296,0.000000) (300,0.000000) (304,0.000000) (308,0.000000) (312,0.000000) (316,0.000000) (320,0.000000) (324,0.000000) (328,0.000000) (332,0.000000) (336,0.000000) (340,0.020751) (344,0.078894) (348,0.130711) (352,0.176954) (356,0.218367) (360,0.255658) (364,0.223479) (368,0.188831) (372,0.158754) (376,0.132784) (380,0.110485) (384,0.091445) (388,0.075284) (392,0.061645) (396,0.050205) (400,0.040666) (404,0.032760) (408,0.026234) (412,0.020908) (416,0.016572) (420,0.013063) (424,0.010241) (428,0.007986) (432,0.006193) (436,0.004777) (440,0.003666) (444,0.002797) (448,0.002124) (452,0.001603) (456,0.000000) (460,0.000000) (464,0.000000) (468,0.000000) (472,0.000000) (476,0.000000) (480,0.000000) (484,0.000000) (488,0.000000) (492,0.000000) (496,0.000000) (500,0.000000) (504,0.000000) (508,0.000000) };
\addplot[black, thick, dashed] coordinates { (0,0.102740) (4,0.110251) (8,0.118179) (12,0.126537) (16,0.135335) (20,0.144585) (24,0.154295) (28,0.164474) (32,0.175131) (36,0.186270) (40,0.197899) (44,0.210019) (48,0.222635) (52,0.235746) (56,0.249352) (60,0.265054) (64,0.280161) (68,0.295903) (72,0.312313) (76,0.329430) (80,0.347302) (84,0.365986) (88,0.385552) (92,0.406085) (96,0.427684) (100,0.450465) (104,0.474566) (108,0.501729) (112,0.529698) (116,0.559835) (120,0.592481) (124,0.628034) (128,0.666959) (132,0.709795) (136,0.757162) (140,0.809767) (144,0.868412) (148,0.933999) (152,1.000000) (156,1.000000) (160,1.000000) (164,1.000000) (168,1.000000) (172,1.000000) (176,1.000000) (180,1.000000) (184,1.000000) (188,1.000000) (192,1.000000) (196,1.000000) (200,1.000000) (204,1.000000) (208,1.000000) (212,1.000000) (216,1.000000) (220,1.000000) (224,1.000000) (228,1.000000) (232,1.000000) (236,1.000000) (240,1.000000) (244,1.000000) (248,1.000000) (252,1.000000) (256,1.000000) (260,1.000000) (264,1.000000) (268,1.000000) (272,1.000000) (276,1.000000) (280,1.000000) (284,1.000000) (288,1.000000) (292,1.000000) (296,1.000000) (300,1.000000) (304,1.000000) (308,1.000000) (312,1.000000) (316,1.000000) (320,1.000000) (324,1.000000) (328,1.000000) (332,1.000000) (336,1.000000) (340,1.000000) (344,1.000000) (348,1.000000) (352,1.000000) (356,1.000000) (360,1.000000) (364,0.933999) (368,0.868412) (372,0.809767) (376,0.757162) (380,0.709795) (384,0.666959) (388,0.628034) (392,0.592481) (396,0.559835) (400,0.529698) (404,0.501729) (408,0.474566) (412,0.450465) (416,0.427684) (420,0.406085) (424,0.385552) (428,0.365986) (432,0.347302) (436,0.329430) (440,0.312313) (444,0.295903) (448,0.280161) (452,0.265054) (456,0.249352) (460,0.235746) (464,0.222635) (468,0.210019) (472,0.197899) (476,0.186270) (480,0.175131) (484,0.164474) (488,0.154295) (492,0.144585) (496,0.135335) (500,0.126537) (504,0.118179) (508,0.110251) };
\end{axis}
\end{tikzpicture}

%% file: figures/combined_experiments/n3_cyclic_cross_section_PowermLaw_vm1.tex
\begin{tikzpicture}
\begin{axis}[width=\cellsize,height=\cellsize,scale only axis,yticklabels={},xmin=0,xmax=511,ymin=0,ymax=1.1,grid=major,xtick=\empty,ytick={0,0.25,0.5,0.75,1}]
\addplot[red, thick] coordinates { (0,0.000000) (4,0.000000) (8,0.000000) (12,0.000000) (16,0.000000) (20,0.000000) (24,0.000000) (28,0.000000) (32,0.000000) (36,0.000000) (40,0.000000) (44,0.000000) (48,0.000000) (52,0.000000) (56,0.000000) (60,0.000000) (64,0.000000) (68,0.000000) (72,0.000000) (76,0.000000) (80,0.000000) (84,0.000000) (88,0.000000) (92,0.000000) (96,0.000000) (100,0.000000) (104,0.000000) (108,0.001065) (112,0.001534) (116,0.002187) (120,0.003089) (124,0.004318) (128,0.005976) (132,0.008189) (136,0.011109) (140,0.014921) (144,0.019841) (148,0.026121) (152,0.034047) (156,0.043937) (160,0.056135) (164,0.071005) (168,0.088922) (172,0.110251) (176,0.135335) (180,0.164474) (184,0.197899) (188,0.235746) (192,0.278037) (196,0.324652) (200,0.375311) (204,0.429557) (208,0.486752) (212,0.546074) (216,0.606531) (220,0.666977) (224,0.726149) (228,0.782705) (232,0.835270) (236,0.882497) (240,0.923116) (244,0.955997) (248,0.980199) (252,0.995012) (256,1.000000) (260,0.995012) (264,0.980199) (268,0.955997) (272,0.923116) (276,0.882497) (280,0.835270) (284,0.782705) (288,0.726149) (292,0.666977) (296,0.606531) (300,0.546074) (304,0.486752) (308,0.429557) (312,0.375311) (316,0.324652) (320,0.278037) (324,0.235746) (328,0.197899) (332,0.164474) (336,0.135335) (340,0.110251) (344,0.088922) (348,0.071005) (352,0.056135) (356,0.043937) (360,0.034047) (364,0.026121) (368,0.019841) (372,0.014921) (376,0.011109) (380,0.008189) (384,0.005976) (388,0.004318) (392,0.003089) (396,0.002187) (400,0.001534) (404,0.001065) (408,0.000000) (412,0.000000) (416,0.000000) (420,0.000000) (424,0.000000) (428,0.000000) (432,0.000000) (436,0.000000) (440,0.000000) (444,0.000000) (448,0.000000) (452,0.000000) (456,0.000000) (460,0.000000) (464,0.000000) (468,0.000000) (472,0.000000) (476,0.000000) (480,0.000000) (484,0.000000) (488,0.000000) (492,0.000000) (496,0.000000) (500,0.000000) (504,0.000000) (508,0.000000) };
\addplot[green!70!black, thick] coordinates { (0,0.102740) (4,0.110251) (8,0.118179) (12,0.126537) (16,0.135335) (20,0.144585) (24,0.154295) (28,0.164474) (32,0.175131) (36,0.186270) (40,0.197899) (44,0.210019) (48,0.222635) (52,0.235746) (56,0.249352) (60,0.263451) (64,0.278037) (68,0.293106) (72,0.308647) (76,0.324652) (80,0.341108) (84,0.358000) (88,0.375311) (92,0.393022) (96,0.411112) (100,0.429557) (104,0.448332) (108,0.467905) (112,0.487501) (116,0.507447) (120,0.527755) (124,0.548448) (128,0.569568) (132,0.591181) (136,0.613383) (140,0.636307) (144,0.660136) (148,0.685113) (152,0.711562) (156,0.739905) (160,0.770700) (164,0.804683) (168,0.842841) (172,0.886523) (176,0.864665) (180,0.835526) (184,0.802101) (188,0.764254) (192,0.721963) (196,0.675348) (200,0.624689) (204,0.570443) (208,0.513248) (212,0.453926) (216,0.393469) (220,0.333023) (224,0.273851) (228,0.217295) (232,0.164730) (236,0.117503) (240,0.076884) (244,0.044003) (248,0.019801) (252,0.004988) (256,0.000000) (260,0.004988) (264,0.019801) (268,0.044003) (272,0.076884) (276,0.117503) (280,0.164730) (284,0.217295) (288,0.273851) (292,0.333023) (296,0.393469) (300,0.453926) (304,0.513248) (308,0.570443) (312,0.624689) (316,0.675348) (320,0.721963) (324,0.764254) (328,0.802101) (332,0.835526) (336,0.864665) (340,0.886523) (344,0.842841) (348,0.804683) (352,0.770700) (356,0.739905) (360,0.711562) (364,0.685113) (368,0.660136) (372,0.636307) (376,0.613383) (380,0.591181) (384,0.569568) (388,0.548448) (392,0.527755) (396,0.507447) (400,0.487501) (404,0.467905) (408,0.448332) (412,0.429557) (416,0.411112) (420,0.393022) (424,0.375311) (428,0.358000) (432,0.341108) (436,0.324652) (440,0.308647) (444,0.293106) (448,0.278037) (452,0.263451) (456,0.249352) (460,0.235746) (464,0.222635) (468,0.210019) (472,0.197899) (476,0.186270) (480,0.175131) (484,0.164474) (488,0.154295) (492,0.144585) (496,0.135335) (500,0.126537) (504,0.118179) (508,0.110251) };
\addplot[blue, thick] coordinates { (0,0.000000) (4,0.000000) (8,0.000000) (12,0.000000) (16,0.000000) (20,0.000000) (24,0.000000) (28,0.000000) (32,0.000000) (36,0.000000) (40,0.000000) (44,0.000000) (48,0.000000) (52,0.000000) (56,0.000000) (60,0.001845) (64,0.002494) (68,0.003363) (72,0.004528) (76,0.006090) (80,0.008192) (84,0.011035) (88,0.014912) (92,0.020273) (96,0.027853) (100,0.038965) (104,0.056347) (108,0.088840) (112,0.178514) (116,0.490366) (120,0.469157) (124,0.447234) (128,0.424456) (132,0.400630) (136,0.375508) (140,0.348772) (144,0.320023) (148,0.288766) (152,0.254391) (156,0.216158) (160,0.173166) (164,0.124312) (168,0.068237) (172,0.003226) (176,0.000000) (180,0.000000) (184,0.000000) (188,0.000000) (192,0.000000) (196,0.000000) (200,0.000000) (204,0.000000) (208,0.000000) (212,0.000000) (216,0.000000) (220,0.000000) (224,0.000000) (228,0.000000) (232,0.000000) (236,0.000000) (240,0.000000) (244,0.000000) (248,0.000000) (252,0.000000) (256,0.000000) (260,0.000000) (264,0.000000) (268,0.000000) (272,0.000000) (276,0.000000) (280,0.000000) (284,0.000000) (288,0.000000) (292,0.000000) (296,0.000000) (300,0.000000) (304,0.000000) (308,0.000000) (312,0.000000) (316,0.000000) (320,0.000000) (324,0.000000) (328,0.000000) (332,0.000000) (336,0.000000) (340,0.003226) (344,0.068237) (348,0.124312) (352,0.173166) (356,0.216158) (360,0.254391) (364,0.288766) (368,0.320023) (372,0.348772) (376,0.375508) (380,0.400630) (384,0.424456) (388,0.447234) (392,0.469157) (396,0.490366) (400,0.178514) (404,0.088840) (408,0.056347) (412,0.038965) (416,0.027853) (420,0.020273) (424,0.014912) (428,0.011035) (432,0.008192) (436,0.006090) (440,0.004528) (444,0.003363) (448,0.002494) (452,0.001845) (456,0.000000) (460,0.000000) (464,0.000000) (468,0.000000) (472,0.000000) (476,0.000000) (480,0.000000) (484,0.000000) (488,0.000000) (492,0.000000) (496,0.000000) (500,0.000000) (504,0.000000) (508,0.000000) };
\addplot[black, thick, dashed] coordinates { (0,0.102740) (4,0.110251) (8,0.118179) (12,0.126537) (16,0.135335) (20,0.144585) (24,0.154295) (28,0.164474) (32,0.175131) (36,0.186270) (40,0.197899) (44,0.210019) (48,0.222635) (52,0.235746) (56,0.249352) (60,0.265296) (64,0.280531) (68,0.296469) (72,0.313175) (76,0.330743) (80,0.349300) (84,0.369035) (88,0.390223) (92,0.413296) (96,0.438965) (100,0.468522) (104,0.504678) (108,0.557810) (112,0.667549) (116,1.000000) (120,1.000000) (124,1.000000) (128,1.000000) (132,1.000000) (136,1.000000) (140,1.000000) (144,1.000000) (148,1.000000) (152,1.000000) (156,1.000000) (160,1.000000) (164,1.000000) (168,1.000000) (172,1.000000) (176,1.000000) (180,1.000000) (184,1.000000) (188,1.000000) (192,1.000000) (196,1.000000) (200,1.000000) (204,1.000000) (208,1.000000) (212,1.000000) (216,1.000000) (220,1.000000) (224,1.000000) (228,1.000000) (232,1.000000) (236,1.000000) (240,1.000000) (244,1.000000) (248,1.000000) (252,1.000000) (256,1.000000) (260,1.000000) (264,1.000000) (268,1.000000) (272,1.000000) (276,1.000000) (280,1.000000) (284,1.000000) (288,1.000000) (292,1.000000) (296,1.000000) (300,1.000000) (304,1.000000) (308,1.000000) (312,1.000000) (316,1.000000) (320,1.000000) (324,1.000000) (328,1.000000) (332,1.000000) (336,1.000000) (340,1.000000) (344,1.000000) (348,1.000000) (352,1.000000) (356,1.000000) (360,1.000000) (364,1.000000) (368,1.000000) (372,1.000000) (376,1.000000) (380,1.000000) (384,1.000000) (388,1.000000) (392,1.000000) (396,1.000000) (400,0.667549) (404,0.557810) (408,0.504678) (412,0.468522) (416,0.438965) (420,0.413296) (424,0.390223) (428,0.369035) (432,0.349300) (436,0.330743) (440,0.313175) (444,0.296469) (448,0.280531) (452,0.265296) (456,0.249352) (460,0.235746) (464,0.222635) (468,0.210019) (472,0.197899) (476,0.186270) (480,0.175131) (484,0.164474) (488,0.154295) (492,0.144585) (496,0.135335) (500,0.126537) (504,0.118179) (508,0.110251) };
\end{axis}
\end{tikzpicture}

%% file: src7_results.tex
\section{Results}
\label{sec:results}

We evaluate our non-exponential transmittance models in two experimental settings: (i)~\emph{reconstruction from scratch}, where a 3D Gaussian Splatting (3DGS) representation is optimized from COLMAP point clouds and multi-view images, and (ii)~\emph{refinement of existing assets}, where a pre-trained 3DGS scene produced by a conventional rasterization-based renderer is fine-tuned under our ray-tracing pipeline. 
In both cases, we compare three proposed transmittance decay functions --- \texttt{superlinear}, \texttt{linear}, and \texttt{sublinear} --- against the baseline \texttt{exponential} transmittance in the original 3DGS formulation~\cite{kerbl20233d}. The two non-linear transmittance functions are implemented using the \texttt{quadratic} transmittance with $c=-0.5$ (\texttt{sublinear}) and $c=0.5$ (\texttt{superlinear}) respectively. 
%

\subsection{Implementation Details}
\label{sec:implementation}

\paragraph{Renderer.}
Our pipeline is implemented on top of Mitsuba~\cite{Mitsuba3}. Unlike the tile-based rasterizer of \citet{kerbl20233d}, our renderer evaluates Gaussian primitives via ray tracing, following the approach of~\citet{3dgrt2024}. For each pixel, a ray is cast through the scene's bounding volume hierarchy and collects all intersected Gaussians. Each intersected Gaussian $i$ is rendered as a billboard at a single depth $t_i$, located at the peak location of the Gaussian along the ray. The intersected Gaussians are composited front-to-back, with early termination when the accumulated transmittance $\widebar{T}_i$ saturates to zero. 
The maximum number of Gaussians along the ray is set to $N=128$ and all experiments were conducted on a single NVIDIA RTX A6000 GPU. 

For fairness, all results computed using the baseline \texttt{exponential} transmittance are computed through our same pipeline and renderer. Since ray-tracing is usually slower than rasterization-based pipelines, the provided numbers with the baseline are slower than previous works´. However, they serve as a baseline reference in terms of both quality and speed-ups.

\paragraph{Loss function.}
Similar to \citet{kerbl20233d} we use a image-space loss in sRGB space combining the $\ell_1$ and the structural similarity (SSIM) losses, following
\begin{equation}
  \mathcal{L} = (1 - \lambda)\,\mathcal{L}_1 + \lambda\,\mathcal{L}_{\text{SSIM}}\,.
  \label{eq:loss}
\end{equation}

\paragraph{Optimization. }
We use the bounded Adam optimizer~\citep{condor2025don} with per-parameter learning rates modulated by an exponential decay schedule from an initial rate of $1.0$ down to $0.1$ (see~\cref{tab:hyperparams} for all hyperparameters used in the optimization). For the backward phase, we use path-replay backpropagation \citep{vicini2021path}, with the recursive adjoints  derived in \cref{sec:adjoints}. 

\paragraph{Coarse-to-fine scheduling.}
We use a coarse-to-fine strategy to accelerate convergence at early steps. Training starts at $\frac{1}{4}$ resolution and is upsampled to $\frac{1}{2}$ resolution at iteration~$64$, then to full resolution
at iteration~$200$. The refinement pipeline operates at full resolution throughout, since the Gaussians are already well-initialized from the pretrained model.

\paragraph{Adaptive density control.}
We adopt the densification strategy of \citet{kerbl20233d}. Every $100$ iterations, Gaussians with accumulated positional gradient magnitude exceeding $5 \times 10^{-6}$ are candidates for splitting. We split along the longest axis and disable cloning, as we found split-only densification to be more stable for our transmittance models.
Pruning is applied every $200$ iterations, removing Gaussians with opacity below $0.008$ or maximum scale below $10^{-4}$. 
Densification is active only during the first $3{,}000$ iterations, and the maximum primitive count is capped at $200{,}000$.
In the refinement pipeline, both densification and pruning are \emph{disabled}; the Gaussian count is inherited from the pretrained model ($\sim$3.4M primitives for realistic scenes).

\subsection{Reconstruction}
\label{sec:results_recon}
\paragraph{Initialization.}
Gaussians are initialized from a sparse
points obtained using COLMAP \citep{schoenberger2016sfm}, with scales set to each point's nearest-neighbor distance (clamped to maximum scale of $0.02$), identity quaternion rotations, and initial opacity of $0.1$. Colors are initialized from the COLMAP point colors. We additionally scatter $15{,}000$ random Gaussians to help filling unobserved regions.

\paragraph{Training budget.}
To highlight the practical impact of rendering speed on optimization throughput, we impose a \emph{wall-clock time budget} rather than a fixed iteration count.
For reconstruction, each run is limited to $5{,}000$\,s (${\sim}83$\,min) with a maximum of $15{,}000$ iterations.
Within this budget, the \texttt{superlinear} model completes roughly $13{,}000$ iterations, while the \texttt{exponential} baseline manages only ${\sim}3{,}300$---a ${\sim}4\times$ difference in optimizer steps, directly attributable to the lower per-iteration rendering cost.

\begin{figure*}
    \centering
    \includegraphics[width=\linewidth]{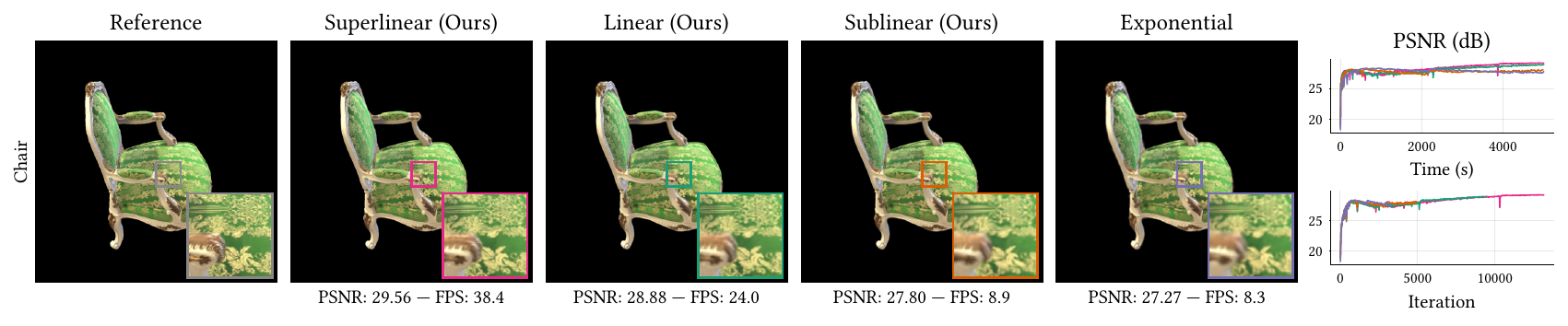}
    \\ \vspace{0.05cm}
    \includegraphics[width=\linewidth]{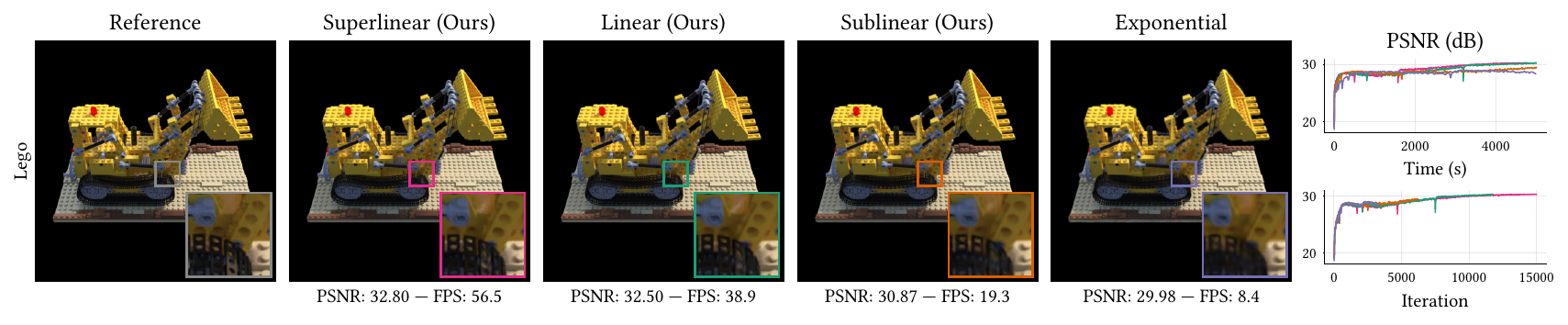}
    \\ \vspace{0.05cm}
    \includegraphics[width=\linewidth]{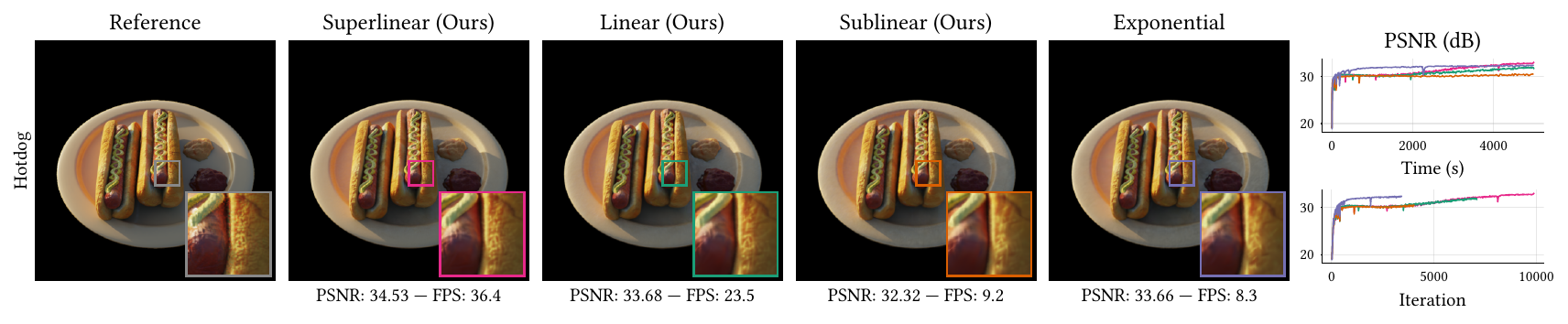}
    \\ \vspace{0.05cm}
    \includegraphics[width=\linewidth]{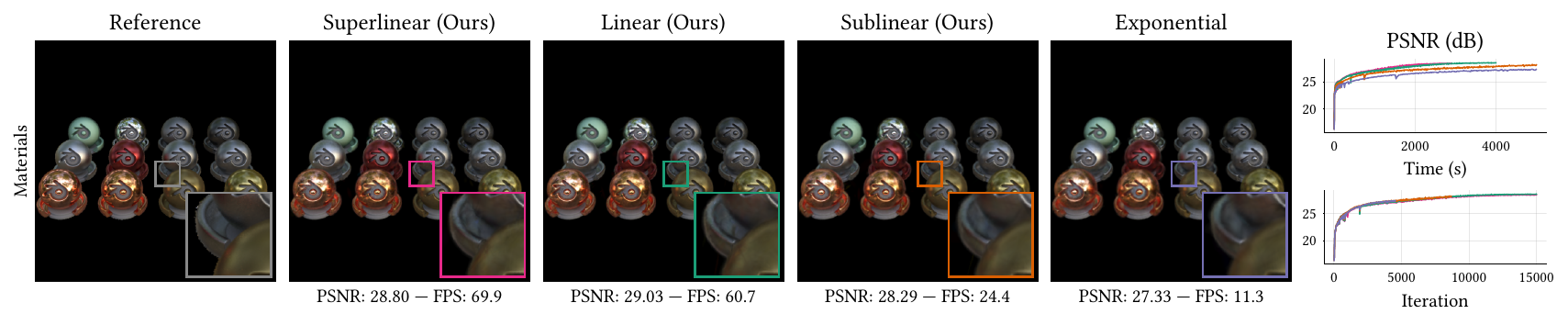}
      \caption{%
        \textbf{Reconstruction results on NeRF synthetic scenes.}
        Each row shows one scene (\emph{Chair}, \emph{Hotdog}, \emph{Lego} and \emph{Materials}), with the ground-truth reference on the left and the reconstructions using each of the four image formation models to the right.
        The rightmost column plots PSNR convergence over wall-clock time (top) and iteration count (bottom).
        Under a fixed time budget, the non-exponential models run significantly more iterations and converge to a higher PSNR, while per-iteration convergence remains comparable across all models.
      }
  \label{fig:reconstruction}
\end{figure*}

\paragraph{Results.}
We evaluate reconstruction quality on four scenes from the NeRF synthetic dataset \citep{mildenhall2020nerf} (\emph{Chair}, \emph{Hotdog}, \emph{Lego}, \emph{Materials}).
Quality and performance metrics are summarized in the top half of \cref{tab:quality,tab:performance}, respectively, and a visual comparison is shown in~\cref{fig:reconstruction}.

As reported in~\cref{tab:quality}, the \texttt{superlinear} model achieves the highest average PSNR ($30.13$\,dB) and SSIM ($0.9705$), and the lowest MSE ($0.0010$), outperforming the exponential baseline ($28.99$\,dB, SSIM $0.9547$, MSE $0.0014$) by over $1$\,dB on average.
The \texttt{linear} model follows closely ($29.88$\,dB, SSIM $0.9677$), while even the \texttt{sublinear} variant matches the exponential baseline in PSNR while producing substantially less overdraw.
This quality advantage is a direct consequence of the reduced rendering and optimization cost: as shown in~\cref{tab:performance}, the \texttt{superlinear} model renders at $50.3$\,FPS on average compared to $9.1$\,FPS for the \texttt{exponential} model---a $\mathbf{5.5\times}$ speedup---with an average overdraw of only $16.2$ ($72.1$ for the baseline).
Under a fixed $5{,}000$\,s training budget, this speed-up allows the optimizer to complete ${\sim}4\times$ more steps, leading to a substantially better-converged reconstruction.

\Cref{fig:reconstruction} provides a qualitative comparison. Each scene is shown as a row, with the ground-truth reference on the left and the four transmittance models to the right. The rightmost column shows PSNR convergence over both wall-clock time and iterations: the non-exponential models converge faster in wall-clock time due to their lower per-iteration cost, while their per-iteration convergence rate remains comparable to the exponential baseline. PSNR and FPS values are annotated below each image.

\subsection{Refinement}
\label{sec:results_refine}

\paragraph{Initialization.}
We start from a pre-trained 3DGS asset, obtained through the implementation of \citet{kerbl20233d} with the standard image formation model, including the parameters of the Gaussians, their opacity and SH color coefficients. 
%
%

\paragraph{Training budget.}
For refinement, the budget is $1{,}800$\,s ($30$\,min) with a maximum of $1{,}000$ iterations; the faster methods comfortably finish all iterations, while the exponential baseline may not.

\begin{figure*}
    \centering
    \includegraphics[width=\linewidth]{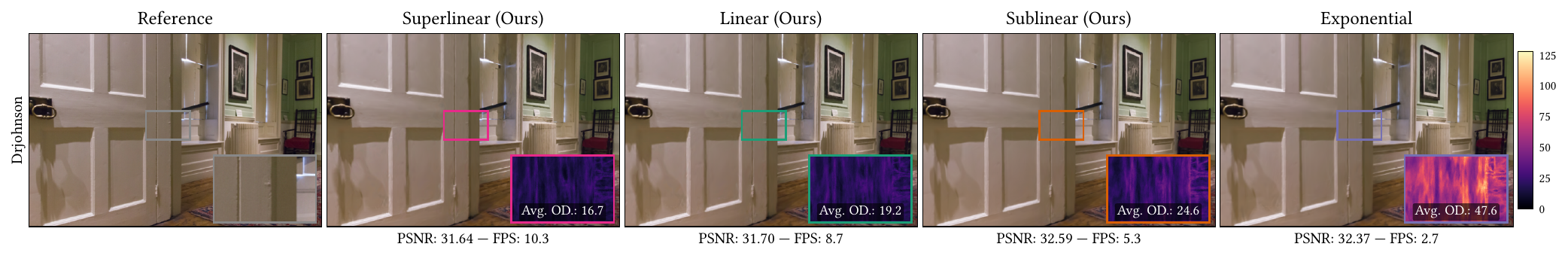}
    \\ \vspace{0.05cm}
    \includegraphics[width=\linewidth]{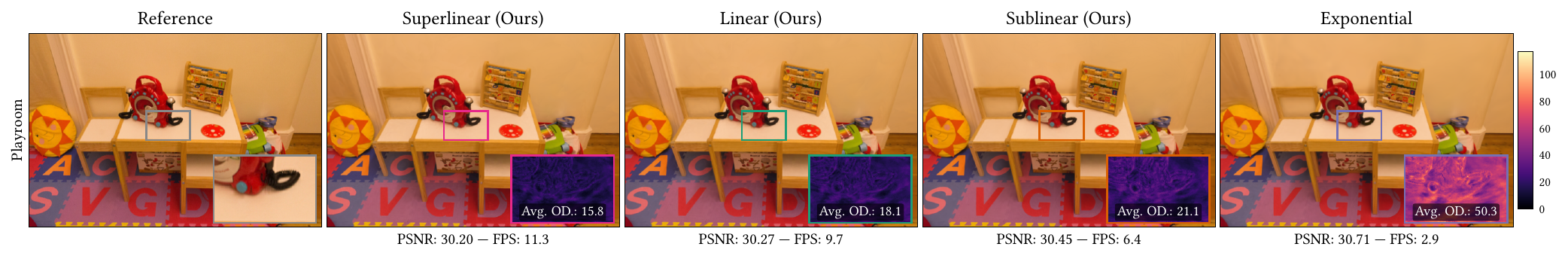}
    \\ \vspace{0.05cm}
    \includegraphics[width=\linewidth]{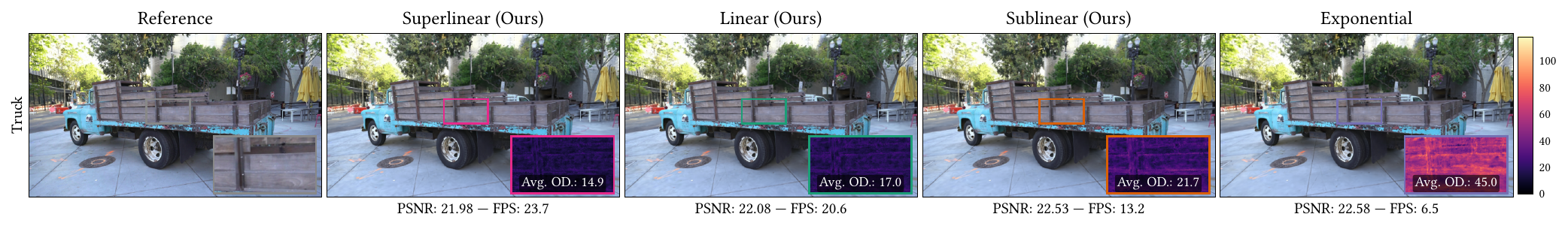}
    \\ \vspace{0.05cm}
    \includegraphics[width=\linewidth]{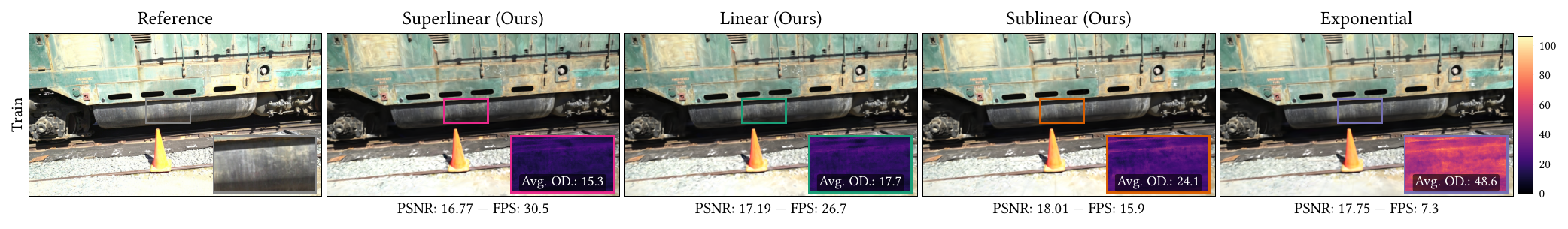}
    \caption{%
      \textbf{Refinement results on real scenes.}
      Each row shows one scene (\emph{Dr\,Johnson}, \emph{Playroom},
      \emph{Train} and \emph{Truck}), with the ground-truth reference on the left and the reconstructions using each of the four image formation models to the right.
      Insets show the per-pixel overdraw count. The non-exponential models maintain comparable visual quality comparable to the
      exponential baseline while substantially reducing overdraw,
      which results into $3$--$4\times$ rendering speedups.
    }
  \label{fig:refinement}
\end{figure*}

\paragraph{Results. }
For the refinement experiment, we use two indoor scenes (\emph{Dr\,Johnson} and \emph{Playroom} \cite{barron2022mipnerf360}), and two outdoor scenes (\emph{Train} and \emph{Truck} \cite{Knapitsch2017}).
We start from pretrained models using the original splatting-based renderer~\cite{kerbl20233d} ($30{,}000$ iterations, exponential transmittance). 
We fine-tune these assets using our ray-tracing pipeline, continuing the optimization for up to $1{,}000$ iterations (or $30$\,min). Densification and pruning are disabled; only the existing Gaussian parameters are fine-tuned. \Cref{tab:quality,tab:performance} (bottom half) show the reconstruction quality and performance metrics for each scene, while \cref{fig:refinement} presents a qualitative comparison.

The quality metrics in~\cref{tab:quality} confirm that our non-exponential models maintain image fidelity after refinement. 
The \texttt{sublinear} model achieves the best average PSNR ($25.90$\,dB)
and SSIM ($0.8407$), closely followed by the exponential baseline
($25.85$\,dB, SSIM $0.8399$). The \texttt{linear} and \texttt{superlinear} variants are less than $1$\,dB worse on average, demonstrating that the Gaussians originally trained under exponential transmittance can be successfully adapted to non-exponential models with minimal quality loss.

The performance gains remain substantial even in this refinement
setting (\cref{tab:performance}). The \texttt{superlinear} model achieves $18.9$\,FPS on average versus $4.8$\,FPS for the exponential baseline---a $\mathbf{3.9\times}$ speedup---while reducing average overdraw from $47.9$ to $15.7$. \Cref{fig:refinement} shows that the refined images are visually similar across methods, while the overdraw insets clearly reveal the efficiency advantage of the non-exponential models.





\begin{table}[t]
  \centering
  \caption{Hyperparameters used for the reconstruction and refinement experiments.}
  \label{tab:hyperparams}
  \small
  \setlength{\tabcolsep}{4pt}
  \begin{tabular}{lcc}
    \toprule
    Parameter & Recon. & Refine. \\
    \midrule
    Max iterations & 15\,000 & 1\,000 \\
    Wall-clock budget & 5\,000\,s & 1\,800\,s \\
    Max primitives & 200\,000 & (inherited) \\
    Centers LR & 0.13 & 0.015 \\
    Scales LR & 0.08 & 0.03 \\
    Quaternions LR & 0.45 & 0.1 \\
    Opacity LR & 1.0 & 1.0 \\
    SH coefficients LR & 2.0 & 0.2 \\
    LR schedule & Exp.\ decay & Exp.\ decay \\
    SSIM weight $\lambda$ & 0.2 & 0.2 \\
    Coarse-to-fine & $4\!\times\!\to\!2\!\times\!\to\!1\!\times$ & Disabled \\
    Densification interval & 100 it. & Disabled \\
    Pruning interval & 200 it. & Disabled \\
    SH degree & 1 & (inherited) \\
    \bottomrule
  \end{tabular}
\end{table}

\begin{table*}[t]
  \centering
  \caption{Image quality metrics across reconstruction and refinement experiments.}
  \label{tab:quality}
  \small
  \setlength{\tabcolsep}{4pt}
  \begin{tabular}{l|ccc|ccc|ccc|ccc}
    \toprule
     & \multicolumn{3}{c}{Superlinear (Ours)} & \multicolumn{3}{c}{Linear (Ours)} & \multicolumn{3}{c}{Sublinear (Ours)} & \multicolumn{3}{c}{Exponential} \\
    \cmidrule(lr){2-4} \cmidrule(lr){5-7} \cmidrule(lr){8-10} \cmidrule(lr){11-13}
    Scene & PSNR$\uparrow$ & MSE$\downarrow$ & SSIM$\uparrow$ & PSNR$\uparrow$ & MSE$\downarrow$ & SSIM$\uparrow$ & PSNR$\uparrow$ & MSE$\downarrow$ & SSIM$\uparrow$ & PSNR$\uparrow$ & MSE$\downarrow$ & SSIM$\uparrow$ \\
    \midrule
    \multicolumn{13}{l}{\textit{Reconstruction}} \\
    \midrule
    Chair & \textbf{29.16} & \textbf{0.0012} & \textbf{0.9721} & 28.85 & 0.0013 & 0.9654 & 28.13 & 0.0015 & 0.9530 & 27.72 & 0.0017 & 0.9471 \\
    Hotdog & \textbf{32.77} & \textbf{0.0005} & \textbf{0.9777} & 32.01 & 0.0006 & 0.9735 & 30.52 & 0.0009 & 0.9669 & 32.39 & 0.0006 & 0.9731 \\
    Lego & \textbf{30.26} & \textbf{0.0009} & \textbf{0.9748} & 30.19 & 0.0010 & 0.9731 & 29.44 & 0.0011 & 0.9622 & 28.61 & 0.0014 & 0.9536 \\
    Materials & 28.33 & 0.0015 & 0.9573 & \textbf{28.47} & \textbf{0.0014} & \textbf{0.9589} & 28.04 & 0.0016 & 0.9532 & 27.26 & 0.0019 & 0.9448 \\
    \textit{Average} & \textbf{30.13} & \textbf{0.0010} & \textbf{0.9705} & 29.88 & 0.0011 & 0.9677 & 29.03 & 0.0013 & 0.9588 & 28.99 & 0.0014 & 0.9547 \\
    \midrule\midrule
    \multicolumn{13}{l}{\textit{Refinement}} \\
    \midrule
    Drjohnson & 31.64 & 0.0007 & 0.9490 & 31.70 & 0.0007 & 0.9496 & \textbf{32.59} & \textbf{0.0006} & \textbf{0.9578} & 32.37 & 0.0006 & 0.9574 \\
    Playroom & 30.20 & 0.0010 & 0.9096 & 30.27 & 0.0009 & 0.9109 & 30.45 & 0.0009 & 0.9152 & \textbf{30.71} & \textbf{0.0008} & \textbf{0.9195} \\
    Train & 16.77 & 0.0210 & 0.6332 & 17.19 & 0.0191 & 0.6433 & \textbf{18.01} & \textbf{0.0158} & \textbf{0.6839} & 17.75 & 0.0168 & 0.6741 \\
    Truck & 21.98 & 0.0063 & 0.7778 & 22.08 & 0.0062 & 0.7829 & 22.53 & 0.0056 & 0.8062 & \textbf{22.58} & \textbf{0.0055} & \textbf{0.8087} \\
    \textit{Average} & 25.15 & 0.0073 & 0.8174 & 25.31 & 0.0067 & 0.8217 & \textbf{25.90} & \textbf{0.0057} & \textbf{0.8407} & 25.85 & 0.0059 & 0.8399 \\
    \bottomrule
  \end{tabular}
\end{table*}

\begin{table*}[t]
  \centering
  \caption{Rendering performance across reconstruction and refinement experiments.}
  \label{tab:performance}
  \small
  \setlength{\tabcolsep}{4pt}
  \begin{tabular}{l|ccc|ccc|ccc|ccc}
    \toprule
     & \multicolumn{3}{c}{Superlinear (Ours)} & \multicolumn{3}{c}{Linear (Ours)} & \multicolumn{3}{c}{Sublinear (Ours)} & \multicolumn{3}{c}{Exponential} \\
    \cmidrule(lr){2-4} \cmidrule(lr){5-7} \cmidrule(lr){8-10} \cmidrule(lr){11-13}
    Scene & FPS$\uparrow$ & Speedup$\uparrow$ & Avg.\ OD$\downarrow$ & FPS$\uparrow$ & Speedup$\uparrow$ & Avg.\ OD$\downarrow$ & FPS$\uparrow$ & Speedup$\uparrow$ & Avg.\ OD$\downarrow$ & FPS$\uparrow$ & Speedup$\uparrow$ & Avg.\ OD$\downarrow$ \\
    \midrule
    \multicolumn{13}{l}{\textit{Reconstruction}} \\
    \midrule
    Chair & \textbf{38.4} & \textbf{4.63$\times$} & \textbf{17.8} & 24.0 & 2.89$\times$ & 20.8 & 8.9 & 1.07$\times$ & 34.2 & 8.3 & 1.00$\times$ & 79.0 \\
    Hotdog & \textbf{36.4} & \textbf{4.39$\times$} & \textbf{18.7} & 23.5 & 2.83$\times$ & 21.0 & 9.2 & 1.11$\times$ & 35.4 & 8.3 & 1.00$\times$ & 73.2 \\
    Lego & \textbf{56.5} & \textbf{6.73$\times$} & \textbf{13.9} & 38.9 & 4.63$\times$ & 18.2 & 19.3 & 2.30$\times$ & 28.0 & 8.4 & 1.00$\times$ & 70.3 \\
    Materials & \textbf{69.9} & \textbf{6.19$\times$} & \textbf{14.6} & 60.7 & 5.37$\times$ & 16.8 & 24.4 & 2.16$\times$ & 26.6 & 11.3 & 1.00$\times$ & 65.9 \\
    \textit{Average} & \textbf{50.3} & \textbf{5.49$\times$} & \textbf{16.2} & 36.8 & 3.93$\times$ & 19.2 & 15.5 & 1.66$\times$ & 31.0 & 9.1 & 1.00$\times$ & 72.1 \\
    \midrule\midrule
    \multicolumn{13}{l}{\textit{Refinement}} \\
    \midrule
    Drjohnson & \textbf{10.3} & \textbf{3.81$\times$} & \textbf{16.7} & 8.7 & 3.22$\times$ & 19.2 & 5.3 & 1.96$\times$ & 24.6 & 2.7 & 1.00$\times$ & 47.6 \\
    Playroom & \textbf{11.3} & \textbf{3.90$\times$} & \textbf{15.8} & 9.7 & 3.34$\times$ & 18.1 & 6.4 & 2.21$\times$ & 21.1 & 2.9 & 1.00$\times$ & 50.3 \\
    Train & \textbf{30.5} & \textbf{4.18$\times$} & \textbf{15.3} & 26.7 & 3.66$\times$ & 17.7 & 15.9 & 2.18$\times$ & 24.1 & 7.3 & 1.00$\times$ & 48.6 \\
    Truck & \textbf{23.7} & \textbf{3.65$\times$} & \textbf{14.9} & 20.6 & 3.17$\times$ & 17.0 & 13.2 & 2.03$\times$ & 21.7 & 6.5 & 1.00$\times$ & 45.0 \\
    \textit{Average} & \textbf{18.9} & \textbf{3.89$\times$} & \textbf{15.7} & 16.4 & 3.35$\times$ & 18.0 & 10.2 & 2.09$\times$ & 22.9 & 4.8 & 1.00$\times$ & 47.9 \\
    \bottomrule
  \end{tabular}
\end{table*}

%% file: src6_discussion.tex
\section{Discussion}

We have seen experimentally that our faster-than-exponential transmittance models consistently achieve $3$--$5\times$ rendering speed-ups with small-to-negligible impact on image quality, even improving quality in the reconstruction case by enabling more optimization iterations within the same time budget.
The quadratic \texttt{superlinear} model offers the best trade-off between speed and quality: It consistently delivers the highest FPS and lowest overdraw while remaining highly competitive in terms of image quality. 
The \texttt{linear} model provides a balanced middle ground, while the \texttt{sublinear} variant sacrifices some speed for the closest quality match to the exponential baseline---making it the safest drop-in replacement while still resulting in a significant speed-up.

\paragraph{Limitations and Future Work}
One limitation of our work is that we have only experimented our transmittance models with ray-tracing, which limits the comparison with state-of-the-art Gaussian splatters. However, our new non-exponential models can be directly ported to rasterization, which would still benefit from the significant reduction in overdraws using faster-than-exponential transmittance models, though sorting would still be required. Note however that this performance gain is only possible if early termination when transmittance saturates is available. 

More conceptually, order-independent stochastic transparency techniques \citep{kheradmand2025stochasticsplats} are only available for exponential transmittance, where the binary Russian roulette applied on the splats differential extinction probability is independent from previous splats. See \cref{sec:stochastic_3dgs} for a more thorough explanation. This is a similar problem to delta tracking in heterogeneous non-exponential media \citep{bitterli2018radiative}, and requires further investigation. 

Finally, we have assumed a fixed pre-defined transmittance during optimization. Including the transmittance parameters in the optimization, even adaptively per Gaussian following the approach of \citet{vicini2021non}, might increase the reconstruction quality, by giving additional degrees of freedom in the image formation model. 





\paragraph{Relationship with alpha blending}
Our image formation model defined by \cref{eq:generalized_gaussian,eq:nexp_p} offers a physically-grounded form for deriving new alpha blending operators, with any arbitrary decay function. This directly relates with the physical interpretation provided by \citet{glassner2015interpreting} of alpha and compositing, and in particular on how blending two semitransparent layers would depend on the correlation between these two layers. For example, as described in \cref{fig:correlation}, the linear attenuation is a form of perfectly negative correlation between scatterers. Further exploring physically-based transmittance functions capable to model the whole range of inter-matter correlation (from negative to positive) is still an open problem.

\paragraph{Conclusion}
We have generalized 3D Gaussian splatting to support non-exponential transmittance regimes. From a theoretical point of view, this has helped to link 3DGS with continuous volume rendering. From the family of non-exponential 3DGS models we have proposed, we have explored the usability of the \texttt{quadratic} transmittance, analyzing the performance of \texttt{superlinear}, \texttt{linear} and \texttt{sublinear} decays on radiance field reconstruction. Our results show that these new 3DGS models achieve comparable quality to their classic exponential counterpart, while achieving $3-4\times$ less overdraws, directly translated on significant speed-up during rendering and training.

%% file: srcA_3dgs2rte.tex
\section{3D Gaussian splatting as the RTE}
\label{sec:3dgs2rte}
Here we derive the radiative transfer equation (RTE) for emissive media (\cref{eq:rte}) from the Gaussian splatting image formation model (\cref{eq:gaussian_splatting}). For that, we will start from the inner product term, that accounts for the transparency from previous splats, and define it as the transmittance from $\px$ to the point at splat $\px_i$ as
\begin{equation}
T_i(\px, \px_i) = \prod_{j<i} \left(1-\alpha_j(\px_j)\right).  
\end{equation}
Now, using the continuous definition of the medium density in \cref{eq:sigma_and_q}, and by taking $M$ small steps, we can rewrite $T_i(\px,\px_i)$ as
\begin{equation}
    T(\px, \px_i) = \prod_{k=0}^{M} (1-\sigma(\px_k) \cdot\Delta s),
    \label{eq:3dgs_transmittance}
\end{equation}
with $\px_k = \px + \bomega \cdot \Delta s\cdot k$, and $\Delta s = \frac{t_i}{M}$ with $t_i=|\px_i - \px|$.
Now, following the derivations from Kostinski~\cite{kostinski2001extinction}, we apply logarithms to both sides and use a first order approximation on the right to get
\begin{align*}
    \log\left(T(\px, \px_i)\right) & = \log\left(\prod_{k=0}^{M} \sigma(\px_k)\cdot\Delta s\right) \\
    & \approx -\sum_{k=0}^{M} \sigma(\px_k)\cdot\Delta s,
\end{align*}
which by taking the limit $\Delta s \to 0$, applying exponentials to both sides gives
\begin{equation*}
    T(\px, \px_i) = \exp \left(-\int_0^{t_i} \sigma(\px_s)\,d s\right),
\end{equation*}
which is the heterogeneous transmittance predicted by the Beer-Boulder-Lambert law.

Now, using the source and density definitions in \cref{eq:sigma_and_q}, plugging in $T(\px, \px_i)$, and taking $M$ small steps until a maximum distance $t_{\max}$, we can rewrite \cref{eq:gaussian_splatting} as
\begin{equation*}
L_0(\px, \bomega) = \sum_{j=0}^{M} Q(\px_k,\omega)\cdot \sigma(\px_k) \cdot T(\px,\px_k)\cdot\Delta s,
\end{equation*}
which by again taking the limit $\Delta s \to 0$ and setting $t_{\max}=\infty$ results into the \cref{eq:rte}:
\begin{equation}
L_0(\px, \bomega) = \int_{0}^{\infty} Q(\px_k,\omega)\cdot \sigma(\px_k) \cdot T(\px,\px_k) \,ds.
\end{equation}

\section{Transmittance as a Russian-roulette process}
\label{sec:stochastic_3dgs}
\citet{kheradmand2025stochasticsplats} leveraged stochastic transparency \citep{enderton2010stochastic} for avoiding sorting in 3DGS at the cost of introducing variance. For that, they transformed \cref{eq:gaussian_splatting} into a recursive estimator:
\begin{equation}
    L_i = E_i \cdot \alpha_i + (1-\alpha_i) \cdot L_{i+1},
    \label{eq:oi_3dgs}
\end{equation}
which can be solved stochastically via Russian roulette, by choosing whether we draw splat $i$ with probability $\alpha_i$, or if on the other hand $L_{i+1}$ is drawn. This allows to avoid sorting splats when rasterizing. 

This approach is still valid within the framework of the generalized non-exponential light transport: By writing \cref{eq:3dgs_transmittance} using the differential extinction probability (\cref{eq:gbe_sigma}) we get
\begin{equation}
    T(\px, \px_i) = \prod_{k<0}^M \left(1-\sigma(\px_k,\tau_k) \cdot \Delta s\right) = \prod_{k<0}^M \left(1-\frac{p(\tau_k)}{T(\tau_k)} \cdot \Delta s\right).
    \label{eq:gen3dgs_transmittance}
\end{equation}

\begin{proof}
Starting with \cref{eq:gen3dgs_transmittance} and applying logarithms in both sides, using a first-order approximation of the logarithm, we get
\begin{align*}    
\log(T(\px,\px_i)) & = \log\left( \prod_{k<0}^M \left(1-\frac{p(\tau_k)}{T(\tau_k)}\cdot \Delta s\right)\right) \\
&= \sum_{k<0}^M \left(\log\left( 1-\frac{p(\tau_k)}{T(\tau_k)}\cdot \Delta s\right)\right) \\
&\approx \sum_{k<0}^M \left(-\frac{p(\tau_k)}{T(\tau_k)}\cdot \Delta s\right).
\end{align*}
Finally, taking the limit $\Delta s\to0$, and using the relationship between $p(\tau)$ and $T(\tau)$ (\cref{eq:gbe_sigma}), and integrating we get
\begin{align*}    
\log(T(\px,\px_i)) & = \int_0^{t_i} -\frac{p(\tau_s)}{T(\tau_s)} ds \\
&=  \int_0^{t_i} -\frac{-\frac{d}{ds}T(\tau_s)}{T(\tau_s)} ds \\
&= \log(T(\px,\px_i)).
\end{align*}
\end{proof}

Now, to build the recursive form of our generalized 3DGS from the transmittance defined in \cref{eq:gen3dgs_transmittance}, we discretize the medium in individual splats, which gives us
\begin{equation}
    L_i = E_i \cdot \frac{\widebar{p}_i}{\widebar{T}_i} + \left(1-\frac{\widebar{p}_i}{\widebar{T}_i}\right) \cdot L_{i+1}.
    \label{eq:oi_3dgs}
\end{equation}
Unfortunately, except for the case of exponential transmittance where $\frac{\widebar{p}_i}{\widebar{T}_i} = \alpha_i$, the differential extinction probability defined by $\frac{\widebar{p}_i}{\widebar{T}_i}$ is a function of the traversed optical depth (i.e., of all previous splats). Thus, while stochastically we can early terminate the recursion, sorting and evaluating the previous splats is still required.

%% file: srcB_adjoints.tex
\section{Adjoints}
\label{sec:adjoints}
In order to use path-replay backpropagation for efficient front-to-back differentiable rendering, we cannot completely rely on automatic differentiation for computing the adjoint phase recursively. Here we derive the adjoints for both the \texttt{linear} and \texttt{quadratic} image formation models. 
Note that the former is a special case of the \texttt{quadratic} transmittance: We include it explicitly since given its simpler form it allows for a more efficient implementation. 

In order to compute the adjoints, we will first plug \cref{eq:nexp_p} in \cref{eq:generalized_gaussian}, and rewrite $L_0$ as
\begin{align*}
    L_0 & = \sum_{i<k} E_i\cdot p_i + E_k \cdot(1-\sum_{i<k} p_i) + L_b \cdot (1-\sum_{i=1}^N p_i) \\
    & = \sum_{i<k} (E_i - E_k - L_b)\cdot p_i + E_k + L_b.
\end{align*}
Note that for the background contribution $L_b$ to be visible, transmittance cannot saturate so in practice we can assume that when $k=N$, then $E_k = L_b$, and we can remove $L_b$ from $L_0$, leaving
\begin{align}
L_0 & = \sum_{i<k} (E_i - E_k)\cdot p_i + E_k \nonumber \\
&=\sum_{i<k} \Delta E_i \cdot p_i + E_k.
\label{eq:generalized_gaussians_gen}
\end{align}

\subsection{Linear transmittance}
Plugging the $p_{i<k}$ term for linear transmittance into \cref{eq:generalized_gaussians_gen} we get
\begin{align}
L_0 =\sum_{i<k} \Delta E_i \cdot p_i + E_k,
\label{eq:linear_3dgs}
\end{align}
with derivative
\begin{align*}
    \delta L_0 
    &= \sum_{i<k} \delta \left(\Delta E_i\cdot\alpha_i\right) + \delta E_k \\
    &= \sum_{i<k} \left(\delta E_i \cdot\alpha_i + E_i\cdot\delta \alpha_i - \delta E_k\cdot\alpha_i - E_k \cdot \delta \alpha_i \right) + \delta E_k \\
    &= \sum_{i<k} \bigg(\delta (E_i \cdot \alpha_i) - E_k \cdot \delta \alpha_i\bigg) + \delta E_k \cdot(1 - \sum_{i<k} \alpha_i),
\end{align*}
which can be written recursively as
\begin{equation}
\delta L_i = \begin{cases}
      \bigg(\delta (E_i \cdot \alpha_i) - E_k \cdot \delta \alpha_i\bigg) + \delta L_{i+1} & i < k \\
      \delta E_k \cdot \widebar{T}_k & i = k \\
      0  & \text{otherwise,} \\
    \end{cases}
\end{equation}
where $\widebar{T}_k$ can be computed from previous iterations, and $E_k$ needs to be computed in the forward pass and passed as input to the adjoint pass.

\subsection{Quadratic transmittance}
Plugging the $p_{i<k}$ term for quadratic transmittance into \cref{eq:generalized_gaussians_gen} we get
\begin{align}
\label{eq:quadratic_3dgs}
L_0 &= \sum_{i<k} \Delta E_i \cdot \alpha_i \cdot (1 + c\cdot\sum_{j<i}\alpha_j) + E_k \\
& = \sum_{i<k} \Delta E_i \cdot \alpha_i + c \cdot \sum_{i<k} \left(\Delta E_i \cdot \alpha_i \cdot\sum_{j<i}\alpha_j\right) + E_k. \nonumber
\end{align}
The second sum can be expanded as
\begin{align*}
    \label{eq:quadratic_second_sum}
    \sum_{i=0}^N  \Delta E_i \cdot \alpha_i \cdot \sum_{j=0}^{i-1} \alpha_j 
        &= 0 \cdot \Delta E_0 \cdot \alpha_0 + \alpha_0 \cdot \Delta E_1 \cdot \alpha_1 \\
        & + (\alpha_0 + \alpha_1) \cdot \Delta E_2 \cdot \alpha_2 + ... \\
        &= \alpha_0 \cdot (\Delta E_1 \cdot \alpha_1 + \Delta E_2 \cdot \alpha_2 + ...) \\
        & + \alpha_1 \cdot (\Delta E_2 \cdot \alpha_2  + \Delta E_3 \cdot \alpha_3 + ...) + ...\\
        &= \sum_{i=0}^N \left( \alpha_i \cdot \sum_{j=i+1}^{N} \Delta E_j \cdot \alpha_j \right).
\end{align*}
%
%
%
%
%
%
Now taking the derivative using the produce rule, and using the definition of the discrete optical depth $\widebar{\tau}_i = \sum_{j<i} \alpha_j$ we get
\begin{align*}
    \delta L_0
        &= \sum_{i<k} \delta (\Delta E_i \cdot \alpha_i) + c\cdot \sum_{i<k} \left( \delta (\Delta E_i \cdot \alpha_i) \cdot \sum_{j=0}^{i-1} \alpha_j \right) \\
        & \quad + c\cdot \sum_{i<k} \left( \Delta E_i \cdot \alpha_i \cdot \sum_{j=0}^{i-1} \delta \alpha_j \right) + \delta E_k \\
        &= \sum_{i<k} \delta (\Delta E_i \cdot \alpha_i) + c\cdot\sum_{i<k} \left( \delta (\Delta E_i \cdot \alpha_i) \cdot \sum_{j=0}^{i-1} \alpha_j \right) \\
        & \quad + c\cdot\sum_{i<k} \left( \delta \alpha_i \cdot \sum_{j=i+1}^{k-1} \Delta E_j \cdot \alpha_j \right) + \delta E_k \\
        &= \sum_{i<k} \big( \delta (\Delta E_i \cdot \alpha_i) + c\cdot\delta (\Delta E_i \cdot \alpha_i) \cdot \widebar{\tau}_i \\
        & \qquad + c\cdot\delta \alpha_i \cdot \sum_{j=i+1}^{k-1} \Delta E_j \cdot \alpha_j \big) + \delta E_k \\
        &= \sum_{i<k} \bigg( \delta (\Delta E_i \cdot \alpha_i) \cdot \left(1 + c\cdot\widebar{\tau}_i\right) + c\cdot \delta \alpha_i \cdot \Theta_i \bigg) + \delta E_k,
\end{align*}
with $\Theta_i$ defined recursively as
\begin{equation*}
\Theta_i =
\begin{cases}
  \sum_{j=1}^{k-1} \Delta E_j \cdot \alpha_j
  & i = 0 \\
  \Theta_{i-1} - \Delta E_i \cdot \alpha_i
  & i > 0. \\
\end{cases}
\end{equation*}
Further expanding $\Delta E_i$ in $\delta L_0$ we get
\begin{align*}
\delta L_0
        &= \sum_{i<k} \bigg( \big(\delta (E_i \cdot \alpha_i) - \delta E_k\cdot \alpha_i - E_k \cdot \delta\alpha_i\big)\cdot \left(1 + c\cdot\widebar{\tau}_i\right) 
        \\ &\qquad+ c\cdot \delta \alpha_i \cdot \Theta_i \bigg) + \delta E_k \\
        &= \sum_{i<k} \bigg( \big(\delta (E_i \cdot \alpha_i) - E_k \cdot \delta\alpha_i\big)\cdot \left(1 + c\cdot\widebar{\tau}_i\right) 
        - \delta E_k\cdot \alpha_i \cdot \left(1 + c\cdot\widebar{\tau}_i\right) 
        \\ &\qquad + c\cdot \delta \alpha_i \cdot \Theta_i \bigg) + \delta E_k \\
        &= \sum_{i<k} \bigg( \big(\delta (E_i \cdot \alpha_i) - E_k \cdot \delta\alpha_i \big) \cdot \left(1 + c\cdot\widebar{\tau}_i\right) 
        + c\cdot \delta \alpha_i \cdot \Theta_i \bigg) \\ &\quad + \delta E_k \cdot\left(1-\sum_{i<k} \alpha_i \cdot \left(1 + c\cdot\widebar{\tau}_i\right)\right) \\
        &= \sum_{i<k} \bigg( \big(\delta (E_i \cdot \alpha_i) - E_k \cdot \delta\alpha_i \big) \cdot \left(1 + c\cdot\widebar{\tau}_i\right) 
        + c\cdot \delta \alpha_i \cdot \Theta_i \bigg) + \delta E_k \cdot p_k.
\end{align*}
With this, we can obtain a recursive formula for the adjoint:
\begin{equation}
    \delta L_i =
    \begin{cases}
      \big(\delta (E_i \cdot \alpha_i) - E_k \cdot \delta\alpha_i \big) \cdot \left(1 + c\cdot\widebar{\tau}_i\right) + c\cdot \delta \alpha_i \cdot \Theta_i + \delta L_{i+1} & i < k \\
      \delta E_k \cdot \widebar{T}_k & i = k \\
      0  & \text{otherwise.} \\
    \end{cases}
\end{equation}
Both $\widebar{\tau}_i$ and $\widebar{T}_k$ can be computed recursively from previous iterations in the backward pass, while $\Theta_i$ requires $\Theta_0$, which can be precomputed in the forward pass, and passed as input to the adjoint pass with $E_k$.